\documentclass[useAMS,usenatbib]{mn2e}
\usepackage{graphicx}
\usepackage{mathptmx}
\newcommand{\threetotwo}{$J=3\to 2$}
\newcommand{\twotoone}{$J=2\to 1$}
\newcommand{\onetozero}{$J=1\to 0$}
\newcommand{\thirteenco}{$^{13}$CO}
\newcommand{\twelveco}{$^{12}$CO}
\newcommand{\ceighteeno}{C$^{18}$O}
\newcommand{\ammonia}{NH$_3$}
\newcommand{\ngc}{NGC\,1333}

\newcommand{\htwo}{H$_2$}
\newcommand{\msun}{M$_\odot$}
\newcommand{\mjybeam}{mJy\,beam$^{-1}$}
\newcommand{\kkms}{K\,km\,s$^{-1}$}
\newcommand{\kms}{km\,s$^{-1}$}

\newcommand{\wfcam}{WFCAM}

\newcommand{\jcmt}{JCMT}

\newcommand{\scuba}{SCUBA}

\newcommand{\iram}{IRAM}
\newcommand{\harp}{HARP}

\newcommand{\sed}{SED}
\newcommand{\lte}{LTE}
\newcommand{\tdyn}{$\tau_\mathrm{d}$}
\newcommand{\tbol}{$T_\mathrm{bol}$}
\newcommand{\fco}{$F_\mathrm{CO}$}
\newcommand{\fcostar}{$F_\mathrm{CO}^*$}
\newcommand{\vmax}{$v_\mathrm{max}$}
\newcommand{\fcounit}{\msun\,\kms\,yr$^{-1}$}
\newcommand{\apj}{ApJ}
\newcommand{\apjl}{ApJ}
\newcommand{\apjs}{ApJS}
\newcommand{\apjss}{ApJS}
\newcommand{\aap}{A\&A}
\newcommand{\aaps}{A\&AS}
\newcommand{\mnras}{MNRAS}
\newcommand{\aj}{AJ}
\newcommand{\araa}{ARA\&A}

\title[Molecular outflows in the Perseus molecular cloud]{A submillimetre survey of the kinematics of the Perseus molecular
cloud -- II. Molecular outflows}
\author[E. I. Curtis et al.]{Emily~I.~Curtis,$^{1,2}$\thanks{E-mail:
    e.curtis@mrao.cam.ac.uk}, John~S.~Richer$^{1,2}$\thanks{E-mail: jsr10@cam.ac.uk},
  Jonathan~J.~Swift$^{3}$ and Jonathan~P.~Williams$^{3}$\\
$^{1}$Astrophysics Group, Cavendish Laboratory, 19 J. J. Thomson
Avenue, Cambridge, CB3 0HE\\
$^{2}$Kavli Institute for Cosmology, c/o Institute of Astronomy,
  University of Cambridge, Madingley Road, Cambridge, CB3 0HA\\
$^{3}$Institute for Astronomy, 2680 Woodlawn Drive, Honolulu, HI
96822-1897, USA}
\begin{document}

\date{Accepted 2010 June 16; Received 2010 June 15; in original form 2010 May 7}

\pagerange{\pageref{firstpage}--\pageref{lastpage}} \pubyear{2010}

\maketitle

\label{firstpage}

\begin{abstract}

We present a census of molecular outflows across four active regions
of star formation in the Perseus molecular cloud (\ngc, IC348/HH211,
L1448 and L1455), totalling an area of over 1000\,arcmin$^2$. This is one of
the largest surveys of outflow evolution in a single molecular cloud published to date. We analyse large-scale, sensitive CO
\threetotwo\ datasets from the James Clerk Maxwell Telescope,
including new data towards \ngc. Where possible we make use of our
complementary \thirteenco\ and \ceighteeno\ data to correct for the
\twelveco\ optical depth and measure ambient cloud properties. Of the 65 submillimetre cores in our
fields, we detect outflows towards 45. 24 of these are marginal
detections where the outflow's shape is unclear or could be confused
with the other outflows. We compare various parameters
between the outflows from Class 0 and I protostars, including
their mass, momentum, energy and momentum flux. Class 0 outflows are
longer, faster, more massive and have more energy than Class I
outflows. The dynamical time-scales we derive from these outflows are
uncorrelated to the age of the outflow driving source, computed from the
protostar's bolometric temperature. We confirm the results of
\citeauthor{bontemps96}, that outflows decrease in force as they
age. There is a decrease in momentum flux from the Class 0 to I
stage: $\langle F_\mathrm{CO} \rangle =(8\pm 3) \times 10^{-5}$ compared to $(1.1\pm
  0.3) \times 10^{-5}$\,\msun\,\kms\,yr$^{-1}$, suggesting a decline in
  the mass accretion rate assuming the same entrainment fraction for
  both classes of outflow. If $F_\mathrm{rad}=L_\mathrm{bol}/c$ is the
  flux expected in radiation from the central source, then $F_\mathrm{CO}(\mathrm{Class\,I})\sim
100F_\mathrm{rad}$ and $F_\mathrm{CO}(\mathrm{Class\,0})\sim
1000F_\mathrm{rad}$. Furthermore, we confirm there are additional
sources of mass loss from protostars. If a core's mass is only lost
from outflows at the current rate, cores would endure a few
million years, much longer than current estimates for the duration of
the protostellar stage. Finally, we note that the total energy
contained in outflows in \ngc, L1448 and L1455 is greater than the
estimated turbulent energy in the respective regions, which may have
implications for the regions' evolution. 
\end{abstract}


\section{Introduction}

Molecular outflows are ubiquitous in star-forming regions, with every
protostar thought to undergo a period of outflow activity
(e.g.\ \citealp*{shu87}). Outflow observations provide unique
information about the mass-loss history of the
parent protostar and thereby the accretion process which
determines its evolution. Recently there has been renewed
interest in the suggestion that outflows may drive turbulence in
molecular clouds \citep{norman80}, given that the total energy they contain may be a
significant portion of a molecular cloud's turbulent energy
and they can cause much disruption of their natal environment
(e.g.\ \citealp{knee00}; \citealp*{wolf-chase00};
\citealp{stanke07}). Theoreticians are divided about whether
an efficient mechanism exists to feed energy from a small quantity of
fast-moving material in an outflowing jet into random isotropic
motions on molecular cloud scales
(e.g.\ \citealp{cunningham06,cunningham08} versus
\citealp*{banerjee07}). However, analyses of data from regions of
molecular clouds with high densities of outflows suggest that outflows are not the dominant driving
mechanism (\citealt*{brunt09}; \citealt{padoan09}). If they were, then
cloud lifetimes should be short (consistent with the rapid formation
scenario proposed by \citealt*{hartmann01}) or the star formation rate
very high (see \citealt*{williams03}). In addition,
the detection of outflows from candidate star-forming cores, located
through say (sub)millimetre continuum surveys (e.g.\ \citealt{gbs}),
offers a way to classify such sources as either starless or
protostellar (e.g.\ \citealt*{hatchell07b};
hereafter\defcitealias{hatchell07b}{HFR} \citetalias{hatchell07b}),
which complements methods based on their spectral energy distributions
(SEDs, e.g.\ \citealt{hatchell07a,enoch09}).  

Molecular outflows have diverse structures,
reflecting both their driving sources and environments. Such diversity
has lead to a wide range of plausible models (see \citealp{arce07} for
a summary) with no single model accounting for all the features of
outflows observed to date. Young, collimated outflows point to
jet-driven models (e.g.\ \citealp{raga93b,masson93}), yet older,
wider-angle outflows evoke wide-angle wind models (e.g.\ \citealp{shu91}). Some
observational studies even suggest a two-component wind is necessary
(e.g.\ \citealp*{yu99}). 

In this paper we present a survey for molecular outflows towards the
Perseus molecular cloud (hereafter simply Perseus) using our \twelveco\ \threetotwo\ datasets (\citealp*{paper1};
hereafter\defcitealias{paper1}{Paper\,I} \citetalias{paper1}) and data
presented here for the first time towards \ngc, the most active
regions of star formation in Perseus. This outflow survey towards a single cloud is one of the largest to date, allowing us to look for
statistical variations in outflow properties. This work is the second
in a series of papers surveying the kinematics of molecular gas in
Perseus. \citetalias{paper1} presented the data acquisition and
large-scale kinematic properties, whilst \defcitealias{paper2}{Paper
  III}\citetalias{paper2} \citep{paper2} explores the detailed
gas motions inside star-forming cores. This
study is arranged as follows: Section \ref{sec:observations} describes
the observations we analyse, in particular the new data towards \ngc,
and presents an overview of the survey before listing the naming
conventions we apply to star-forming cores in our fields. In
Section \ref{sec:outflow_detection_rates}, we look for an outflow detection towards the position of
every submillimetre core identified with
\scuba\ \citep{hatchell05,hatchell07a} and analyse the corresponding
detection rates and core classifications. We set out the methods we
have chosen to analyse the detected outflows in Section \ref{sec:outflowparams_method} (the
outflows themselves are described in Appendix \ref{appendix:outflows}) before exploring
the evolution of these parameters in Section \ref{sec:outflowevolution}. Finally, we summarize our results in Section \ref{sec:summary}.

\section{Observations}
\label{sec:observations}

\begin{figure*}
\includegraphics[angle=270,width=\textwidth]{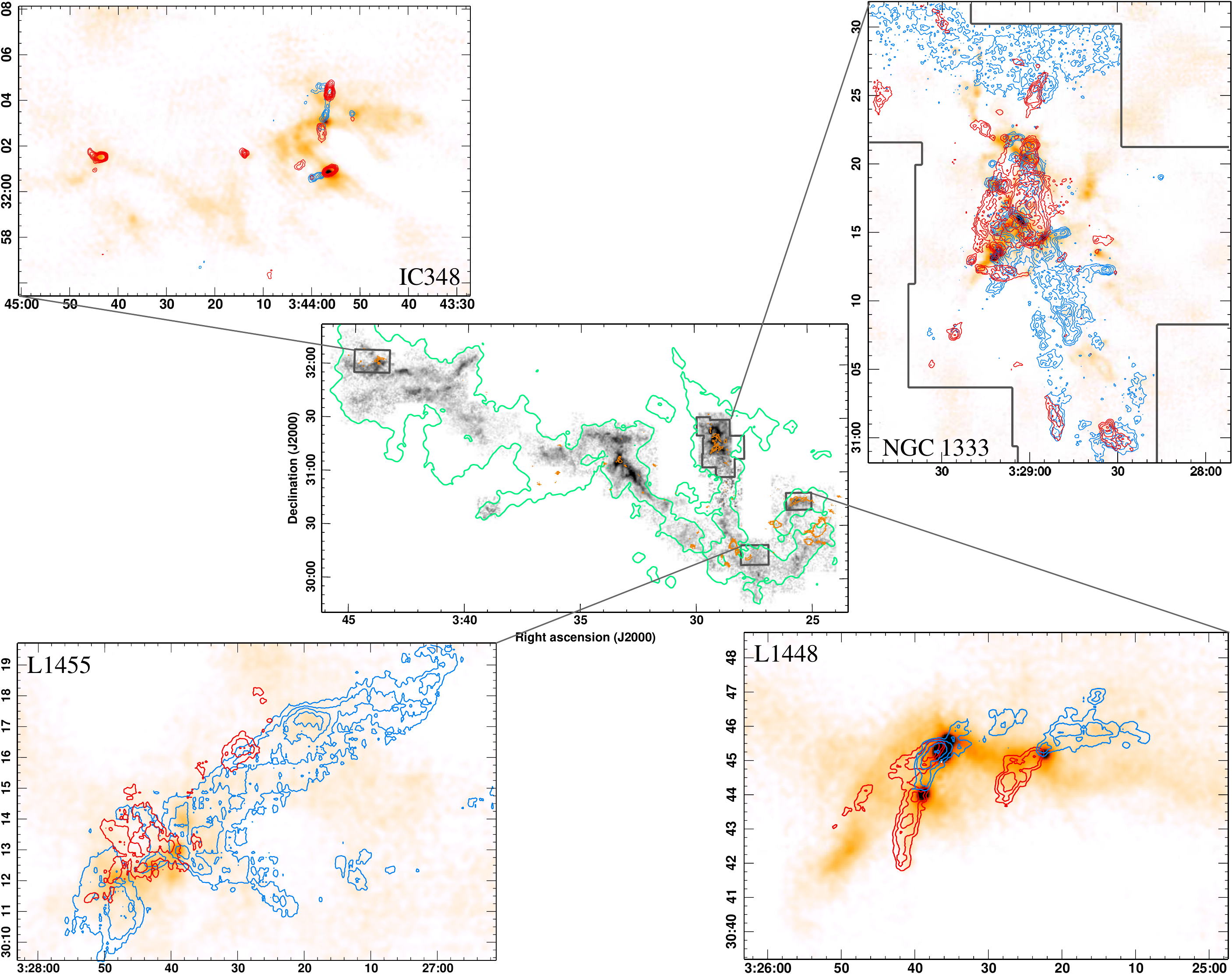}
\caption{Overview of our outflow data across the Perseus molecular
  cloud. The central image of the entire cloud is 
  \ceighteeno\ \onetozero\ integrated intensity, scaled from 0 to 4\,\kkms\, from the
  FCRAO survey \citep{hatchell05}. The green contour encloses areas
  where the visual extinction is $>3$ using data from the COMPLETE team
  \citep{ridge06}, whilst the orange contours signify the presence of
  high column densities of dust, marking where the SCUBA
  850\,\micron\ flux density is 200, 800 and
  3200\,\mjybeam\ \citep{hatchell05}. Our four survey regions (\ngc,
  IC348/HH211, L1448 and L1455) are marked with dark grey
  boxes. The enlargements of our four fields present
  \twelveco\ \threetotwo\ integrated intensity, blue- or red- shifted with respect
  to the ambient cloud velocity, contoured on SCUBA
  850\,\micron\ images, scaled from 0 to
  1600\,\mjybeam\ \citep{hatchell05}. The velocity limits of the integrations
  are the same as for Figs.\ \ref{fig:ngc1333_outflows}, \ref{fig:ic348_outflows} (middle), \ref{fig:l1448_outflows} (top)
and \ref{fig:l1455_outflow}. Except for IC348, the contours are at 5, 10, 20, 40,
80 and $160 \sigma$, where $\sigma$ is the rms noise on the integrated
intensity image (computed per pixel). Towards IC348, where we have plotted lower-velocity
outflows than the other regions (which trace the most spatial
structure), the contours are at 3, 4, 5, 6, 7, 8, 9 and $10 \sigma$.}
\label{fig:overview}
\end{figure*}

The \twelveco\ \threetotwo\ data\footnote{The transition frequency is
345.796\,GHz.} we analyse are towards the four largest clusters of
star-forming cores in Perseus (see Table \ref{table:regiondetails}
and Fig.\ \ref{fig:overview} for an overview of the data). The datacubes of IC348/HH211, L1448 and
L1455 have been published previously \citepalias{paper1} where we refer the
reader for details of the reduction. Briefly, the data were taken as
`basket-weaved' scan maps using HARP (Heterodyne Array Receiver Project; \citealt{buckle09}) on the James Clerk Maxwell Telescope (JCMT) over
nine nights between December 2007 and January 2008. The data were
taken in good to excellent observing conditions with a median system
temperature of $T_\mathrm{sys}=325$\,K. The final data products are sampled on a
3\,arcsec grid and are smoothed slightly spatially and spectrally to
an equivalent full-width half maximum (FWHM) beam size of 16.8\,arcsec
(0.020\,pc at 250\,pc, our assumed distance to Perseus) and velocity channel width
of 1\,\kms\ respectively. Smoothing to 1\,\kms\ resolution provides additional sensitivity in the \twelveco\ linewings. 

\begin{table}
\caption{Details of regions included in the survey. The data towards
  IC348, L1448 and L1455 have been published previously \citepalias{paper1}.}

\begin{tabular}{lllrrr}
\hline
Region & \multicolumn{2}{c}{Centre\,$^\mathrm{a}$} & Area &
$\sigma_\mathrm{rms}$\,$^\mathrm{b}$ & $v_0$\,$^\mathrm{c}$\\
& (h m s) & ($^\circ$ $'$ $''$) & (arcmin$^2$) & (K) & (\kms)\\
\hline
\ngc\ & 03:28:56 & $+$31:17:30 & 612 & 0.07 & 7.8\\
IC348 & 03:44:14 & $+$31:49:49 & 199 & 0.09 & 8.9\\
L1448 & 03:25:30 & $+$30:43:45 & 111 & 0.04 & 4.3\\
L1455 & 03:27:27 & $+$30:14:30 & 127 & 0.07 & 4.5\\
\hline
\end{tabular}
\label{table:regiondetails}\\
$^\mathrm{a}$\,Position of the map centre in J2000 coordinates.\\
$^\mathrm{b}$\,Median rms noise across the map measured on spectra
with 1\,\kms\ channel widths.\\ 
$^\mathrm{c}$\,Characteristic velocity in the region, taken as the
median value of the \ceighteeno\ \threetotwo\ line centre velocity
derived from Gaussian fits to every spectrum where the line is
detected with a peak brightness temperature $>3\sigma_\mathrm{rms}$ \citepalias{paper1}.\\
\end{table}

\begin{figure}
\includegraphics[width=0.49\textwidth]{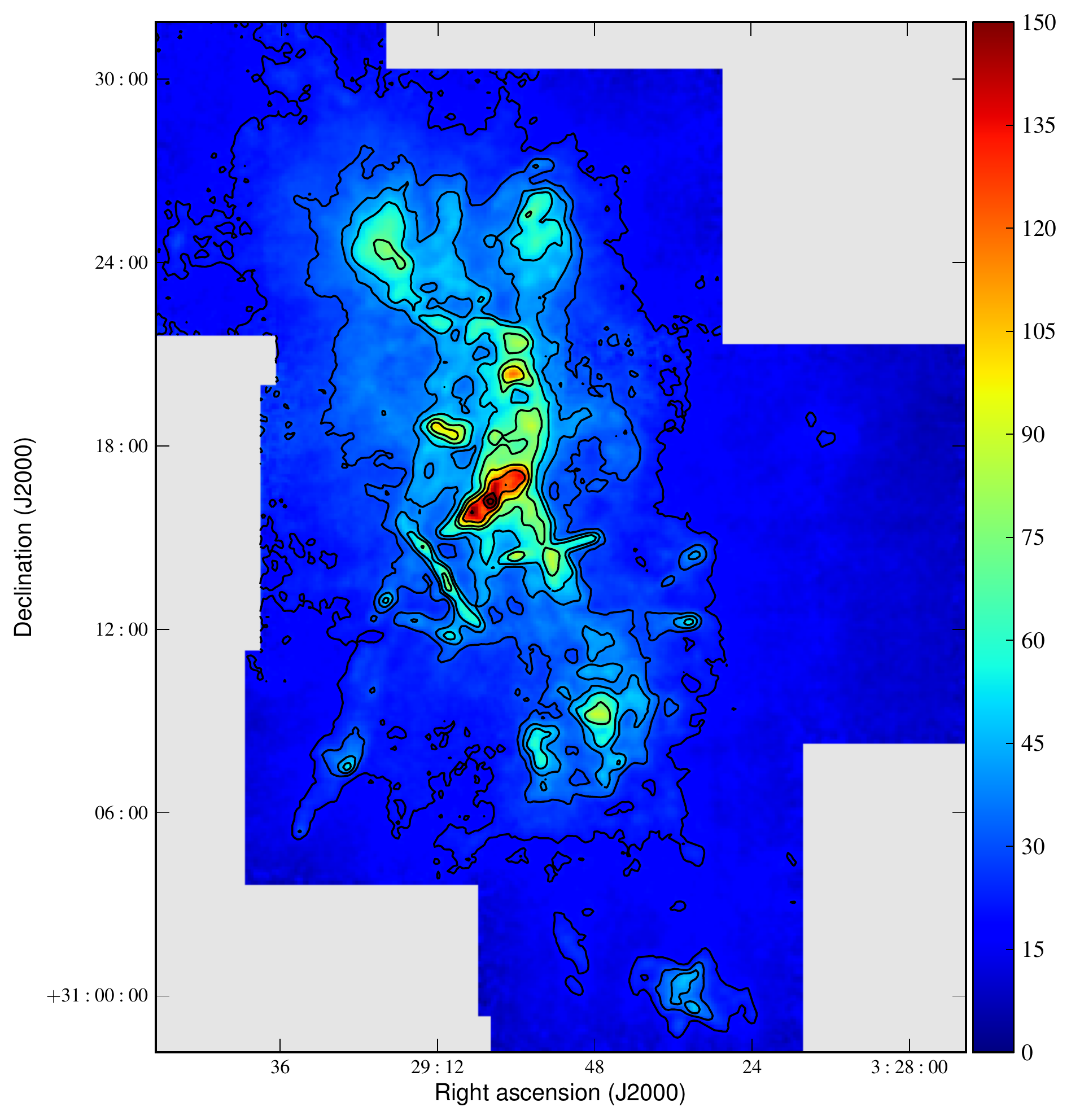}
\caption{Overview of the new \twelveco\ \threetotwo\ data towards
  \ngc. The colour-scale is integrated intensity, $\int T_\mathrm{A}^*
  \mathrm{d}v$ from $-24$ to 40\,\kms\ with contours at 20, 30, 40,
  50, 70, 90, 120, 150, 180\,\kkms.}
\label{fig:ngc1333_integ}
\end{figure}

Our survey area is more than doubled by including new \twelveco\
\threetotwo\ HARP data towards the young infrared (IR) cluster
\ngc. In Fig.\ \ref{fig:ngc1333_integ} we plot an integrated intensity
image of the new data towards \ngc, which shows a wealth of outflows
criss-crossing the region. The largest previous CO
\threetotwo\ dataset was presented by \citeauthor{knee00}
(\citeyear{knee00}; hereafter\defcitealias{knee00}{KS00}
\citetalias{knee00})\footnote{This was subsequently expanded by a
  number of surveys \citep{hatchell07b,hatchell09}.} and was only
towards a small region (63\,arcmin$^2$) in the centre of our
field. Our current dataset is nearly ten times larger in
area (see Table \ref{table:regiondetails}). The data were taken over
four nights between 13th and 16th January 2007 and comprise around
27\,hr of observing time in total. The sky opacity recorded at
225\,GHz varied from $\tau_{225}=0.025$ to 0.125, with a median for
all the scans of $\tau_{225}=0.04$, which contributed to the median system
temperature of $T_\mathrm{sys}=210$\,K. The final mosaic is composed
of multiple smaller tiles of various dimensions, which were chosen to
cover not only the brightest \twelveco\ emission, but also all the
`green' nebulous features in the \emph{Spitzer} IRAC map of the
region. This explains why the map extends so far to the west. The idea
was to make sure all of the outflow activity associated with \ngc\ was
covered in the survey. Each tile is itself a basket-weaved scan map. The
$4\times4$ detector array was inclined at 14\,deg to the scan
direction as is customary to maximize the efficiency with HARP
observations \citep{buckle09}. Unusually, the perpendicular spacing
between adjacent scan rows was set to be 6\,arcsec to yield Nyquist
sampled observations ($\lambda/2D$ is 6.0\,arcsec at 345\,GHz) and ensure
more of the 16 detectors cross the same spatial point in the sky,
providing redundancy and smoothing-out variations in detector
performance (i.e.\ noise). The backend correlator, ACSIS
\citep{buckle09}, was configured to supply 1\,GHz of bandwidth, split
into 2048 spectral channels, each 488\,kHz (0.42\,\kms) wide. The final
\ngc\ datacube analysed is on a 6\,arcsec spatial grid (distributed using
a nearest-neighbour algorithm) and has been smoothed spectrally to
a velocity resolution of 1\,\kms\ as in the other regions. The
resultant rms noise is listed in Table \ref{table:regiondetails}. All
the standard observing procedures were followed at the telescope, in
particular calibration spectra were taken frequently towards the
CRL\,618 standard to verify its intensity was within the expected
calibration tolerance. 

Any HARP data we present in figures is on the antenna temperature
scale ($T_\mathrm{A}^*$; \citealt{kutner81}), which we convert to main
beam brightness temperature, $T_\mathrm{mb}$, where necessary in our calculations
using $T_\mathrm{mb}=T_\mathrm{A}^*/\eta_\mathrm{mb}$. We use an
efficiency, $\eta_\mathrm{mb}=0.66$, as measured during the
commissioning of HARP. 

\subsection{Survey overview and naming conventions}

Our \twelveco\ datacubes contain a large sample of outflows across
Perseus, in total 23 Class 0 and 14 Class I protostars are situated in
our fields \citep{hatchell07a}. The CO \threetotwo\ line is arguably a better
probe of outflows than lower $J$ transitions
(e.g.\ \citealp{takahashi08}) as it requires warmer temperatures for
excitation ($E_\nu/k=33.2$\,K above ground) that better match the gas
temperatures observed in outflows (e.g.\ \citealp*{hatchell99};
\citealp{nisini00}). Detailed images and a description of the outflows
we find are presented in Appendix \ref{appendix:outflows}. In those
descriptions and elsewhere we refer extensively to the submillimetre
cores identified by \citep{hatchell05,hatchell07a} from
850\,\micron\ SCUBA data. We label such cores with the prefix HFR,
i.e.\ HFR1 is core 1 in \citet{hatchell07a}. These cores can be
found in the SIMBAD database as [HRF2005]{\it nn} or [HRF2007]{\it
  nn}. A number of cores were originally identified in the Perseus Bolocam
survey \citep{enoch06} and we refer to them using their original
numbers with the prefix Bolo, e.g.\ Bolo101.  

The largest survey for outflows in Perseus to date was undertaken by Hatchell and coworkers (\citetalias{hatchell07b};
\citealt{hatchell09}). They surveyed 83 star-forming
cores (compared to our 65) for CO \threetotwo\ outflows. Our survey is highly
complementary to the Hatchell et al.\ survey since we focus on outflow
properties whereas they concentrate on source classification. Furthermore, we map larger, contiguous areas
which encompass a greater amount of each individual outflow's emission, thereby 
increasing the accuracy with which we can estimate various outflow
parameters. In addition, we use our complementary data from
\twelveco's isotopologues \citepalias{paper1}. We correct our outflow mass estimates for the \twelveco\ optical depth which we
calculate from our \thirteenco\ \threetotwo\ data and derive estimates of the velocity of the driving source/ambient
emission from our \ceighteeno\ data of the same transition.

\section{Outflow detection rates and source classification} 
\label{sec:outflow_detection_rates}

Submillimetre continuum surveys select the earliest stages of star formation, when cores are
identified as being either starless or young protostars. An IR detection of a core or the presence of an outflow are thought to be robust
indicators of a protostellar source. The star-forming cores located by
\citet{hatchell05} with \scuba\ were subsequently classified from their SEDs (including \emph{Spitzer} data,
\citealt{hatchell07a}) and by the presence/absence of outflows
(\citetalias{hatchell07b}; \citealt{hatchell09}). A major conclusion of \citetalias{hatchell07b} is that
outflow mapping is as good as \emph{Spitzer} at identifying
protostars. 

We use an objective criterion to search for outflows, as applied by
\citetalias{hatchell07b} and \citet{hatchell09}: are the
\twelveco\ linewings greater in intensity than $3\sigma_\mathrm{rms}$
at velocities $\pm 3$\,\kms\ from the ambient cloud/driving-source
velocity? This definition of detectable outflowing gas was chosen as it is
simple, unbiased, uses our ancillary information and allows direct
comparison with the \citeauthor{hatchell07a} surveys. Often, authors
rely on more subjective criteria to identify outflows, such as a
definite bipolar structure. When we look at measured outflow
properties in Section \ref{sec:outflowevolution}, we too will only examine outflows
with definite extended structure or bipolarity to ensure we are
measuring true outflow characteristics. However, the aim of this
section is to look in an \emph{unbiased} manner at the classification of
SCUBA cores. Our criterion may find a number of false positives, due
to multiple velocity components along the line of sight or other
unusual velocity structures in the ambient cloud, but we have
tried to minimize such contamination by inspecting every spectrum and
point out cores with dubious detections in Table \ref{table:co_outflow_detections}.  

As the noise on our \harp\ data varies,
we apply this criterion at every spatial point individually. We use
our existing velocity maps \citepalias{paper1}, computed by fitting a
Gaussian function\footnote{A single Gaussian was fitted as the
  \ceighteeno\ \threetotwo\ line has a single peak over the majority
  of the map where it is well detected. In \citetalias{paper2} we note
only 2 out of 58 spectra, inspected towards the peak of known
SCUBA cores, have double peaked \ceighteeno\ \threetotwo\ line profiles.} to the
\ceighteeno\ \threetotwo\ line profile at every spatial point, as an
estimate of the ambient cloud velocity, $v_0$. For 11 of the 65 cores
surveyed (10 cores in \ngc\ and one in L1455) there was no estimate of
$v_0$ at the same spatial position as the core peak. This is because
our \twelveco\ data cover a larger area than our \ceighteeno\ data in
\ngc\ and the \ceighteeno\ line is too weak in one of the cores in
L1455. For these sources we take $v_0$ to be a characteristic
velocity for the region, listed in Table \ref{table:regiondetails}. Therefore we identify blue(red)-shifted outflowing gas
at a position if the \twelveco\ data is greater than
$3\sigma_\mathrm{rms}$ (where $\sigma_\mathrm{rms}$ is calculated for
\twelveco\ at each position) at velocities $v_0-(+)3$\,\kms. Almost the entire map around \ngc\ satisfies this
criterion so we increased the offset here to 4\,\kms\ to produce more
plausible outflow identifications.

Table \ref{table:co_outflow_detections} lists the results of the
outflow search towards the \scuba\ cores identified by
\citet{hatchell07a}. Spectra from 

\begin{table*}
\caption{Results of the outflow search towards the submillimetre cores
  in our fields \citep{hatchell07a}.}
\begin{small}
\begin{tabular}{lccccccccc}
\hline
Source\,$^\mathrm{a}$ & Region & Name & Class\,$^\mathrm{b}$ &
\multicolumn{2}{c}{Outflow?\,$^\mathrm{c}$} & Confusion?\,$^\mathrm{d}$
& Verdict\,$^\mathrm{e}$ & \multicolumn{2}{c}{Hatchell et al.\ verdict\,$^\mathrm{f}$} \\
 & & & & R & B & & & B & HARP\\
\hline
HRF41 & NGC\,1333 & IRAS4A & 0 & y & y & n & Out & y & --\\ 
HRF42 & NGC\,1333 & IRAS4B & 0 & y & y & n & Out & y & --\\
HRF43 & NGC\,1333 & SVS13 & I & y & y & n & Out & y & --\\
HRF44 & NGC\,1333 & IRAS2A & 0 & y & y & n & Out & y & --\\
HRF45 & NGC\,1333 & SK24 & I & y & y & n & Out & y & -- \\
HRF46 & NGC\,1333 & SK20/21 & 0 & y & y & n & Out & y & --\\
HRF47 & NGC\,1333 & SK31 & 0 & y & y & n & Out & y & --\\
HRF48 & NGC\,1333 & IRAS4C & 0 & y & n & y & M & y?c41,42 & --\\
HRF49 & NGC\,1333 & SK6 & I & y & y & n & Out & -- & y\\
HRF50 & NGC\,1333 & SK15 & I & y & y & y & M & y?c43,51 & --\\
HRF51 & NGC\,1333 & SK16 & S & y & y & y & M & y?c43,50 & --\\
HRF52 & NGC\,1333 & SK14 & 0 & y & y & y & M & y?c43,50 & -- \\
HRF53 & NGC\,1333 & SK26 & S & y & y & y & M & y?c45 & --\\
HRF54 & NGC\,1333 & SK28 & I & y & n & n & M & -- & y?c56\\
HRF55 & NGC\,1333 & & 0 & n & n & n & No & -- & y?c45\\
HRF56 & NGC\,1333 & SK29 & I & y & y & y & M & -- & y\\
HRF57 & NGC\,1333 & SK33 & S & n & n & n & No & -- & n\\
HRF59 & NGC\,1333 & & S & y & y? & y & M & -- & y?c43\\
HRF60 & NGC\,1333 & & S & n & n & n & No & -- & y?c45\\
HRF61 & NGC\,1333 & & 0 & n & n & n & No & n & n\\
HRF62 & NGC\,1333 & SK18 & 0 & n & y & y & M & y?c46 & --\\
HRF63 & NGC\,1333 & SK32 & I & y & y? & n & M & -- & y?\\
HRF64 & NGC\,1333 & & S & n & n & n & No & n & n\\
HRF65 & NGC\,1333 & SK1 & 0 & y & y & n & Out & -- & y\\
HRF66 & NGC\,1333 & SK30 & S & y & y & y & M & -- & y?c56\\
HRF67 & NGC\,1333 & & I & y & y? & n & M & -- & y\\
HRF68 & NGC\,1333 & & 0 & y & y & y & M & y?c45 & --\\
HRF69 & NGC\,1333 & & I & n & y & n & M & -- & y?c44\\
HRF70 & NGC\,1333 & SK22 & 0 & n & y? & n & No & -- & y?c46\\
HRF71 & NGC\,1333 & & 0 & y & n & n & M & -- & y?\\
HRF72 & NGC\,1333 & & S & n & n & n & No & -- & y?c43\\
HRF74 & NGC\,1333 & & I & n & n & n & No & -- & y\\
HRF75 & NGC\,1333 & & 0 & n & n & n & No & -- & y\\
HRF85 & NGC\,1333 & & S & y & y & n & Out & -- & y?\\
Bolo26 & NGC\,1333 & & S & n & n & n & No & -- & y?c65\\
Bolo44 & NGC\,1333 & & S & y & y & y & M & -- & --\\
\hline
HRF12 & IC348 & HH211 & 0 & y & y & n & Out & y & y \\ 
HRF13 & IC348 & IC348-MMS & 0 & y & y & n & Out & y & -- \\  
HRF14 & IC348 & & I & y & y & n & Out & y & -- \\  
HRF15 & IC348 & IC348-SMM3 & 0 & y & y & n & Out & y & y\\  
HRF16 & IC348 & & S & y? & n & n & M & n & n\\ 
HRF17 & IC348 & & S & y? & y? & y & M & y?c13? & -- \\
HRF18 & IC348 & & S & y? & n & n & No & n & n\\
HRF19 & IC348 & & S & n & n & n & No & n & n\\
HRF20 & IC348 & & S & n & n & n & No & n & n\\
HRF21 & IC348 & & S & y? & n & n & No & n & n\\
HRF23 & IC348 & & S & n & n & n & No & n & n\\
HRF24 & IC348 & & S & n & n & n & No & n & n\\
HRF25 & IC348 & & S & n & n & n & No & n & n\\
HRF26 & IC348 & & S & n & n & n & No & n & --\\
HRF101 & IC348 & & I & n & y & n & M & -- & y\\ 
Bolo111 & IC348 & & S & n & n & n & No & -- & --\\
Bolo113 & IC348 & & S & n & n & n & No & -- & --\\ 
\hline
\end{tabular}
\end{small}\\
\label{table:co_outflow_detections}
\begin{flushleft}
$^\mathrm{a}$~\citet{hatchell07a} source number. \\
$^\mathrm{b}$~Source classification from \citet{hatchell07a}: S=starless,
0=Class 0 and I=Class I. \\
$^\mathrm{c}$~Is red (R) or blue (B) outflowing gas detected according
to the outflow criterion (see text)? Sources are marked `?' if only a
limited region has outflowing gas. \\
$^\mathrm{d}$~Is the outflow confused with others? \\
$^\mathrm{e}$~Out=definite detection, M=marginal and No=no detection. \\
$^\mathrm{f}$~Outflow verdict of the Hatchell et al.\ surveys using either
receiver B \citetalias{hatchell07b} or HARP \citep{hatchell09} on the
JCMT. Question marks denote potentially confused sources, with any
numbers recording the sources of such confusion.\\
\end{flushleft}
\end{table*}

\begin{table*}
\begin{center}
\contcaption{}
\begin{small}
\begin{tabular}{lccccccccc}
\hline
Source & Region & Name & Class &
\multicolumn{2}{c}{Outflow?} & Confusion? & Verdict &
\multicolumn{2}{c}{Hatchell et al.\ verdict} \\
 & & & & R & B & & & B & HARP\\
\hline
HRF27 & L1448 & L1448NW & 0 & y & y & y & Out & y & --\\   
HRF28 & L1448 & L1448N:A/B & 0 & y & y & y & Out & y & --\\
HRF29 & L1448 & L1448C & 0 & y & y & n & Out & y & --\\
HRF30 & L1448 & IRS2 & 0 & y & y & n & Out & y & --\\
HRF31 & L1448 & & 0 & y & y & y & Out & y?c30? & --\\
HRF32 & L1448 & & S & y & n & y & M & -- & n?c27\\
Bolo11 & L1448 & & S & y & n & y & M & -- & --\\ 
\hline
HRF35 & L1455 & RNO15-FIR & I & y & y & n & Out & y & --\\
HRF36 & L1455 & & 0 & y & y & n & Out & y & --\\
HRF37 & L1455 & & I & y & y & y & M & y? & --\\
HRF39 & L1455 & & I & y & y & y & M & y? & --\\
HRF40 & L1455 & & S & y & n & y & M & y?c35,36 & --\\
\hline
\end{tabular}
\end{small}
\end{center}
\end{table*}

\noindent each source are
presented in Appendix \ref{appendix:outflowspectra}. We record an outflow
detection if all the positions surrounding a source meet the above
criterion. If there is the possibility of confusion with other flows
or the outflow structure is unrealistic (with small lobes that do not
resemble outflows) we mark it as a marginal detection unless the source is particularly well-known. Our fields cover 65 \citet{hatchell07a}
cores: 26 starless, 25 Class 0 and 15 Class I protostars. We detect
outflows towards 45 (69\,per cent) of these (21 firm and 24 marginal
detections, see Table \ref{table:detectionsbytype}). 

Nearly all of our sources (61 out of 65) were also examined by either
\citetalias{hatchell07b} or \citet{hatchell09}. We disagree in our
outflow identifications towards 12 cores (18 per cent). Of these
twelve cores, $\sim 7$ of our
disagreements result from our more conservative definition of an
outflow, i.e.\ it has to resemble a standard bow shock shape and/or be
bipolar for us to record a definite over marginal detection. Of the
remaining five cores (all in \ngc), at least three have outflowing gas in the
vicinity, which we attribute to other driving sources. 

All but one of our definite outflow detections are protostellar. The
exception is HRF85, in the south of \ngc\ (see Fig. \ref{fig:ngc1333_outflows}), along the line of
blue-shifted gas from IRAS2A via HH13. The outflow structure is not
highly collimated and bipolar, anticipated from the source's young
age and is therefore perhaps more likely to be emanating from
IRAS2A. \citet{hatchell09} also note the unclear CO outflow structure
around this source which makes its classification ambiguous.  

Around three quarters of the definite detections are Class 0 sources,
which is not unexpected considering the greater degree of
collimation in young outflows (e.g.\ \citealt{arce06}). Just five of the catalogued protostars
(all in \ngc) do not have any detection. HRF61 is a
prominent non-detection for \citetalias{hatchell07b} (and again in
HARP data, \citealt{hatchell09}), which
they explain might be caused by radiation and winds from nearby
luminous stars blowing away its surrounding gas. The other non-detections in HRF55, HRF70, HRF74 and HRF75 might be
explained similarly or by having extremely weak and/or small
flows. Outflows close to the plane of the sky will also be hard to
detect. Furthermore, orientation and a non-uniform environment possibly explain the protostars with only one prominent
outflow lobe.  

Of the marginal outflow detections towards starless cores, all but one (HRF85) are potentially confused
with strong outflows from neighbouring protostars. Therefore we
exclude them from the following analysis. Our outflow detection rates are comparable
to \citetalias{hatchell07b}'s; the extra sensitivity
of this survey (over \citetalias{hatchell07b}; compare their
sensitivity of $\sim 0.3$\,K in 1\,\kms\ to Table
\ref{table:regiondetails}) has not overturned any of the starless classifications of
\citet{hatchell07a} as found by \citet{hatchell09} as well. Indeed, we
are probably limited more by resolution -- many of the single
\scuba\ clumps are well-known binaries with multiple outflows not
resolved with the \jcmt. Additionally, confusion (due to resolution and
crowding) is a severe hindrance in classifying sources in this
way. The original continuum survey is incomplete to low-mass
protostars: \citet{hatchell07a} calculate the $3 \sigma$
completeness limit is 0.3\,\msun\ for a 10\,K protostar. This estimate
of the mass limit is uncertain by up to a factor of $\sim 10$, given the uncertainties in
the properties of the dust. By searching for outflows only towards identified SCUBA cores we miss outflows driven by low-mass
protostars, which were undetected in the continuum survey.  

\section{Estimation of outflow parameters} \label{sec:outflowparams_method}

In this section, we describe the methods we will subsequently employ to
  compute the physical parameters of extended outflows, identified around the protostars in our survey
  fields in Section \ref{sec:outflowevolution}. These properties for individual outflows are
  detailed in Tables \ref{table:outflow_params1} and
  \ref{table:outflow_params2} with the totals listed for all the outflows in each
  region in Table \ref{table:outflow_totals}.

\subsection{Mass} \label{sec:massestimates}

We follow a standard method (e.g.\ \citealp{goldsmith84}) and
correct for the optical depth of \twelveco\ using our complementary
\thirteenco\ \threetotwo\ data. At moderate outflow velocities (close to that of the ambient cloud emission),
\twelveco\ is optically thick but any detectable \thirteenco\ is
optically thin. Assuming \twelveco\ is optically thick,
\thirteenco\ optically thin, local thermodynamic equilibrium (LTE), identical beam-filling factors
and excitation temperatures for both isotopologues, the
following relation holds (e.g.\ \citealp{hatchell99}): 
\begin{equation}
  \frac{T_\mathrm{A}^*(\mathrm{^{12}CO})}{T_\mathrm{A}^*(\mathrm{^{13}CO})}
  = \left( \frac{\nu_{12}}{\nu_{13}}\right)^2
  \frac{[\mathrm{^{12}CO}]}{[\mathrm{^{13}CO}]}\,\frac{1-e^{-\tau_{12}}}{\tau_{12}}\mathrm{,}
  \label{eqn:opticaldepth} \end{equation}
where $\nu_\mathrm{12}=345.796$\,GHz and $\nu_\mathrm{13}=330.558$\,GHz are the transition
frequencies of the \twelveco\ and \thirteenco\ lines respectively, $\tau_{12}$
is the optical depth of the \twelveco\ gas and the abundance ratio,
$[\mathrm{^{12}CO}]/[\mathrm{^{13}CO}]=62$ \citep{langer93}. $\tau_{12}$ can be evaluated
  numerically and the correction, $\tau_{12}/(1-\exp(-\tau_{12}))$, applied to the \twelveco\ mass. 
\citet{cabrit90} showed that this method was sufficient to
correct the mass for optical depth effects in typical outflows. We
apply the correction to all the velocity channels where there is
detectable \thirteenco\ emission above the noise (i.e.\ where it is
$>\sigma_\mathrm{rms}$). The \thirteenco\ data do not cover the entire
area observed in \twelveco\ for \ngc, so some of the outflow masses
are not adjusted. A majority of outflows are corrected for the optical
depth of \twelveco\ (28 out of the 32 outflows whose parameters we report
in Tables \ref{table:outflow_params1} and
\ref{table:outflow_params2}). The  masses are increased by a factor
ranging from 1.8 to 14.3 with a median of 3.8.  

Without an opacity correction, the column density of CO,
$N(\mathrm{CO})$, can be calculated from the corresponding \threetotwo\
emission assuming \lte\ thus:
\begin{equation} \left( \frac{N(\mathrm{CO})}{\rmn{cm^{-2}}} \right) = 4.72 \times 10^{12}
  \frac{T_\mathrm{ex}/\rmn{K}}{\exp(-33.7\,\mathrm{K}/T_\mathrm{ex})} \left( \frac{\int
    T_\mathrm{mb}(v)\,\mathrm{d}v}{\mathrm{K\,km\,s^{-1}}}
    \right)\mathrm{,}\label{eqn:N_12co} \end{equation}
where $T_\mathrm{ex} $ is the excitation temperature, which following
\citetalias{hatchell07b} we take as 50\,K. For an outflow in Perseus
(250\,pc away) containing $N_\mathrm{pix}=100$ pixels, each of area
$A_\mathrm{pix}$ and average integrated antenna temperature, $\left<\int T_\mathrm{A}^*(v)\,\mathrm{d}v\right>=1$\,\kkms, its mass is:
\begin{eqnarray} \left( \frac{M_\mathrm{CO}}{\rmn{M_\odot}} \right)= 2.1\times 10^{-4}\,\left(
  \frac{X_\mathrm{CO}}{10^{-4}}\right)^{-1}\left(
  \frac{\eta_\mathrm{mb}}{0.66} \right)^{-1} \left(
  \frac{A_\mathrm{pix}}{9\,\mathrm{sq.\,arcsec}} \right) \nonumber\\
  \times \left(
  \frac{N_\mathrm{pix}}{100}\right) \left(\frac{ \left<\int T_\mathrm{A}^*(v)\,\mathrm{d}v\right>}{\mathrm{1\,K\,km\,s^{-1}}}\right)\mathrm{,}
\end{eqnarray} using the value of the abundance of CO relative to
\htwo, $X_\mathrm{CO}=10^{-4}$, from \citet*{frerking82}. In our
analysis we integrate the emission to find the mass over the total
observable velocity extent of the
outflow, i.e.\ from 2\,\kms\ away from the ambient cloud velocity to
the velocity offset, above which the outflow emission falls to below the
noise ($v_\mathrm{max}$; see also Sections \ref{sec:pande} and \ref{sec:fco}).

In addition to any uncertainties associated with the above equations,
masses derived from CO observations may not represent the true mass of
outflowing gas for a number of reasons: (i) Strong shocks can
dissociate molecular gas into atomic material
(e.g.\ \citealt{downes99}) so we may miss a significant fraction of
the total mass. Fortunately, outflow mass estimates are dominated by low-velocity
material which is unlikely to have passed through strong shocks; (ii)
Some of the low-velocity outflow emission may be missed as it is
swamped by signal from the ambient cloud material. If we consider
material moving at a modest offset ($<2$\,\kms) from the ambient
cloud velocity to belong to the outflow, \citet{downes07} estimate
that the mass underestimate could be as large as a factor of 2--3; and
(iii) In reality outflows will not be at a single constant
temperature. \citet{downes07} show that assuming a temperature
of 10\,K for mass estimates from the CO \twotoone\ transition
introduces only a small error into the outflow momentum and none into the
mass. We expect similar results with our higher assumed temperature
for the \threetotwo\ line. 

\subsection{Momentum and energy} \label{sec:pande}

\begin{table}
\caption{Number of
  outflow detections towards SCUBA cores.}
\begin{tabular}{lcccc}
\hline
Core type & Outflow & Marginal & Non-detection & Total\\
\hline
Starless & 1 & 10 & 15 & 26\\
Class 0 & 15 & 5 & 4 & 24\\
Class I & 5 & 9 & 1 & 15\\
Total & 21 & 24 & 20 & 65 \\ 
\hline
\end{tabular}
\label{table:detectionsbytype}
\end{table}

We define the observed outflow momentum along the jet axis:
\begin{equation} p_\mathrm{out} = \int m(v)|v-v_0|\,\mathrm{d}v\mathrm{,} \end{equation} 
where $v_0$ is the outflow centre velocity, approximately the
velocity of the driving source, which we derive from our \ceighteeno\
\threetotwo\ observations, and $m(v)$ is the observed \twelveco\ mass
corrected for the \twelveco\ optical depth. Inclination effects can greatly reduce the amount of
momentum observed (by $1/\cos i$ for uni-directional jets, where $i$
is the angle of inclination to the line of sight). We do not make any
inclination corrections until Section \ref{sec:globaloutfloweffect}, as the inclination angles are mostly unknown. The largest contribution to the momentum comes from the
highest-velocity material, that most affected by
dissociation. Therefore, any outflow momentum derived from CO
observations further underestimates the true momentum,
with the estimate getting poorer as the density contrast between the
jet and ambient material increases \citep{downes07}. 

An outflow's kinetic energy is similarly:
\begin{equation} E_\mathrm{out} = \frac{1}{2}\int m(v)|v-v_0|^2\,\mathrm{d}v. \end{equation}
This is even more affected by inclination ($1/\cos^2 i$ correction)
and dissociation. \citet{downes07} also point out that the constant
temperature assumption will underestimate the kinetic energy for an
outflow with high density contrast, which can only be
rectified by assuming a higher excitation temperature for the fast
material over the slow. 

We integrate over the full observable velocity-extent
  of the outflow, as above for the mass calculation (see Section
  \ref{sec:massestimates}). This yields the best estimates of
  individual outflow properties and allows us to examine more
  accurately the total impact on
  their natal environment. As the maximum velocity observed
  depends on the noise level, a larger fraction of less massive
  outflows will be missed. An alternative method, employed by
  \citetalias{hatchell07b}, reduces this bias by removing the
  high-velocity contribution to all the outflows through only integrating
  between set velocity offsets. However, depending on the velocity
  limits chosen, this method might severely under-estimate the outflow
  momentum and energy as the largest contribution to these comes from
  the highest outflow velocities.

\subsection{Dynamical time and driving force} \label{sec:fco}

The outflow momentum flux or driving force/thrust, $F_\mathrm{CO}$, is a crucial input into outflow models
and the parameter used to infer that outflows decrease in
strength with age \citep{bontemps96}. Methods to estimate
\fco\ typically fall into two categories: `global' or `local'. Local
measures examine the outflow in the vicinity of the protostar, which
is the only option if the maps do not cover all of the outflow. We use a global measure, looking at the entire outflow, as our datasets are sensitive and
extend over a large portion of the observable outflows. \citet{downes07} evaluated a number of
methods to find \fco\ using a jet-driven outflow model (most suitable
for Class 0 protostars which are the majority of our driving sources)
and we follow the route they find to be most accurate: the so-called `$v_\mathrm{max}$' method
\citep{lada85,cabrit92,beuther02}. We define the dynamical time,
$\tau_\mathrm{d}$, as the time for the bow shock travelling at the maximum velocity in the flow, $v_\mathrm{max}$, to travel the projected lobe length,
$L_\mathrm{lobe}$:
\begin{equation} \tau_\mathrm{d} = \frac{L_\mathrm{lobe}}{v_\mathrm{max}}.\end{equation}
We estimate \vmax\ as the maximum velocity observed across the outflow
lobe (the manner in which we assign spectra to individual
  outflows is described in Section \ref{sec:outflowallocation})
i.e.\ the maximum velocity above which the outflow emission falls below the
noise, $\sigma_\mathrm{rms}$. The force can then be estimated:
\begin{equation} F_\mathrm{CO} = \frac{p_\mathrm{out}}{\tau_\mathrm{d}}.
\end{equation}
The main uncertainty comes from \tdyn. If the expansion rate of the outflow varies, any estimate of \tdyn\ is inaccurate
without detailed knowledge of the mass-loss history of the protostar. \citet*{parker91} suggest \tdyn\ may
underestimate true outflow ages by up to an order of magnitude. This may well be
true for evolved sources where the flow has left the parent cloud
but not young Class 0 sources, where \citet{downes07}
suggest the outflow age is \emph{over}estimated by the dynamical
time-scale. Due to inclination effects, even the maximum outflow velocity does not
necessarily reflect the true bow advance speed. Such inclination effects are
worst for \tdyn\ in outflows close to the plane of the sky, where lobes appear their longest
and $v_\mathrm{max}$ is its smallest. However, we cannot account for
this without accurate inclination angles (calculated
from e.g.\ proper motion studies) for a large number of the outflows.  

\section{Outflow and protostellar evolution}
\label{sec:outflowevolution}
\subsection{Parameters and outflow allocation} \label{sec:outflowallocation}

In Section \ref{sec:outflow_detection_rates}, we extracted the CO
\threetotwo\ spectrum at the peak position of every \scuba\ core \citep{hatchell05,hatchell07a} in
our fields and looked for an outflow. This method produced similar
starless/protostellar core classifications as those based on
\emph{Spitzer} SEDs (see also \citetalias{hatchell07b};
\citealt{hatchell09}). For the rest of this paper we trust that the
SED classifications are accurate and now look for extended outflow
structure around the 34 identified protostars in our fields
\citep{hatchell07a}. We decide if a spatial position has blue-
  or red-shifted outflows based on the objective criterion detailed in
Section \ref{sec:outflow_detection_rates}: is the intensity in the
linewings $>3\sigma_\mathrm{rms}$ $\pm 3$\,\kms\ from the the
driving-source velocity? Contiguous regions adjacent/overlapping a
protostar, whose constituent spatial positions satisfy this
criterion are the ones included in an outflow's blue or red lobe. For
reasonably isolated protostars this easily defines the outflow
structure, however, in more crowded regions (particularly \ngc) the outflows from
different sources overlap and are confused (see Table \ref{table:co_outflow_detections}). For
confused sources we allocate the emission, selected with the outflow
criterion, to different source lobes more subjectively. We were guided
by integrated intensity contours of the outflows (see
Figs.\ \ref{fig:ngc1333_outflows} to
\ref{fig:l1455_outflow}) and constraints from any corresponding red or blue lobe on the
opposite side of the source, which are often not confused (e.g.\ we expect these to be in a straight
line).

For the outflow lobes identified, we estimate
various physical parameters using the prescriptions in the previous
Section (\ref{sec:outflowparams_method}). The physical properties of the outflows are listed for each source in
Tables \ \ref{table:outflow_params1} and \ref{table:outflow_params2} of Appendix
\ref{appendix:outflowspectra}. In addition, Table \ref{table:parametersummary} summarizes the different trends
in parameters for the outflow population. In general the parameters
we derive are comparable to other single-dish surveys (see
\citealt{davis97}; \citetalias{knee00}; \citealt{tafalla06}) but slightly
larger. Small discrepancies are unsurprising as the allocation of emission to
particular outflows tends to be rather subjective, particularly in
confused regions. Our parameters are consistently slightly higher than
previous estimates as we: (i) apply a correction for the
\twelveco\ optical depth; and (ii) have increased sensitivity and
areal coverage in our maps. However, the dynamical time-scales we
estimate are often less than half of estimates using the intensity-weighted velocity. The
$v_\mathrm{max}$ method should always produce comparatively smaller
time-scales. This tendency is reinforced in our estimates as we are
able to follow the outflow emission to higher velocities because our observations are more sensitive. \tdyn\ is particularly inaccurate
for older Class I outflows, which are expected to be longer and
wider. Many of their terminal bow shocks may have left the
gaseous cloud, so \tdyn\ cannot be the outflow age. Additionally, longer, older
flows have had greater opportunity to interact with other outflows and
obstacles, so they are more confused. As an illustration, SVS13, a Class I
source, drives a fast, confused outflow. Using $v_\mathrm{max}$, we find $\tau_\mathrm{d}=4380$\,yr,
younger than many of the Class 0 flows, whereas \citetalias{knee00} estimate $\tau_\mathrm{dyn}=30510$\,yr. SVS13 is an extreme
case but demonstrates the inherent uncertainty in the 
derivation of $\tau_\mathrm{d}$ and therefore $F_\mathrm{CO}$.   

\begin{table*}
\caption{Summary of the trends in physical outflow parameters as
  defined in the text. The outflows are grouped by region or by the classification (0 or I) of
  their associated driving source. All the errors listed are errors on
the mean not sample deviations.}
\label{table:parametersummary}
\begin{tabular}{lllllllllllllll}
\hline
Population& $L$~$^\mathrm{a}$ & $\sigma_L$ & \vmax\ & $\sigma_v$ & $M_\mathrm{out}$ &
$\sigma_M$ & $p_\mathrm{out}$ & $\sigma_P$ & $E_\mathrm{out}$ & $\sigma_E$ & \tdyn\ &
$\sigma_\tau$ & \fco\ & $\sigma_F$ \\ 
          & \multicolumn{2}{c}{(arcsec)} & \multicolumn{2}{c}{(\kms)}
& \multicolumn{2}{c}{(\msun)} & \multicolumn{2}{c}{(\msun\,\kms)} &
\multicolumn{2}{c}{($10^{36}$\,J)} & \multicolumn{2}{c}{(yr)} & \multicolumn{2}{c}{($10^{-5}$\,\fcounit)} \\ 
\hline
NGC\,1333 & 110           & 30         & 15.3 & 1.4 & 0.09  & 0.03  &
0.5 & 0.2 & 9 & 4 & 9800 & 1500 & 7.6 & 3.7\\
IC348     & 50            & 9          & 12   & 3   & 0.009 & 0.005 &
0.05 & 0.02 & 0.7 & 0.4 & 5000 & 500 & 0.9 & 0.4\\
L1448     & 240           & 40         & 24   & 3   & 0.15  & 0.03  &
1.4  & 0.8  & 36 & 17 & 14000 & 5000 & 16.7 & 11.4\\
L1455     & 90            & 30         & 11.5 & 0.6 & 0.032 & 0.006 &
0.15 & 0.03 & 1.7 & 0.3 & 10000 & 3000 & 1.9 & 0.7\\
\hline
Class 0   & 140           & 20         & 17.9 & 1.4 & 0.09 & 0.02 &
0.7 & 0.2 & 14 & 5 & 9800  & 1600 & 7.8 & 3.0\\
Class I   & 88            & 11         & 12.2 & 1.6 & 0.06 & 0.03 &
0.3 & 0.2 & 7  & 6 & 10000 & 2000 & 6.8 & 5.8\\
\hline  
All       & 124           & 14         & 15.9 & 1.1 & 0.08 & 0.02 &
0.54 & 0.17 & 11 & 4 & 9800 & 1200 & 7.4 & 2.8\\
\hline
\end{tabular}\\
\begin{flushleft}
$^\mathrm{a}$~Lobe length.
\end{flushleft}
\end{table*}

\subsection{Length} \label{sec:lobelength}

A simple expectation is that
older flows are longer and those in crowded regions appear shorter as
they intersect with others. There is little difference
in the mean length between blue- and red-shifted lobes, being $(120\pm 20)$\footnote{All the
  uncertainties quoted are errors on the mean, $\sigma_\mathrm{m}$
  formed from the standard deviation of the values $\sigma$ and
  their number, $N$, via $\sigma_\mathrm{m}=\sigma/\sqrt{N}$.} and $(130\pm
20)$\,arcsec respectively. For any particular outflow, the lobes do not
extend symmetrically either side of the driving source. The lobe
length ratio (the larger over smaller lobe length) is on average $2.0\pm
0.6$ with no significant differences between sources of different
types or region. 

Outflows appear to get shorter during source evolution, average lobe lengths are $(140 \pm 20)$ and $(88 \pm
11)$\,arcsec for Class 0 and I sources respectively. There is a
lot of scatter but it is significant that the longest outflows are all
Class 0 (see Table \ref{table:outflow_params1}). However, the apparent
length of an outflow is not necessarily related to its
age. This obviously has major implications for other parameter
estimates such as $\tau_\mathrm{d}$ and $F_\mathrm{CO}$. Flows from
Class 0 sources are typically found to be stronger and more collimated
than those from Class Is \citep{bontemps96,arce06}. Indeed, on further
evolution into the Class II stage, outflows seem to lose all
definite structure \citep{arce06}.  We already noted that it is
more difficult to identify Class I outflows:
three quarters of the definite detections were Class 0. Presumably the same is true for the length of a
Class I flow; if material gets more spread out with time, at lower column
density, then we will not necessarily detect the weaker
parts which may shorten its apparent length. In addition, it is more
likely that older Class I outflows will have broken free from the
parent cloud, so we would underestimate their full extent.  

The different regions also seem to have an effect on the observed
length, although there are small
numbers of outflows everywhere apart from \ngc. The longest outflows
are in L1448, $(240 \pm 40)$\,arcsec, with \ngc\ and L1455's under half their length, at $(110 \pm 20)$ and $(90 \pm
30)$\,arcsec respectively while
those in IC348 are the shortest: $(50\pm 9)$\,arcsec. L1448 has
extremely long outflows, extending perpendicular to the molecular ridge. \citet{bally08} suggest
that the outflows from L1448 and L1455 are huge, stretching towards
each other and possibly interacting. All the outflow sources in L1448
are Class 0, so their long lengths greatly affect the overall region average. The outflows in \ngc, which could presumably
stretch that far if undisturbed, are probably appearing shorter as
they overlap.

\subsection{Maximum observed outflow velocity} \label{sec:maxvelocity}

The trends in \vmax\ affect all the derived outflow parameters. Class 0 sources have higher-velocity outflows;
on average \vmax\ is $(17.9\pm 1.4)$\,\kms\ for Class 0 protostars and $(12.2\pm
1.6)$\,\kms\ for Class Is. This supports the idea that Class 0
outflows are stronger. Interestingly, the trends in lobe length and \vmax\ will 
counter one another in \tdyn. The longer length of Class 0 outflows
increases \tdyn, while their higher velocity decreases \tdyn. Overall,
this may explain a lack of differentiation between \tdyn\ for different
outflow stages. 

\subsection{Mass, momentum and kinetic energy}

\begin{figure}
\begin{center}
\includegraphics[width=0.45\textwidth]{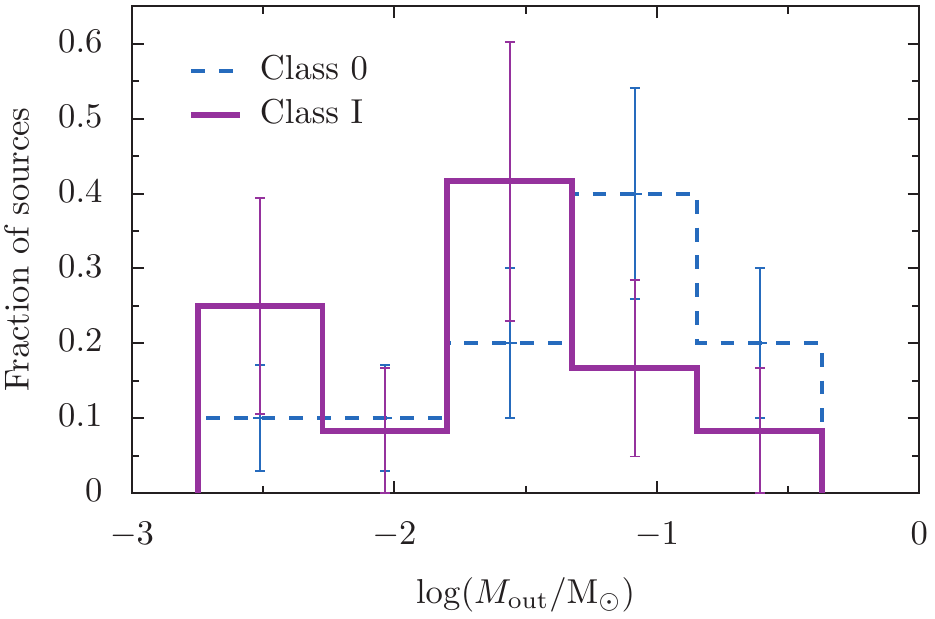}
\includegraphics[width=0.45\textwidth]{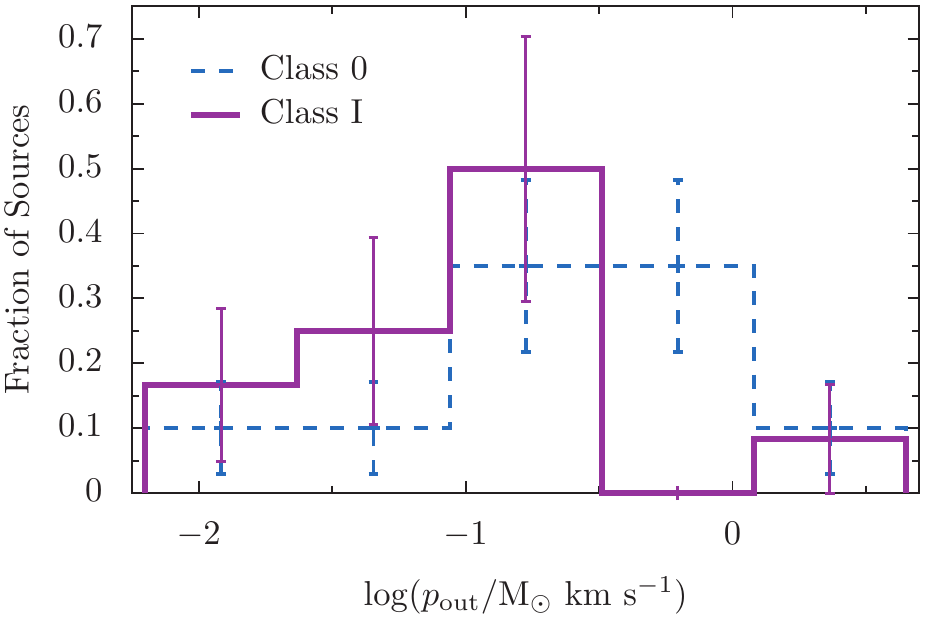}
\includegraphics[width=0.45\textwidth]{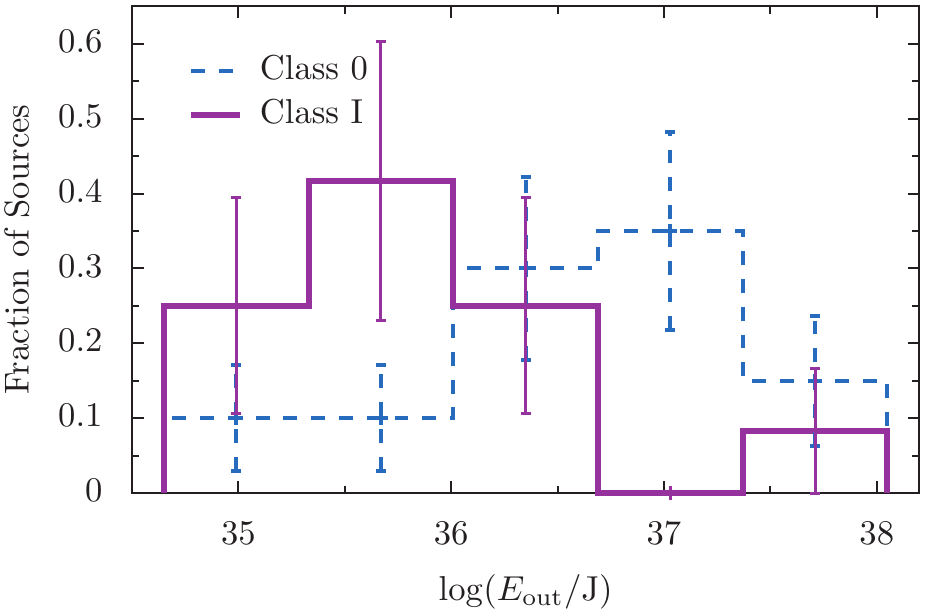}
\caption{Histograms of outflow mass (top), momentum
  (middle) and energy (bottom). The dashed light blue line is for the
  Class 0 outflows, whilst the solid purple one is for Class Is. The
  largest mass, momentum and energy bins for the Class Is only
contain the outflow from SVS13. The bars represent Poisson
errors on the bins.}
\label{fig:outflow_histograms}
\end{center}
\end{figure}

The derived outflow masses range from 0.002\,\msun\ for that from HRF71, a Class I source, to 0.4\,\msun\ for the vast Class 0
outflow from NGC\,1333-IRAS2A. In general, Class 0 outflows are more
massive than Class Is (see Fig. \ref{fig:outflow_histograms}), hardly a
surprise since we found that Class 0 outflows appear
longer (see Section \ref{sec:lobelength}) and span larger
velocity ranges (Section \ref{sec:maxvelocity}). On average, the masses are $\langle M_\mathrm{out}
\rangle = (0.09 \pm 0.02)$\,\msun\ for Class 0
and $(0.06 \pm 0.03)$\,\msun\ for Class I outflows. The distributions are not
significantly different, a Kolmogorov-Smirnov (K-S) test is inconclusive with a
9\,per cent probability that they are drawn from the same
underlying population. Across the different regions, the mean mass is
not very revealing, with most populations severely affected by the
small number of sources. 

The outflow momenta range from 0.007\,\msun\,\kms\ for HRF15, the small
bipolar outflow from IC348-SMM3, to
4.4\,\msun\,\kms\ for HRF30, one of the energetic outflows in L1448. These
two outflows also have the lowest and highest energies: $5\times 10^{34}$\,J and $1\times 10^{38}$\,J respectively. We plot histograms of the
momenta and energies in Fig.
\ref{fig:outflow_histograms}. The pattern for both parameters is
nearly identical. High-velocity outflow material makes an increased contribution to the outflow energy over momentum
\citep{downes07}, assuming a power-law velocity dependence of the mass,
$m(v) \propto v^{-\gamma}$, where $1.5 \geq \gamma \geq 2.0$
(e.g.\ \citealp*{smith97}; \citealp{downes99}). A small difference in the pattern
of values between momentum and energy implies there is little
high-velocity material in the flow i.e.\ these are unusually steep
outflows with high values of $\gamma$ or alternatively the outflows
span only a limited velocity range. Given that the maximum detected
outflow velocity is typically 15\,\kms, the former is probably
favoured. The largest Class I
momentum and energy bins in the histograms both only
contain SVS13, with the second largest bins empty (the largest mass
bin also just SVS13), suggesting SVS13 is
a highly-powerful Class I anomaly. However, it is a particularly complicated source and is
unlikely to harbour a single Class I protostar (see Section
\ref{sec:svs13}). \citetalias{hatchell07b} suggest its outflow may be
driven by a Class 0
protobinary companion. With SVS13 included, the outflow momentum is marginally higher for
Class 0 than Class I sources: $\langle p_\mathrm{out}
\rangle = (0.7 \pm 0.2)$ and $(0.3 \pm
0.2)$\,\msun\,\kms\ respectively. If we leave out SVS13, the Class I
momentum lowers considerably to $(0.10 \pm 0.03)$\,\msun\,\kms, making the
Class I population have significantly less momentum than the Class 0. A
K-S test comparing the two samples (without SVS13) yields a
1\,per cent probability that they are drawn from the same
population. The energy has a similar dependence on SVS13 but is more extreme. Without
SVS13 the Class 0 population has on average over ten times more
kinetic energy than the Class I: $\langle E_\mathrm{out}
\rangle = (1.4 \pm 0.5) \times 10^{37}$
compared to $(1.0\pm 0.3)\times 10^{36}$\,J respectively. There is then an
even smaller K-S probability (0.4\,per cent) that they are drawn from the same
population. 

\subsection{Dynamical time-scale}

\begin{figure}
\begin{center}
\includegraphics[width=0.45\textwidth]{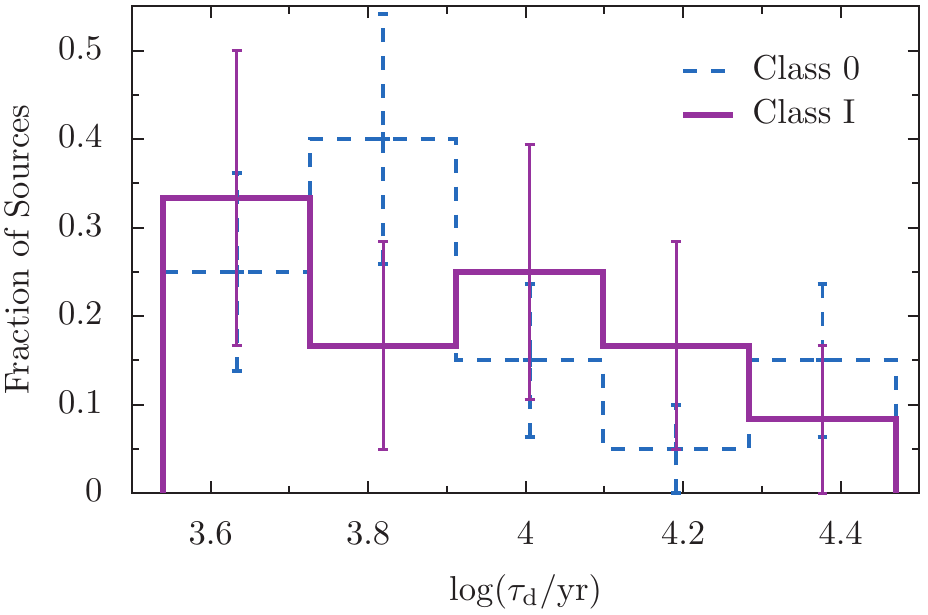}
\includegraphics[width=0.45\textwidth]{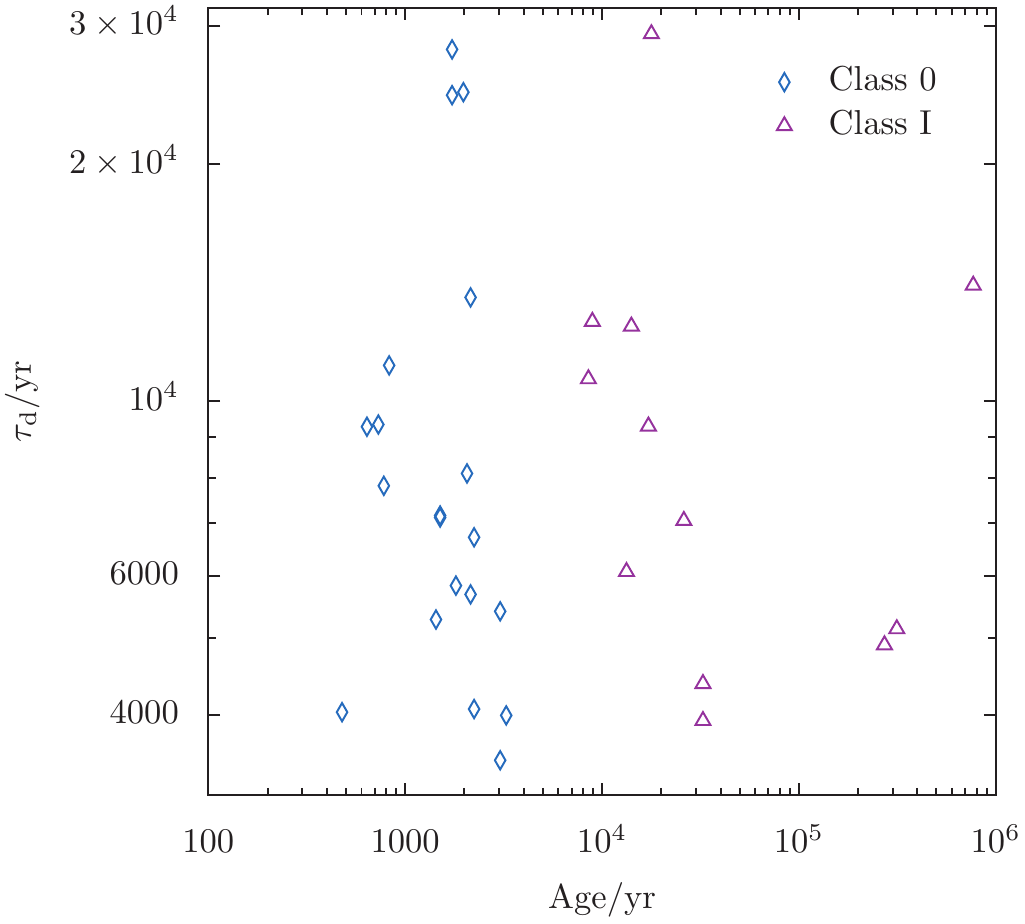}
\caption{Upper panel: Histogram of outflow \tdyn, again the bars
  represent Poisson errors on the bins. Lower panel: \tdyn,
versus nominal driving source age, $t_\mathrm{YSO}$, derived from the bolometric
temperatures in \citet{hatchell07a}.}
\label{fig:outflow_tdyn}
\end{center}
\end{figure}

The dynamical time-scale, does not show the same differentiation between the source
populations (see Fig. \ref{fig:outflow_tdyn}), with averages of $(9800 \pm
1600)$ and $(10000 \pm 2000)$\,yr for Class 0 and I outflows
respectively. Indeed, a K-S test finds an 89\,per cent probability that
they are drawn from the same population. This is not a surprise: Class 0 outflows are
longer \emph{and} faster, effects that act against each other to
produce similar average \tdyn\ to Class Is.  

In Fig. \ref{fig:outflow_tdyn}, we compare \tdyn\ to the
age of the driving source.\footnote{Implicit in all the analysis that
  follows, is that the outflow and protostellar ages are the same,
  i.e.\ outflows endure for the entire embedded phase, which is a
  fairly sound deduction from the high outflow detection rates towards
protostars.} The bolometric temperature, \tbol, is the temperature of a blackbody with the same mean
frequency as the protostar's \sed\. \tbol\ should increase as a protostar clears its circumstellar dust and progresses towards the main sequence
\citep{myers93}. It is thus a measure of the source age, $t_\mathrm{YSO}$. \citet{hatchell07a} use
it to differentiate between Class 0 and I
protostars, with the former below and latter above a division at
$T_\mathrm{bol}=70$\,K \citep{chen95}. \citet*{ladd98} find a
best-fitting relation between $t_\mathrm{YSO}$ and
$T_\mathrm{bol}$: \begin{equation} \log (t_\mathrm{YSO}/\mathrm{yr}) =
  [ 2.4 \log (T_\mathrm{bol}/\mathrm{K}) - 0.9] \pm 0.6 \mathrm{,}\end{equation}
which we use to compute $t_\mathrm{YSO}$ for each driving source, taking $T_\mathrm{bol}$ from
\citet{hatchell07a}. Using this relation the break
between classifications occurs at $t_\mathrm{YSO}=3375$\,yr. There is
no correlation between \tdyn\ and $t_\mathrm{YSO}$ for either
population or overall (see Fig.\ \ref{fig:outflow_tdyn}).

The dynamical time is \emph{not} the age
of an outflow \citep*{padman97}. Statistical outflow lifetimes
deduced by \citet{parker91} and \citet{fukui93} are at least an order
of magnitude larger. \tdyn\ depends on the time
history of the jet velocity and the density of the region between the driving source and terminal
bow shock. This assumes we can identify the terminal
bow shock, i.e.\ it is not confused or lies beyond the
map boundary. A simple calculation for the oldest sources in this sample
with ages of around 10$^{6}$\,yr, implies an outflow length of
$\sim$150\,pc for a jet velocity of 150\,\kms\ -- far larger than the area
we mapped. Even for Class 0 sources, whose shorter,
brighter and more collimated outflows are easier to
accurately trace, there is little correlation between \tdyn\ and age. 

In summary, \tdyn\ is a poor estimate of the outflow age, with little
differentiation between Class 0 and I outflows. However, the problem
for parameters derived from \tdyn\ may actually not be as
serious. Presumably we may underestimate the outflow age from \tdyn\ if we do not detect the terminal bow shock or do not trace the
outflow's extent accurately because of confusion or low column
densities of material. In such a case we also underestimate
the total mass, momentum and energy. To some extent these
underestimates cancel each other out. Other methods that
characterize an outflow by its emission close to the driving source may in fact be a fairer way of estimating
parameters (e.g.\ \citealp{bontemps96}; \citetalias{hatchell07b}) where
there is less confusion and possibility of missing emission, although
we have already noted that some of these methods may underestimate the flux in an entire
outflow. 

\subsection{Momentum flux}

\begin{figure}
\begin{center}
\includegraphics[width=0.45\textwidth]{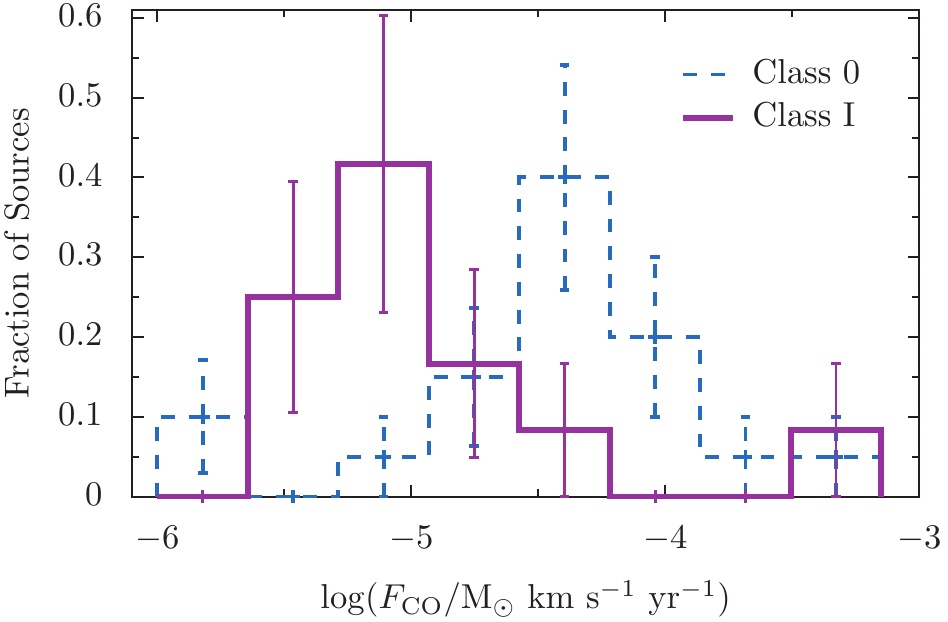}
\includegraphics[width=0.45\textwidth]{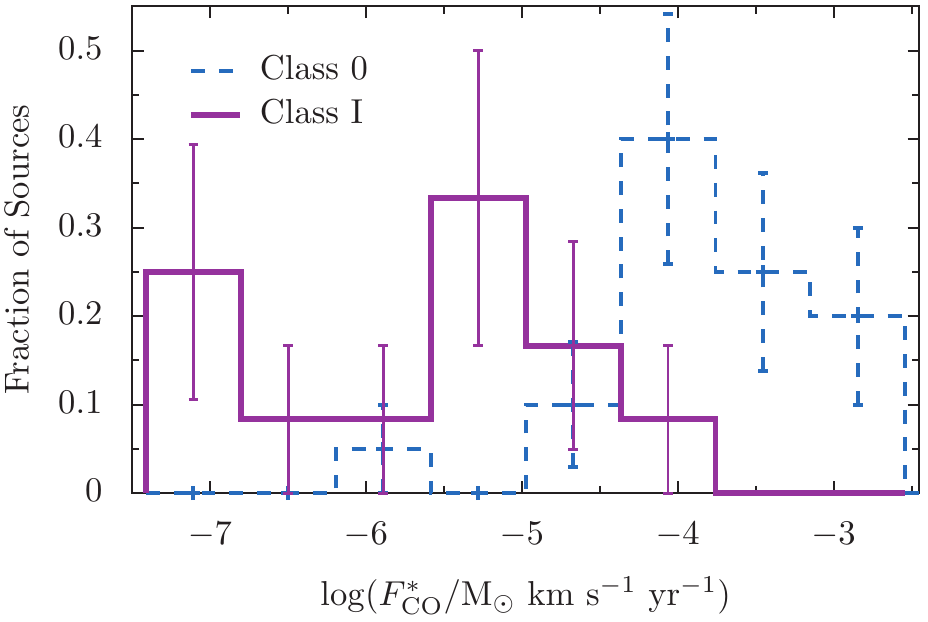}
\caption[]{Histograms of the outflow momentum flux. The momentum flux is derived using a
  time-scale of \tdyn\ (\fco, upper panel) or
the age from \tbol\ (\fcostar, lower panel). The bars represent
Poisson errors on the bins.}
\label{fig:outflow_fco}
\end{center}
\end{figure}

\citet{bontemps96} measured the momentum flux, \fco, for a sample of 9 Class
0 and 36 Class I outflows. They found the momentum flux declined with age, implying a similar decrease in the mass accretion rate
onto the central object ($\dot{M}_\mathrm{acc}$) with this rate directly proportional to the source envelope mass
($M_\mathrm{env}$). We look for a similar trend in Fig.
\ref{fig:outflow_fco}. We estimate the flux in two ways using different
time-scales: (i) \fco\ uses the conventional \tdyn\
($F_\mathrm{CO}=p_\mathrm{out}/\tau_\mathrm{d}$) and (ii) \fcostar\
takes the age of the outflow to be the same as the protostar,
$t_\mathrm{YSO}$ ($F_\mathrm{CO}^*=p_\mathrm{out}/t_\mathrm{YSO}$). The same trends are apparent for both
estimates but \fcostar\ has greater
differentiation between the types. 

Class 0 sources have higher \fco\ than Class Is. Most of the high \fco\
sources are Class 0 apart from
SVS13 which, as already mentioned, has an anomalously large momentum for a Class I
source. The sample averages bear this out if SVS13 is excluded from the
Class I population: $\langle
F_\mathrm{CO} \rangle = (0.8 \pm 0.3) \times
10^{-4}$ for Class 0 outflows and $(1.1 \pm
0.3) \times 10^{-5}$\,\fcounit\ for Class Is ($(0.7 \pm
0.6) \times 10^{-4}$ with SVS13 included). \fcostar\
accentuates the same trends; the Class I average is over
an order of magnitude smaller than the Class 0: $(6 \pm 2) \times
10^{-6}$ ($(1.3 \pm 0.8) \times 10^{-5}$ with SVS13 included) compared to $(4.3 \pm 1.5) \times
10^{-4}$\,\msun\,\kms\,yr$^{-1}$. For both \fco\ and \fcostar\ there is an
extremely small K-S probability that the two protostellar classifications are drawn from the same
population: $\la 0.1$\,per cent. Thus, the outflow momentum flux
declines from the Class 0 to
I stage as found by \citeauthor{bontemps96} Intriguingly, 
\citetalias{hatchell07b} do not find this relation holds. Their
results are influenced by a small number (3-4) of high-flux Class I and low-flux Class 0
outflows. The former they explain are potential binary
sources, containing a more energetic Class 0 companion. After
  SVS13, \citetalias{hatchell07b} find the two most powerful Class I
  sources are HRF45 and HRF50 both in \ngc. We also measure a large
  \fco\ for HRF45 (the second largest), although it is considerably
  smaller than for SVS13 and close to the other Class I protostars. On
  the other hand, we do not evaluate \fco\ for an outflow from HRF50 as it is so
  confused with the one from SVS13 (see Section \ref{sec:svs13}). 

Low-flux Class 0 sources are harder to explain; perhaps some of the youngest Class 0 flows are
just powering up and have not reached their maximum outflow force
(e.g.\ \citealp{smith00}). Furthermore, the source classification
based on SEDs will not be entirely infallible. Inclination effects for
protostars with a disc/bipolar cavity can cause large deviations from
typical SEDs for their class (e.g.\ \citealt{whitney03}) causing them to be mis-classified. 

\begin{figure*}
\begin{center}
\includegraphics[width=\textwidth]{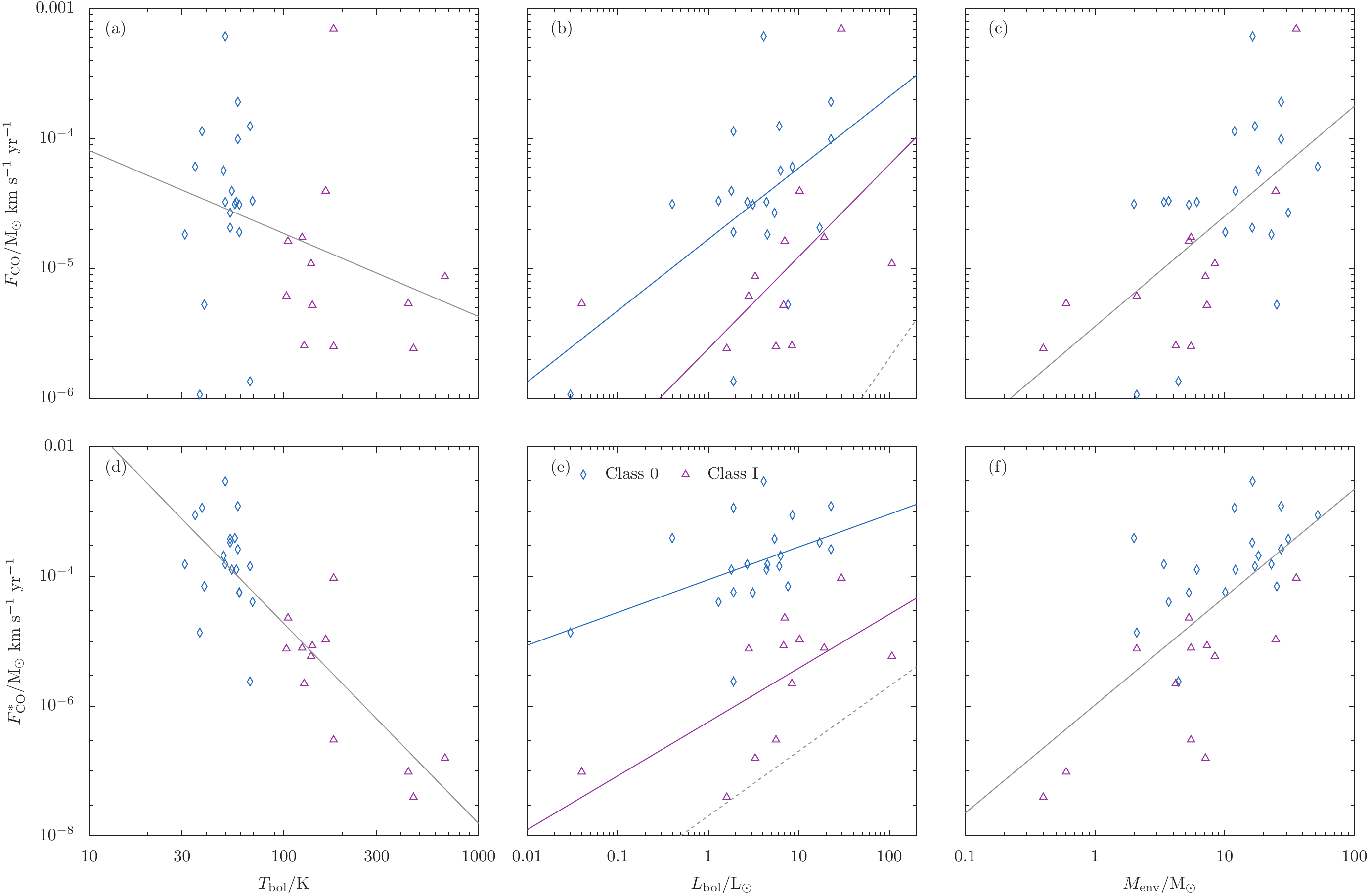}
\caption{Correlations of \fco\ (upper panels) and \fcostar\ (lower panels, see text) with \tbol\ (left), $L_\mathrm{bol}$
  (middle) and $M_\mathrm{env}$ (right) taken from \citet{hatchell07a}. Sources are marked according to their
  different \citet{hatchell07a} classifications. Dashed lines
  represent the force available in stellar photons,
  $F_\mathrm{CO}=L_\mathrm{bol}/c$ (for a single scattering). Solid lines mark
  maximum-likelihood fits to the overall population (unless
  specified): (a)\,$F_\mathrm{CO} \propto {T_\mathrm{bol}}^{-0.64}$,
  (b)\,$F_\mathrm{CO} \propto {L_\mathrm{bol}}^{0.6}$ (blue,
  Class 0) and $F_\mathrm{CO} \propto {L_\mathrm{bol}}^{0.7}$ (purple,
  Class I), (c)\,$F_\mathrm{CO} \propto {M_\mathrm{env}}^{0.9}$, 
  (d)\,$F_\mathrm{CO}^* \propto {T_\mathrm{bol}}^{-3.1}$, (e)\,$F_\mathrm{CO}^* \propto {L_\mathrm{bol}}^{0.51}$ (blue, Class
  0) and $F_\mathrm{CO}^* \propto {L_\mathrm{bol}}^{0.8}$
  (purple, Class I) and (f)\,$F_\mathrm{CO}^* \propto {M_\mathrm{env}}^{1.7}$. }
\label{fig:fco_correlations}
\end{center}
\end{figure*}

The correlations of the momentum flux (both \fco\
and \fcostar) with various properties are shown in Fig.
\ref{fig:fco_correlations}. Since the range of parameters is
small and the fluxes are not corrected for inclination, 
there may be considerable uncertainty. The
first correlation is with \tbol\ and if the outflow force does decrease
with age, we expect there to be a decrease in \fco\ with
\tbol, which is tentatively apparent. Of course
with \fcostar\ there is a much firmer trend, expected as \fcostar\ by
definition is inversely related to \tbol. \citet{hogerheijde98}
also note that an outflow's age is not a good
predictor of the force (the conventional \fco). 

The second correlation, with bolometric source luminosity, $L_\mathrm{bol}$, is
well known (e.g.\ \citealp{cabrit92}). We take $L_\mathrm{bol}$ from \citet{hatchell07a}, who assumed a different
distance to Perseus. Correcting for the distance would
uniformly reduce $L_\mathrm{bol}$ by 40\,per cent, leaving the shape of the
correlation unaffected. In these data there seems to
be two distinct correlations for the different evolutionary
stages in both \fco\ and \fcostar\ with slopes very similar to unity. Both types have forces many times
larger than the momentum supplied in radiation from the driving source
($F_\mathrm{rad}=L_\mathrm{bol}/c$). The
correlations are approximately consistent with Class I sources having
$F_\mathrm{CO}=100F_\mathrm{rad}$ with the flux of Class 0 sources $\sim$10 times higher again. For \fco, maximum-likelihood fits
follow: \begin{eqnarray}
\log (F_\mathrm{CO}/\mathrm{M_\odot\,km\,s^{-1}\,yr^{-1})} = (0.5 \pm
0.2) \log(L_\mathrm{bol}/\mathrm{L_\odot}) \nonumber \\- (4.8 \pm 0.2) 
\rmn{~~Class~0~sources,} \end{eqnarray} \begin{eqnarray}
\log(F_\mathrm{CO}/\mathrm{M_\odot\,km\,s^{-1}\,yr^{-1})} = (0.7 \pm
0.4) \log(L_\mathrm{bol}/\mathrm{L_\odot}) \nonumber\\- (5.6 \pm 0.2) 
\rmn{~~Class~I~sources.}
\end{eqnarray} Class 0 objects
drive systematically more powerful outflows than Class Is for the same
luminosity of object, implying they are more efficient at
driving their outflows. This was also suggested by \citet{bontemps96},
although they had fewer Class 0 sources which did not allow a good
constraint on the correlation.

The final correlation with $M_\mathrm{env}$ is the most
convincing. Again we use
$M_\mathrm{env}$ from \citet{hatchell07a}, who derive them from
\scuba\ fluxes assuming a distance of 320\,pc to Perseus. Like the
$L_\mathrm{bol}$ correlation, to correct for the distance requires a
systematic reduction in $M_\mathrm{env}$ by $\sim$40\,per cent. Maximum-likelihood fitting to the entire
population yields a near direct-proportionality relation:
\begin{eqnarray} \log (F_\mathrm{CO}/\mathrm{M_\odot\,km\,s^{-1}\,yr^{-1})} = (0.85 \pm
0.16) \log(M_\mathrm{env}/\mathrm{M_\odot}) \nonumber\\- (5.4 \pm 0.2)
\end{eqnarray} This correlation again points to a decline in outflow
force with source age, if the envelope mass is inversely related to the outflow age,
i.e.\ Class 0 sources have large masses and Class Is small. 

\begin{figure}
\begin{center}
\includegraphics[width=0.49\textwidth]{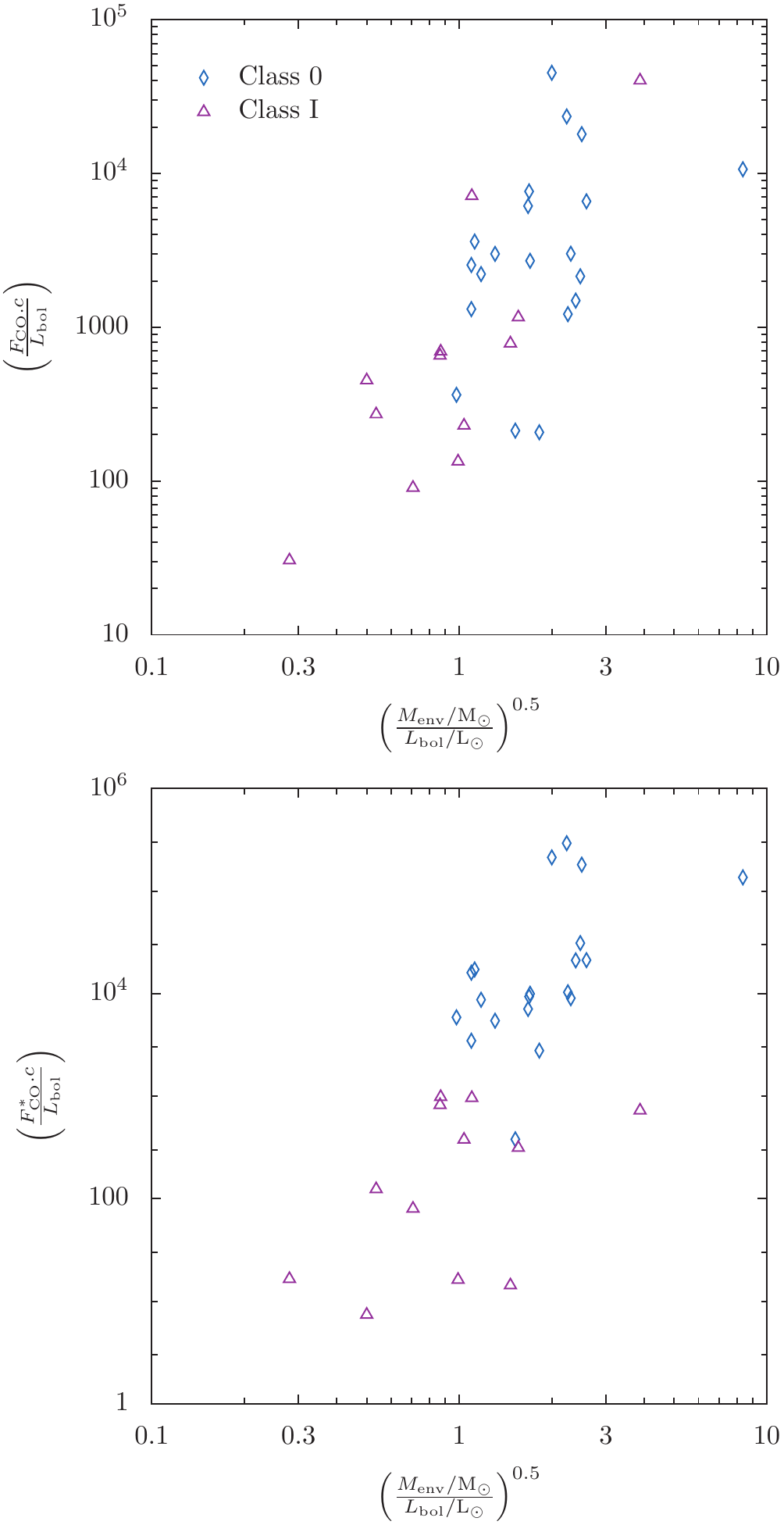}
\caption{Flux
  versus mass relationship corrected for luminosity with sources
  marked according to their \citet{hatchell07a}
  classification. The ordinate measures the ratio of outflow to
  radiative momentum flux (see text) and the abscissa comes from a relation in \citet{hatchell07a}:
  $L_\mathrm{bol} \propto {M_\mathrm{env}}^{1.96}$. See also \citetalias{hatchell07b}.}
\label{fig:fco_menv_corrlbol}
\end{center}
\end{figure}

The common gradient for all sources on a \fco--$L_\mathrm{bol}$
diagram led many authors to conclude that there is a common driving
mechanism for all outflows and their energetics are dominated by
the luminosity of the central source
(e.g.\ \citealp{lada85,levreault88}). However, as \citet{bontemps96}
showed (and we have seen), there are considerable
differences in the energetics of outflows at different evolutionary
stages. To emphasize such trends, \citeauthor{bontemps96} tried to
remove the luminosity dependence from the \fco-$M_\mathrm{env}$
relation (their fig. 7). We produce an equivalent plot in Fig.
\ref{fig:fco_menv_corrlbol}, following the appropriate prescription from
\citetalias{hatchell07b}. Unlike \citetalias{hatchell07b}, there
is a clear evolutionary trend with $F_\mathrm{CO}.c/L_\mathrm{bol}$
declining markedly from Class 0 to I sources. Thus, we confirm the findings of
\citet{bontemps96}, and demonstrate the increased efficiency in Class
0 outflows. \citeauthor{bontemps96}
parameterize the momentum flux: \begin{equation} F_\mathrm{CO} =
  f_\mathrm{ent} \times
  \frac{\dot{M}_\mathrm{w}}{\dot{M}_\mathrm{acc}}v_\mathrm{w} \times
  \dot{M}_\mathrm{acc} \end{equation} where $f_\mathrm{ent}$ is the
entrainment efficiency ($<1$), i.e.\ the fraction of matter from the central
jet/wind fed into the the outflow, $\dot{M}_\mathrm{w}$ the mass-loss rate in
the jet/wind and $v_\mathrm{w}$ the jet/wind velocity. In such a
parametrization they state the only likely cause of a large decline in \fco\ is a
reduction in $\dot{M}_\mathrm{acc}$ as the other parameters do not
vary enough in plausible jet and X-wind models. However, it seems
highly unlikely that the entrainment efficiency will be the same for
Class 0 and I outflows given the differences in their surrounding
environments and degrees of collimation. Nevertheless, if we adopt
similar values for the parameters to \citeauthor{bontemps96},
$f_\mathrm{ent}=1$, $\dot{M}_\mathrm{w}/\dot{M}_\mathrm{acc}=0.1$  and
$v_\mathrm{w}=150$\,\kms, the reduction in \fco\ from Class 0 to I
sources, $(0.8 \pm 0.3) \times 10^{-4}$ to $(1.1 \pm 0.3) \times
10^{-5}$\,\msun\,\kms\,yr$^{-1}$ on average, implies a reduction in the
accretion rate from $\dot{M}_\mathrm{acc}=5\times 10^{-6}$ to
$7\times 10^{-7}$\,\msun\,yr$^{-1}$, somewhat smaller than the
reduction inferred by \citeauthor{bontemps96}   

\subsection{Destruction time-scale}

Outflows act as one means of removing matter from a star-forming
core. The time that it would take an outflow to eject the remaining
mass of its parent core is a limit on the lifetime of
the protostar, although in reality all of this mass will not be lost
via the outflow. \citetalias{hatchell07b} looked at these
limits for their outflow data and with our different
parameters it is worth revisiting some of their findings. We define an
outflow's mass-loss rate, $\dot{M}_\mathrm{out}$, as the outflow mass divided by its \tdyn. The destruction time-scale is then,
$t_\mathrm{des} = M_\mathrm{env}/\dot{M}_\mathrm{out}$. We take
$M_\mathrm{env}$ from \citet{hatchell07a} and correct for the
different assumed distance. 

\begin{figure}
\begin{center}
\includegraphics[width=0.45\textwidth]{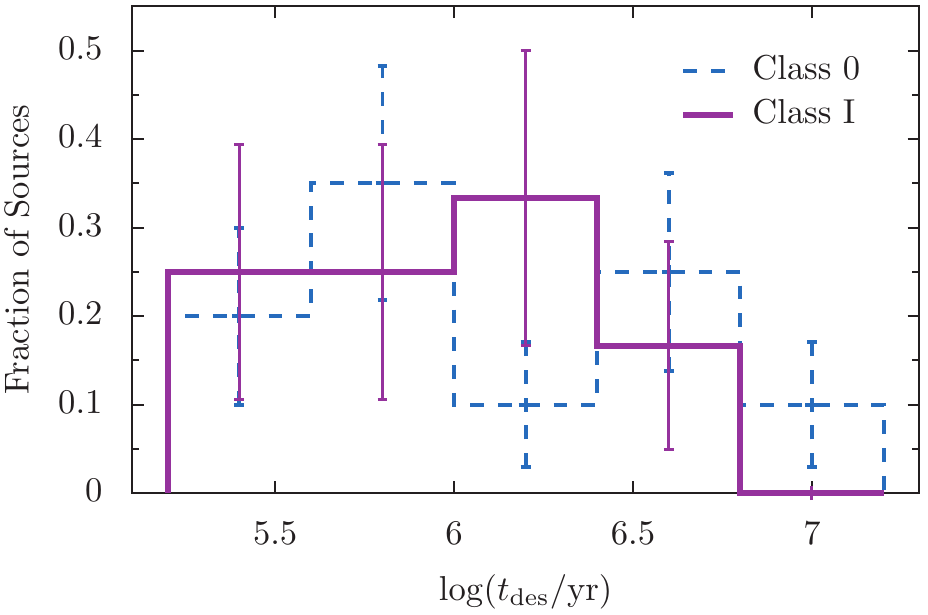}
\caption{Histogram of driving source destruction time-scales. The bars
again represent Poisson errors on the bins.}
\label{fig:outflow_tdest}
\end{center}
\end{figure}

We find destruction time-scales in the range 0.2 to 15\,Myr for Class 0
and 0.3 to 5\,Myr for Class I protostars (see Fig.
\ref{fig:outflow_tdest}). On average the Class 0 time-scale is longer
than the Class I: $\langle t_\mathrm{des} \rangle = (2.5 \pm 0.8)$\,Myr
compared to $\langle t_\mathrm{des} \rangle = (1.6 \pm 0.4)$\,Myr. Class I
sources have both smaller $M_\mathrm{env}$ and
$\dot{M}_\mathrm{out}$, which counteract each other to yield similar
$t_\mathrm{des}$ in both classes. There is a 67\,per cent K-S
probability that the samples from each class are drawn from the same population. These time-scales are smaller
than \citetalias{hatchell07b}'s as we estimate larger mass-loss rates
for the reasons previously discussed. However, they are
still an order of magnitude larger than the mean lifetime of the embedded
phase which is 1.5 to 4$\times 10^{5}$\,yr (\citealt{hatchell07a}
and references therein). This leads us to \citetalias{hatchell07b}'s conclusion: outflows are not the only source of mass-loss in
protostars. Of course accretion from the surrounding circumstellar
material, traced by \scuba, onto the final object may account for some
additional mass-loss but typically the mass-loss rates via accretion
are 10\,per cent of those from outflows
\citep{bontemps96}. \citetalias{hatchell07b} give a thorough discussion of the potential
systematic errors but two points are worth mentioning: (i)
$M_\mathrm{env}$ overestimates the circumstellar mass remaining so we
overestimate $t_\mathrm{des}$. Many apparently single cores are in fact multiple systems. Additionally, it is well
known that looking for clumps in two-dimensional data 
finds more massive sources than in reality exist
(e.g.\ \citealp*{smith_clark08}). 
(ii) If the destruction time-scales are
truly larger than the core lifetimes, the problem is worse in
reality. $t_\mathrm{des}$ was calculated assuming a
constant mass-loss rate, whereas as we have seen it is likely to
decrease with time, making the time-scales longer. 

\subsection{Global effect of the outflows}
\label{sec:globaloutfloweffect}

The outflows within every region in Perseus are energetic and extend
across significant portions of the map. Whilst it is not
possible to estimate accurately how much energy and momentum the
outflows actually impart to the cloud, we can estimate the \emph{maximum} that could
be shared from a simple sum of the contributions
from each outflow. In Table \ref{table:outflow_totals}, we list the total outflow
parameters for the different regions. We cannot correct each
individual outflow for inclination but a correction to the total
parameters is possible, assuming that the outflows are distributed
uniformly, at random inclinations to the line of sight. The average
inclination angle is then $\langle i \rangle =57.3$\,deg. For comparison, Table \ref{table:bindingenergy} presents various estimates of
the momentum and energy from our \ceighteeno\ \threetotwo\ data \citepalias{paper1}. We use the average non-thermal
linewidth of the \ceighteeno\ line,
$\sigma_\mathrm{NT}(\mathrm{C^{18}O})$, as a probe of the turbulence
in the bulk of the gas since it is substantially broadened by
non-thermal motions (see \citetalias{paper1}). The
turbulent energy is then $E_\mathrm{turb} \sim M \sigma_\mathrm{NT}^2$
and momentum $P_\mathrm{turb} \sim M \sigma_\mathrm{NT}$
(\citealp*{walawender05}; \citealp{davis08}). We estimate the mass from the \ceighteeno\
integrated intensity in each region, assuming \lte, $T_\mathrm{ex}=12$\,K and that
the line is optically thin. Using this \ceighteeno\ mass we calculate the gravitational binding energy, $GM/R^2$, with the radius,
$R$, the geometric mean of the major and
minor axes of the cloud `core' taken to be where the integrated
\ceighteeno\ intensity is  $\geq$1\,\kkms. 

\begin{table*}
\begin{center}
\caption[Global Outflow Properties]{Total outflow properties in each
  region. Estimates of the momentum, kinetic energy and
  momentum flux have
  been increased by factors of $\langle 1/\cos i \rangle = 2$,
  $\langle 1/\cos^2 i \rangle = 3$ and $\langle \sin i/\cos^2 i
  \rangle = 4/\pi$ respectively, assuming a random, isotropic
  distributions of outflow inclinations.} 
\begin{tabular}{lcccc}
\hline
Region & Mass & Momentum & Kinetic Energy & Momentum Flux\\
 & (\msun) & (\msun\,\kms) & (J) & (\msun\,\kms\,yr$^{-1}$)\\  
\hline
NGC\,1333 & 1.62 & 19.4 & 4.9$\times 10^{38}$ & 1.8$\times 10^{-3}$\\
IC348     & 0.05 & 0.5  & 1.0$\times 10^{37}$ & 5.5$\times 10^{-5}$\\
L1448     & 0.75 & 13.6 & 5.4$\times 10^{38}$ & 1.1$\times 10^{-3}$\\
L1455     & 0.10 & 0.9  & 1.5$\times 10^{37}$ & 7.1$\times 10^{-5}$\\
\hline
\end{tabular}
\label{table:outflow_totals}
\end{center}
\end{table*}

\begin{table*}
\begin{center}
\caption[\ceighteeno Region Properties]{Properties of the different
  regions from \ceighteeno\ \threetotwo\ observations. Masses are calculated assuming \lte\ and $T_\mathrm{ex}=12$\,K. The radius is the geometric
  mean of the major and minor axes of the main \ceighteeno\ `core'
  (see text). Turbulent momenta and energies are calculated from the
  average \ceighteeno\ non-thermal linewidth, $\sigma_\mathrm{NT}$(\ceighteeno), via
  $P_\mathrm{turb}\sim M \sigma_\mathrm{NT}$ and $E_\mathrm{turb}\sim
  M \sigma_\mathrm{NT}^2$.} 
\begin{tabular}{lcccccc}
\hline
Region & Mass & Radius & Binding Energy &
$\sigma_\mathrm{NT}$(\ceighteeno) & $P_\mathrm{turb}$ &
$E_\mathrm{turb}$\\
 & (M$_\odot$) & (arcsec) & (J) & (\kms) & (M$_\odot$\,\kms) & (J)\\  
\hline
NGC\,1333 & 439 & 775 & $3.5 \times 10^{39}$ & 0.44 & 193 & $1.7 \times 10^{38}$\\
IC348 & 196 & 440 & $1.2 \times 10^{39}$ & 0.26 & 51 & $2.6 \times 10^{37}$\\
L1448 & 59 & 198 & $2.5 \times 10^{38}$ & 0.35 & 21 & $1.4 \times 10^{37}$\\
L1455 & 19 & 90 & $5.5 \times 10^{37}$ & 0.31 & 6 & $3.6 \times 10^{36}$\\
\hline
\end{tabular}
\label{table:bindingenergy}
\end{center}
\end{table*}

Molecular outflows in NGC\,1333 have
sufficient momentum to accelerate the entire region to a few \kms\ over
the cloud's lifetime \citepalias{knee00} and therefore NGC\,1333
provides one of the clearest examples of the self-regulation of star
formation. We find the total mass
of outflowing gas is less than one per cent of the total in the region;
the total momentum is 10 per cent of $P_\mathrm{turb}$ and total energy
is around three times $E_\mathrm{turb}$ but only 14 per cent of the
gravitational binding energy. Thus there is enough energy to drive the
cloud turbulence and regulate the star formation process provided the
outflows can be efficiently coupled to the bulk
motions. Indeed there is a good proportion of the energy required to
destroy the cloud in the outflows, which \citetalias{knee00} suggest
is a likely outcome. The turbulent momentum seems too large to be
provided solely by these outflows, although we probably underestimate
the outflow momentum in CO as most of it is contained in the jet that
CO traces poorly. Alternatively, if our momentum estimates are accurate, \citet{davis08} suggest that
it would take a number of generations of outflows to build up the turbulent
momentum. 

IC348 is at the other end of the scale. Its population contains a
larger proportion of starless cores so we can probably infer it is
a younger region. A tiny proportion
(0.03\,per cent) of its mass is contained in outflows. Additionally, they provide
less than one per cent of the turbulent momentum and cloud binding
energy and only 39\,per cent of the turbulent energy -- the lowest
fractions overall. Only a small fraction of the turbulence in the
region can therefore be attributed to feedback from currently active flows. 

A similar argument for L1448 is probably misleading. Its
outflows clearly burst out of the natal cloud core, making it difficult to understand how extended high-velocity gas,
outside of the cloud, can feedback energy and cause turbulent
motions or large-scale disruption. Nevertheless, in their entirety
the outflows are massive and energetic, containing twice the binding
energy of the cloud (also noted by \citealp{wolf-chase00}), so if only
a fraction of that energy were coupled to the cloud, the core would
probably be broken apart. 

Finally, L1455 is similar to NGC\,1333. Around 0.5\,per cent of the total
cloud mass is held in outflows, with 15 per cent of the turbulent cloud
momentum and four times the turbulent energy. There is still not
quite enough kinetic energy to overcome the binding energy, although
we have not counted the largest northwest-southeast flow in our
calculations \citep{goldsmith84}, which if included may well raise the total
beyond the binding energy. Many of the outflows again stretch outside the
dust and ambient emission, so these totals are upper limits to the energetic input into the cloud.  

Numerous previous studies have explored the total contribution
outflows might provide to the turbulent energy in other molecular
clouds. In the Serpens molecular cloud the kinetic
energy from outflowing gas is $\sim 70$ per cent of the cloud's turbulent
energy and $\sim 62$ per cent of its gravitational binding energy
\citep{graves10}, values perhaps most comparable to IC348 -- although
Serpens is closer to equipartition between its gravitational and
turbulent energies. This estimate for Serpens does apply
a correction for the outflow inclinations but not for the optical
depth of the \twelveco\ \threetotwo\ transition. In Section \ref{sec:massestimates}, we
demonstrated this can increase the outflow mass (and kinetic energy) by a median factor of
3.5. Applying this correction would mean the outflows' energy is more
than double the cloud's turbulent and gravitational binding
energy; a similar situation to that noted in L1448. \citet{stanke07} and \citet{dent09} analysed data
towards two clustered regions forming high-mass stars: L1641-N in
Orion and the Rosette molecular cloud respectively. In L1641-N  a
lower limit for the outflows' energy of $\sim 30$ per cent of the
turbulent was found, again similar to our results in
IC348. Once more, the outflows' kinetic energy may be underestimated by over an order of
magnitude as no correction for inclination nor optical depth was
applied. \citet{dent09} find that their outflows (increased in mass by
an average factor to account for optical depth and inclination)
have similar kinetic energies to their parent stellar cluster's
$E_\mathrm{turb}$, but in total their energy is $\sim 2$ orders of magnitude
smaller than $E_\mathrm{turb}$ for the entire cloud. Our results
also suggest that for some clusters of star-forming cores outflows may
be energetically significant although their contribution is likely to
be far smaller if Perseus is considered in its entirety.   

\section{Summary}
\label{sec:summary}

This paper presented a survey of molecular outflows across the Perseus
molecular cloud. We analysed large-scale
\twelveco\ \threetotwo\ datasets (over 1000\,arcmin$^2$) undertaken with \harp\ on the
\jcmt\, towards four of the most active regions of star formation
(\ngc, IC348, L1448 and L1455),
combining previously published and new data. These new data across
over 600\,arcmin$^2$ of \ngc\ are the largest contiguous CO
\threetotwo\ maps published of \ngc\ to date, extending the mapping of
\citetalias{knee00} by a factor $\sim 10$. For each detected outflow
we calculate various parameters and use our complementary data \citepalias{paper1} to: (i)
correct the outflow mass, momentum and energy for the optical depth of
the \twelveco\ gas; and (ii) estimate the ambient cloud/driving-source
velocity at each spatial position. Our main conclusions can be summarized:
\begin{itemize}
\item[(i)] \textbf{Detection rates}. Of the 65 \scuba\
  cores in our fields \citep{hatchell07a}, we detect outflows towards 45
  (69\,per cent). We disagree with \citetalias{hatchell07b} in outflow
  (non-)detection towards only one of our 40 common
  sources, suggesting that the increased sensitivity of this survey does
  little better at differentiating starless and protostellar
  cores (as found by \citealt{hatchell09}). Indeed, spatial resolution is a greater limitation. 
\item[(ii)] \textbf{Length and maximum velocity}. We find Class 0
  sources drive longer and faster outflows than Class Is. $\langle
  L_\mathrm{lobe} \rangle = (140\pm 20)$\,arcsec and $\langle
  v_\mathrm{max} \rangle = (17.9\pm 1.4)$\,\kms\ compared to $(88\pm 11)$\,arcsec and $(12.2\pm 1.6)$\,\kms\ for Class 0 and I sources
  respectively. This might be explained as Class I outflows have wider
  opening angles than Class 0 \citep{arce06}, so their entrained material is more spread
  out and therefore less bright. Outflows also become less forceful as
  they age (see below and \citealp{bontemps96}) so their emission
  spans smaller velocity ranges and may be more easily masked by
  ambient emission. The ratio of the length of the
  longer to shorter outflow lobe is on average, $2.0 \pm 0.6$. This possibly indicates variations in the environment around the
    protostar. If, for instance, one half of the outflow breaks out of
  the molecular gas closer to the protostar than the other, this
  outflow lobe will correspondingly appear shorter. 
\item[(iii)] \textbf{Mass, momentum and energy evolution}. Class I
  outflows have less observable mass, momentum and energy than Class 0s. The mass
  distributions for each type are not significantly different with
  means of $\langle M_\mathrm{out} \rangle = (0.09\pm 0.02)$\,\msun\ and $(0.06\pm 0.03)$\,\msun\ for Class 0
  and I sources respectively. The momentum and energy
  distributions for the different types are significantly
  different. The Class 0 averages are
  $\langle p_\mathrm{out} \rangle = (0.7\pm 0.2)$\,\msun\,\kms\ and
  $\langle E_\mathrm{out} \rangle = (1.4\pm 0.5)\times
  10^{37}$\,J compared to $(0.10\pm 0.03)$\,\msun\,\kms\ and $(1.0\pm 0.3)\times
  10^{36}$\,J for the Class Is. In the Class I population, SVS13
  (omitted from the previous averages) is an anomalous source, possibly as it has a
  Class 0 protobinary companion (e.g.\ \citetalias{hatchell07b}). 
\item[(iv)] \textbf{Dynamical time-scale and momentum flux evolution}. We compute
   \tdyn\ using the ``$v_\mathrm{max}$'' method over the whole of the outflow. These
  \tdyn\ are unrealistic outflow ages showing little correlation with 
  \tbol\ and little difference between the two source
  classes: $\langle \tau_\mathrm{d} \rangle = (9800\pm 1600)$ and $(10000\pm 2000)$\,yr for the Class 0 and
  I average respectively. For Class I sources in particular it will
  probably underestimate the outflow age considerably, partly as we have
  trouble mapping these outflows accurately. The outflow momentum flux computed
  from \tdyn\ is nevertheless larger for Class 0 sources than Class I: $\langle F_\mathrm{CO} \rangle =(0.8\pm 0.3) \times 10^{-4}$ compared to $(1.1\pm
  0.3) \times 10^{-5}$\,\msun\,\kms\,yr$^{-1}$ (again with SVS13
  omitted). This difference is emphasized if we compute the momentum
  flux using the protostellar age derived from \tbol\ instead of the
  observed outflow dynamical time-scale: $\langle F^*_\mathrm{CO}
  \rangle =(4.3\pm 1.5) \times 10^{-4}$ (Class 0s) compared to $(6\pm
  2) \times 10^{-6}$\,\fcounit\ (Class Is). Class 0 sources are
  systematically more efficient at driving outflows than Class I. If $F_\mathrm{rad}
  =L_\mathrm{bol}/c$ is the flux expected in radiation from the
  central source then
  $F_\mathrm{CO}(\mathrm{Class\,I})\sim 100F_\mathrm{rad}$ and
  $F_\mathrm{CO}(\mathrm{Class\,0})\sim 1000F_\mathrm{rad}$. The
  momentum flux also correlates with the source mass and inversely with
  the source \tbol. Such a decrease in force between stages, with age,
  suggests a decline in the mass accretion rate (see
  \citealp{bontemps96}) or possibly a different entrainment efficiency
  which may occur if for instance a second wind component has started to dominate.
\item[(v)] \textbf{Destruction time-scale}. If mass is only lost from
  the core via its outflow at the current rate, then the
  cores will last too long. The
  average calculated destruction time-scale is a few million years
  compared to the average lifetime of the protostellar stage which is
  a few times $10^5$ years. This is probably because our estimates of the masses
  of the star-forming cores are too large. 
\item[(vi)] \textbf{Global effect of the outflows}. We estimate the
  maximum mass, momentum and kinetic energy available to each region
  from their outflows by summing the contributions from each
  individual outflow and applying an average correction for
  inclination. This is compared to some overall cloud properties: the
  turbulent momentum, energy and binding energy, derived from our
  \ceighteeno\ data. We find there is more energy in outflows than is
  observed in bulk, turbulent motions in \ngc, L1448 and
  L1455. Indeed, the outflows in L1448 are extremely powerful,
  containing twice the binding energy of the region. We note the
  difficulty in feeding-back high-energy outflowing gas, often outside of
  the main cloud core, into turbulent motions. However, if it were
  possible with only a fraction of the outflow energy in L1448, it seems likely the cloud would
  be broken apart.
\end{itemize}

\section{Acknowledgments}

EIC acknowledges studentship support from the STFC. JPW thanks the NSF for support. The authors thank Jane Buckle and Gary Fuller for suggesting
improvements to an early version of this paper and Sarah Graves for
useful discussions and advice. We are grateful to the referee,
Jennifer Hatchell, for comments and suggestions which improved the
clarity of this paper. The James Clerk Maxwell Telescope is operated by The Joint Astronomy
Centre on behalf of the STFC of
the United Kingdom, the Netherlands Organisation for Scientific
Research and the National Research Council of Canada. We have also
made extensive use of the SIMBAD data base, operated at CDS,
Strasbourg, France. The authors acknowledge the data analysis
facilities provided by the Starlink Project which is maintained by JAC
with support from STFC. This research used the facilities of the Canadian
Astronomy Data Centre operated by the National Research Council of
Canada with the support of the Canadian Space Agency. 

\renewcommand{\bibname}{References}

\appendix

\section{Outflow description and background} 
\label{appendix:outflows}

\subsection{NGC\,1333}

\subsubsection{Overview} 

\begin{figure*} 
\begin{center}
\includegraphics[width=\textwidth]{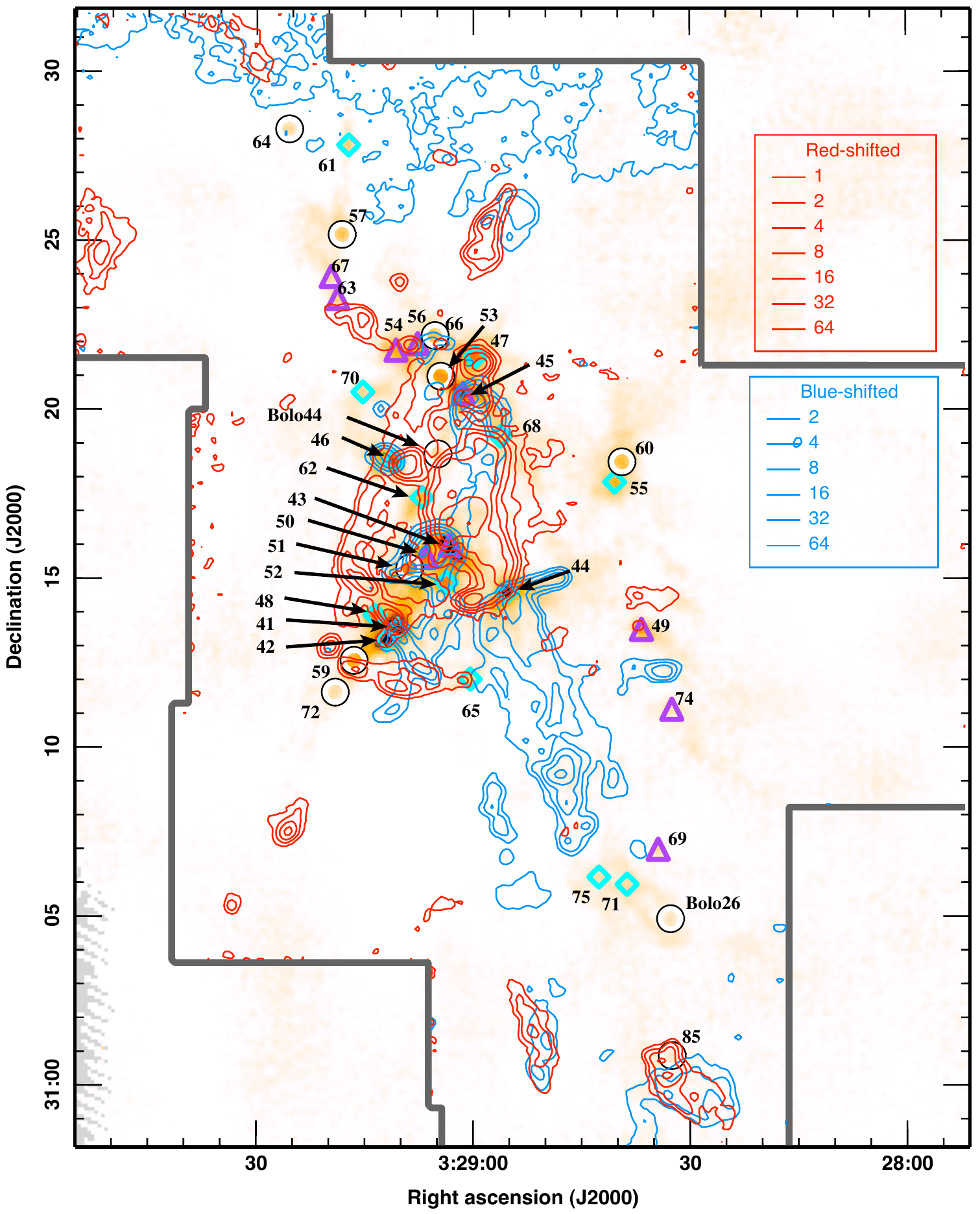}
\caption{High-velocity CO outflows towards \ngc. Red- and blue-shifted
  CO \threetotwo\ emission has been integrated ($\int
T_\mathrm{A}^*\,\mathrm{d}v$) from $-5$ to $3$\,\kms (blue) and from
12 to 18\,\kms\ (red) and is contoured at the levels shown (in
\kkms). The colour-scale is \scuba\ 850\,\micron\ flux density scaled
from 0 to 1600\,\mjybeam. The thick grey line marks the area surveyed
for outflows with HARP.}
\label{fig:ngc1333_outflows} 
\end{center}
\end{figure*}

We present an overview of the outflows in \ngc\ in
Fig.\ \ref{fig:ngc1333_outflows}. Most of the high-velocity gas lies in
the central regions, also mapped by \citetalias{knee00}, although the full extent of
these outflows is better traced here. The blue-shifted emission is dominated by a large-scale flow
north-south from HRF44 and the southeastern blue-shifted lobe from
SVS13 (HRF43). The most prominent red-shifted feature expands into the central
dust cavity, surrounded by the most well-know star-forming cores \citep{sandell01}.     

\clearpage

\subsubsection{NGC\,1333-IRAS2}

HRF44 is better known as the infrared source NGC\,1333-IRAS2, discovered
by \citet{jennings87} in \emph{IRAS} data. In high-resolution
millimetre studies, three sources are apparent: two protostellar (2A
and 2B) and one starless (2C)
(\citealp{blake97}; \citealp*{looney00}). The protostars are also detected at
centimetre wavelengths (e.g.\ \citealp*{rodriguez99}). Two outflows are
associated with the source, one jet-like orientated east-west
\citep{sandell94,bachiller98} and the other shell-like approximately
north-south \citep*{liseau88}. \citetalias{knee00} point out such
different structures may imply they arise from separate components of
IRAS2. Alternatively, a single binary source is possible, supported by the
interferometric observations of \citet{engargiola99} which pinpoint
the origin of both flows at IRAS2A. 

A detailed picture of the outflows around IRAS2 is shown in Fig.
\ref{fig:iras2_outflows}. The two orthogonal flows are clear. The red lobe encompassing
\htwo\ knot 32, extends into the confused area around the central
cavity and \citetalias{knee00} suggest it created part of
this void. High-velocity blue-shifted gas, extends to HH13 in the
south, probably causing a number of knots en route. The lower
velocity gas is even more knotty and supports the continuation of the
flow even further south (\citealp{davis08} suggest it extends
2.2\,pc). However, HRF75 and HRF71 (Class 0) alongside HRF69 (Class I) may drive flows that interfere or contribute to this
lobe, even though there is little \emph{strong} outflow activity around them,
perhaps just an east-west flow from HRF71. 
 
\begin{figure} 
\begin{center}
\includegraphics[width=0.4\textwidth]{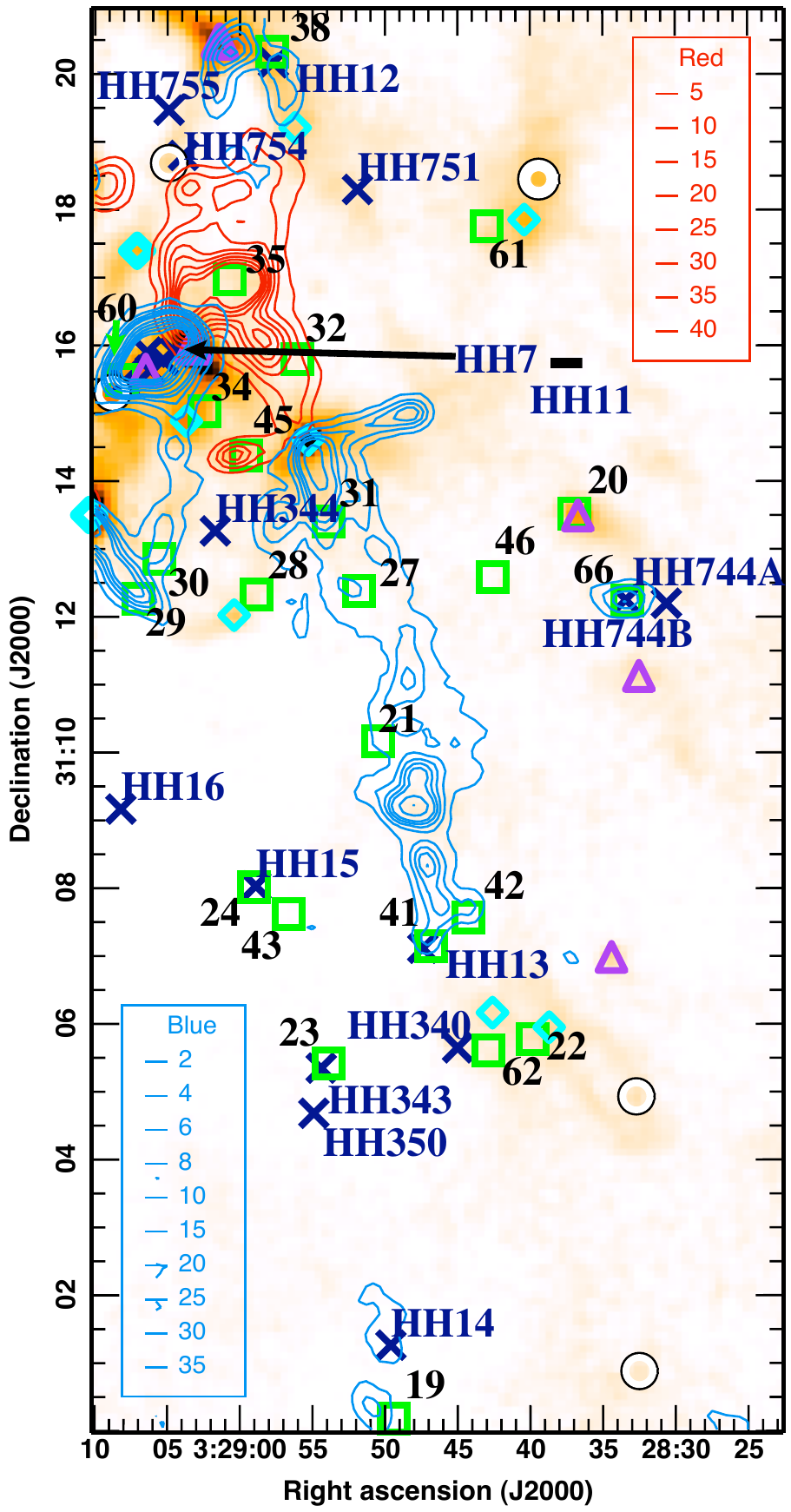}
\caption{High-velocity CO \threetotwo\ outflows towards
  NGC\,1333-IRAS2, overlaid on a \scuba\ 850\,\micron\ data. Symbols mark submillimetre cores classified by
  \citet{hatchell07a} as: starless (black/white circles),
  Class 0 (light blue diamonds) and Class I (purple
  triangles). Green squares are \htwo\ knots \citep{davis08}
and blue crosses are HH objects \citep{walawender05}. $\int
T_\mathrm{A}^*\,\mathrm{d}v$ from: $-10$ to 1\,\kms\
  (blue) and 14 to 25\,\kms\ (red).}
\label{fig:iras2_outflows} 
\end{center}
\end{figure}

\begin{figure} 
\begin{center}
\includegraphics[width=0.4\textwidth]{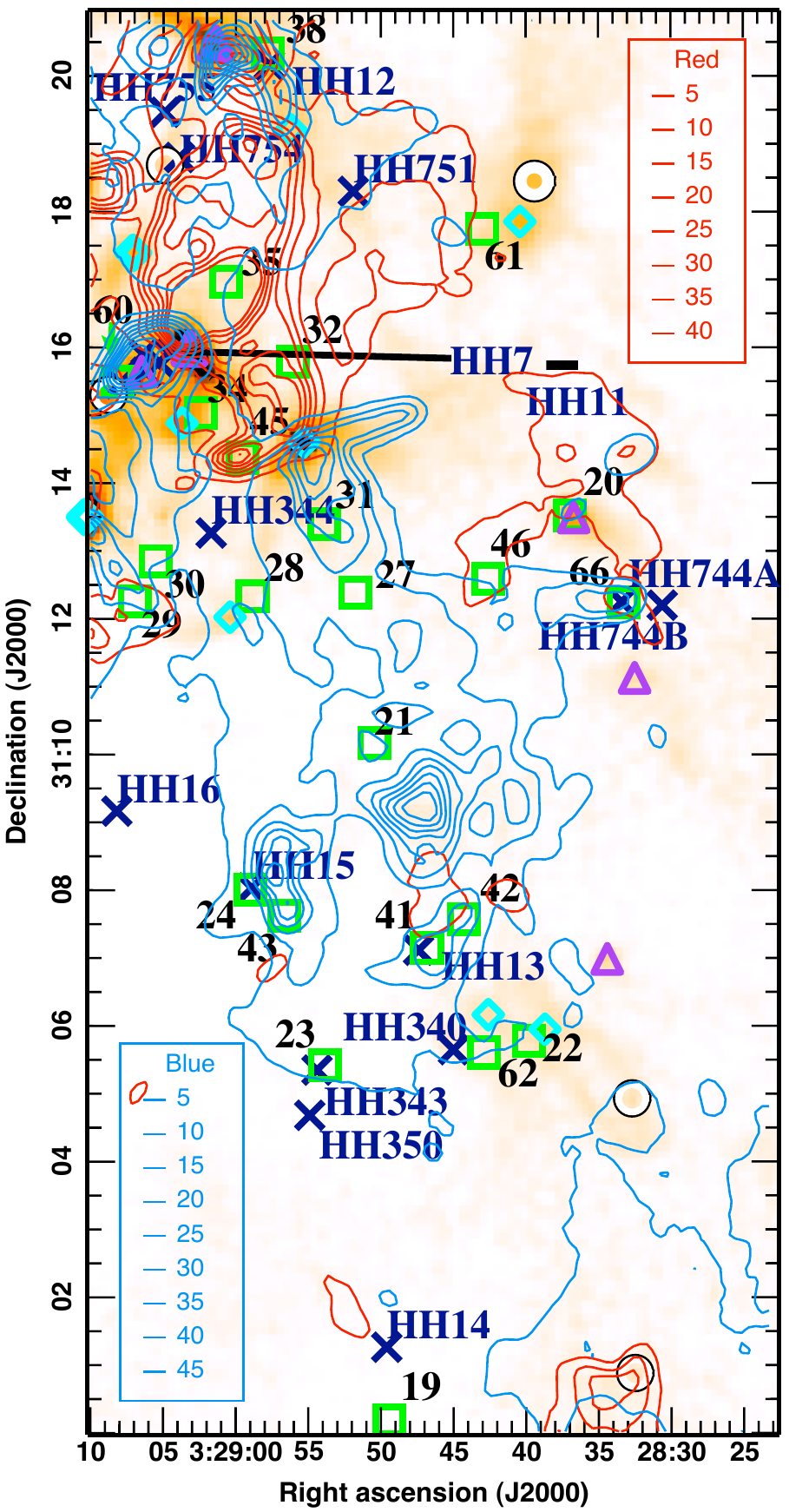}
\contcaption{Low-velocity CO
  gas, $\int
T_\mathrm{A}^*\,\mathrm{d}v$ from: 1 to 5\,\kms\ (blue) and 10 to 14\,\kms\ (red).}
\end{center}
\end{figure}

\subsubsection{NGC\,1333-IRAS4}

IRAS4 \citep{jennings87} contains at
least four distinct sources. \citet{sandell01} resolved two
components: 4A (HRF41) and 4B (HRF42), separated by 30\,arcsec. Each of
these are multiple objects unresolved with \scuba: 4A/4A$'$ \citep*{lay95} and 4B/4B$'$ \citep{looney00}. Another
component 4C (HRF48), $\sim40$\,arcsec north-northeast of the main pair, was
discovered by \citet{rodriguez99}. The outflow was first mapped in CO \threetotwo\ by
\citet{blake95}. They found a change of direction
in the flow from a position angle\footnote{All position
  angles are specified east of north.} (PA) of 45\,deg far away to
0\,deg close to IRAS4A coupled with a string of symmetric features
either side of the source, suggesting some variability in the
ejection. In interferometric data, a small outflow from IRAS4B is
clear (e.g.\ \citealp{jorgensen07b}). 

\begin{figure} 
\begin{center}
\includegraphics[width=0.45\textwidth]{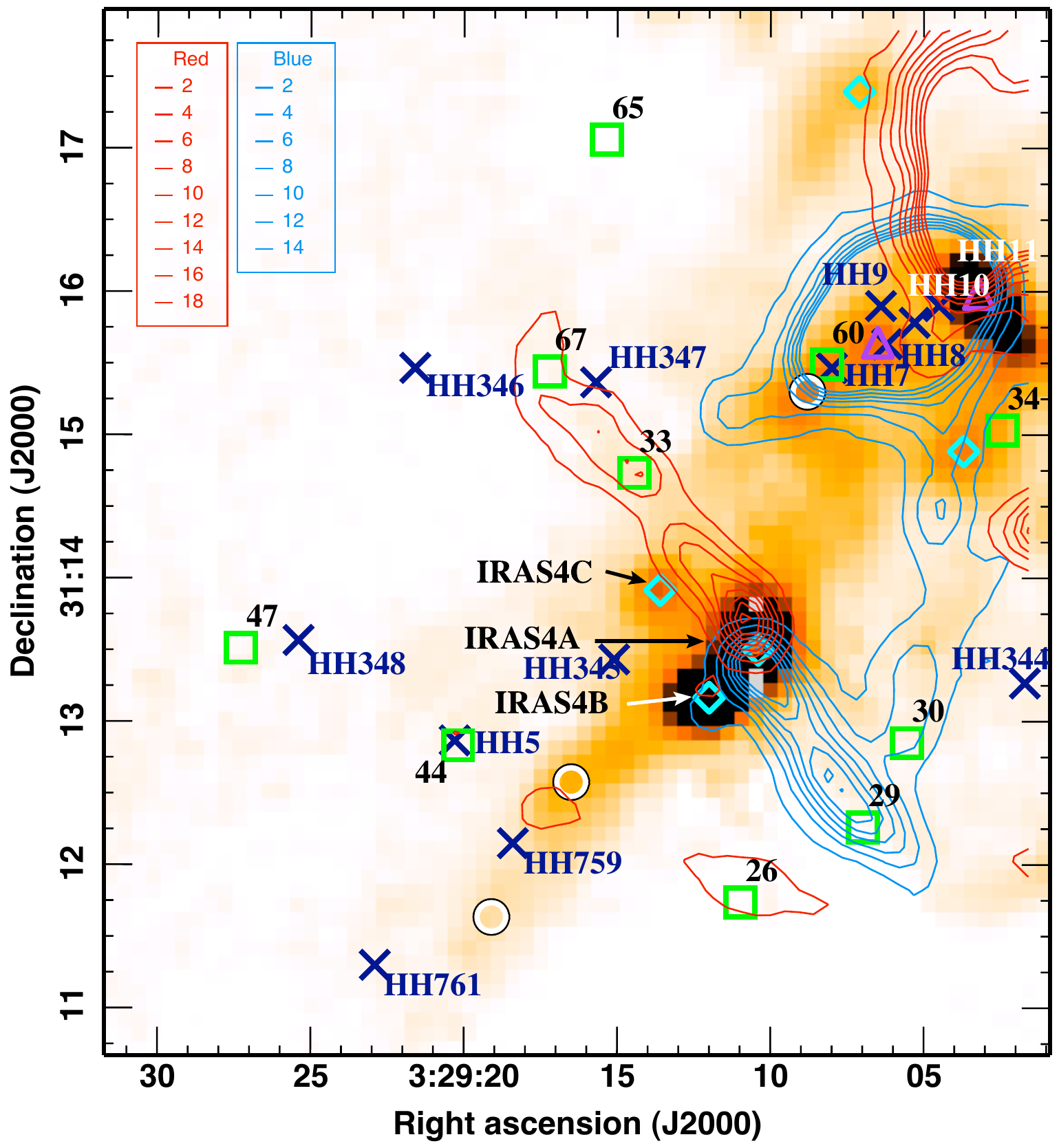}
\includegraphics[width=0.45\textwidth]{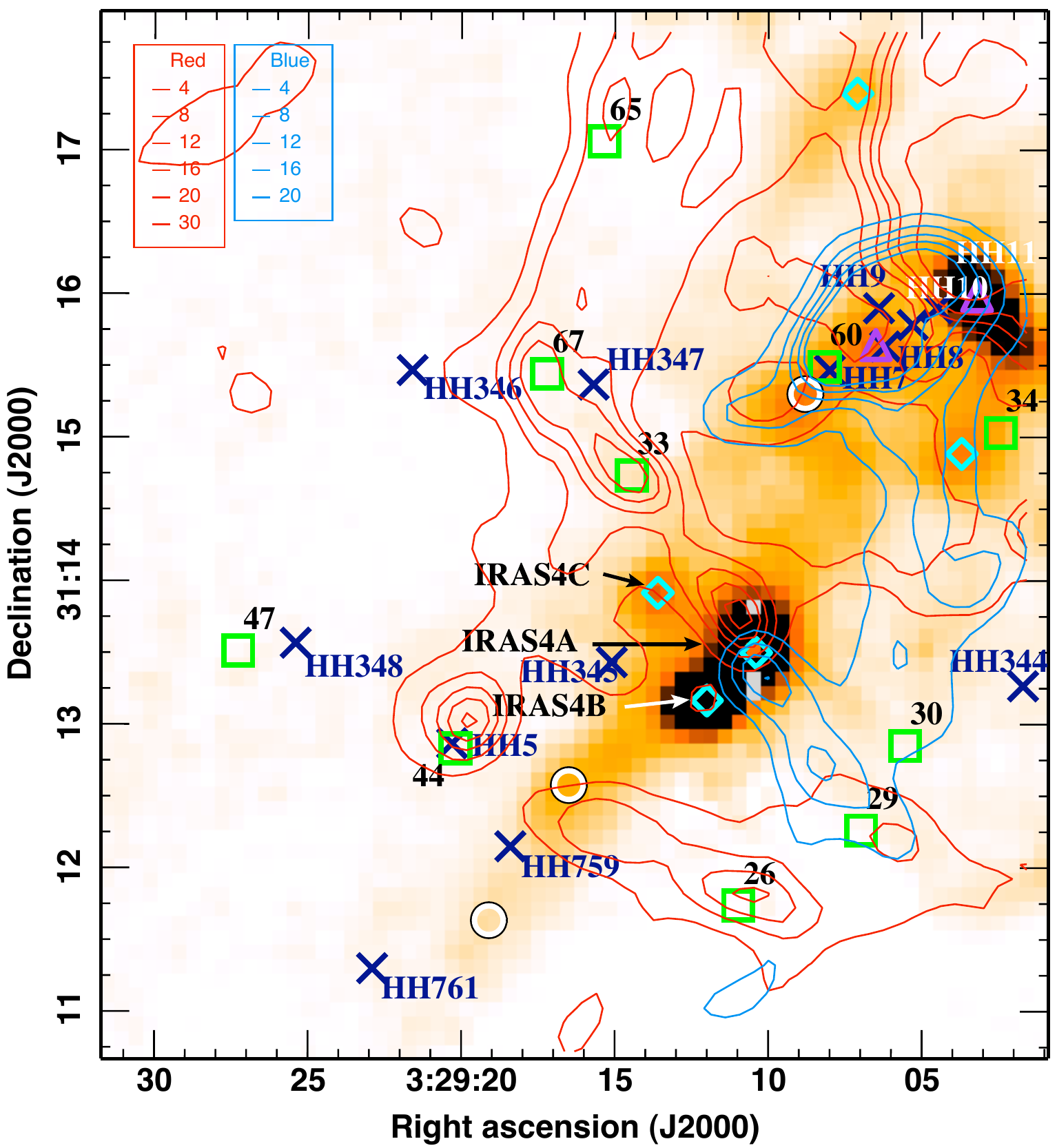}
\caption{NGC\,1333-IRAS4 as
  for Fig. \ref{fig:iras2_outflows}. Upper panel: High-velocity
  CO outflows, integrated from $-10$ to 1\,\kms\ (blue) and 14 to
  25\,\kms\ (red). Lower panel: Low-velocity
  CO outflows, integrated from 1 to 4\,\kms\ (blue) and 10 to 14\,\kms\ (red).}
\label{fig:iras4_outflows} 
\end{center}
\end{figure}

The outflows we detect are plotted in Fig.
\ref{fig:iras4_outflows}. A tiny north-south flow from 4B is
discernible along with the stronger flow from 4A which
encompasses \htwo\ knots 32, 39 and 67. The change in its direction
away from the driving source is clear. Beyond the
high-velocity red lobe, \citet{difrancesco01} suggest the flow
bends to the north to encompass the red-shifted gas
along the eastern side of the entire central region (which
alternatively may originate from a source near HH6), resulting from a variation in the jet direction
caused by the protobinary. SiO observations by
\citet{choi05} show a very straight bipolar flow emanating
from IRAS4A$'$, with a weaker blue lobe in a north-south flow
from IRAS4A. The main north-easterly red lobe has a dramatic change of
direction to the east (just beyond where our CO emission ends) which
\citeauthor{choi05} interprets as evidence of an obstacle in the path
of the flow. Thus, it
seems probable that our CO flow does not veer north but
rather south and the apparent change of direction in the outflow near
to the driving source is an unresolved flow from its close
companion.   

\subsubsection{The SVS13 ridge to HH12} \label{sec:svs13}

The ridge of dust just south of the central cavity is an exceptionally
rich star-forming site, containing the well known SVS13 source (\citealp*{strom76},
HRF43) which is usually associated with NGC\,1333-IRAS3 \citep{jennings87}. The
HRF43 core actually contains numerous sources, in order
from north-east to south-west: (i) SVS13, thought to power the symmetric CO outflow (e.g.\ \citealp{snell81}) associated
with HH7-11 \citep{herbig74}. (ii) SVS13B, a probable Class
0 source \citep{grossman87} and (iii) H$_2$O(B), a maser source
\citep{haschick80,hirota08}. \citet{rodriguez99} additionally found two
centimetre sources, VLA3 and VLA20, interspersing these.

Additional outflows are very confused; both
\citetalias{knee00} and \citet{davis08} propose two further outflows from
the ridge, tentatively suggested by \citetalias{knee00} to be driven
by SVS13B and H$_2$O(B). The first, from SVS13B is almost north-south,
its blue lobe apparent in Fig. \ref{fig:iras4_outflows}, joining the SVS13
ridge to the blue lobe of IRAS4A. The second from H$_2$O(B), passes
close to the long blue lobe of IRAS2 (see Fig.
\ref{fig:iras2_outflows}) and possibly produces the chain of features
south to HH14. There is certainly outflowing gas in the vicinity
  of HRF50 (the Class I source near the peak of the blue-shifted gas, which
  also encompasses HH7-11). However, HH7-11 are commonly associated
  with SVS13 and \emph{not} HRF50. Perhaps some of the nearby
  outflowing gas emanates from HRF50, but it is difficult and
  rather arbitrary to decompose the emission further so we do not
  calculate outflow parameters for HRF50 (i.e.\ in Table
\ref{table:outflow_params1}).

\begin{figure} 
\begin{center}
\includegraphics[width=0.45\textwidth]{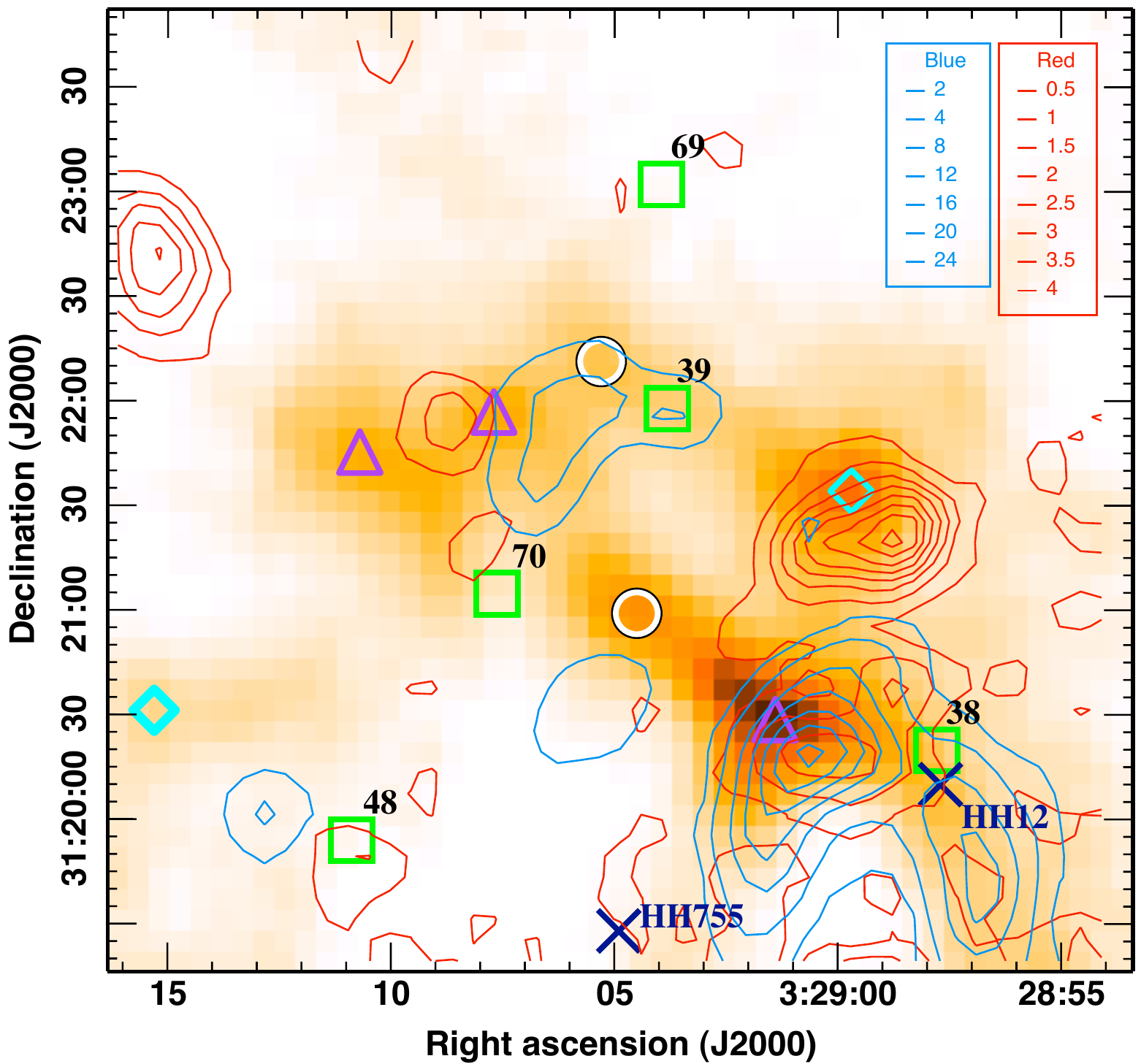}
\includegraphics[width=0.45\textwidth]{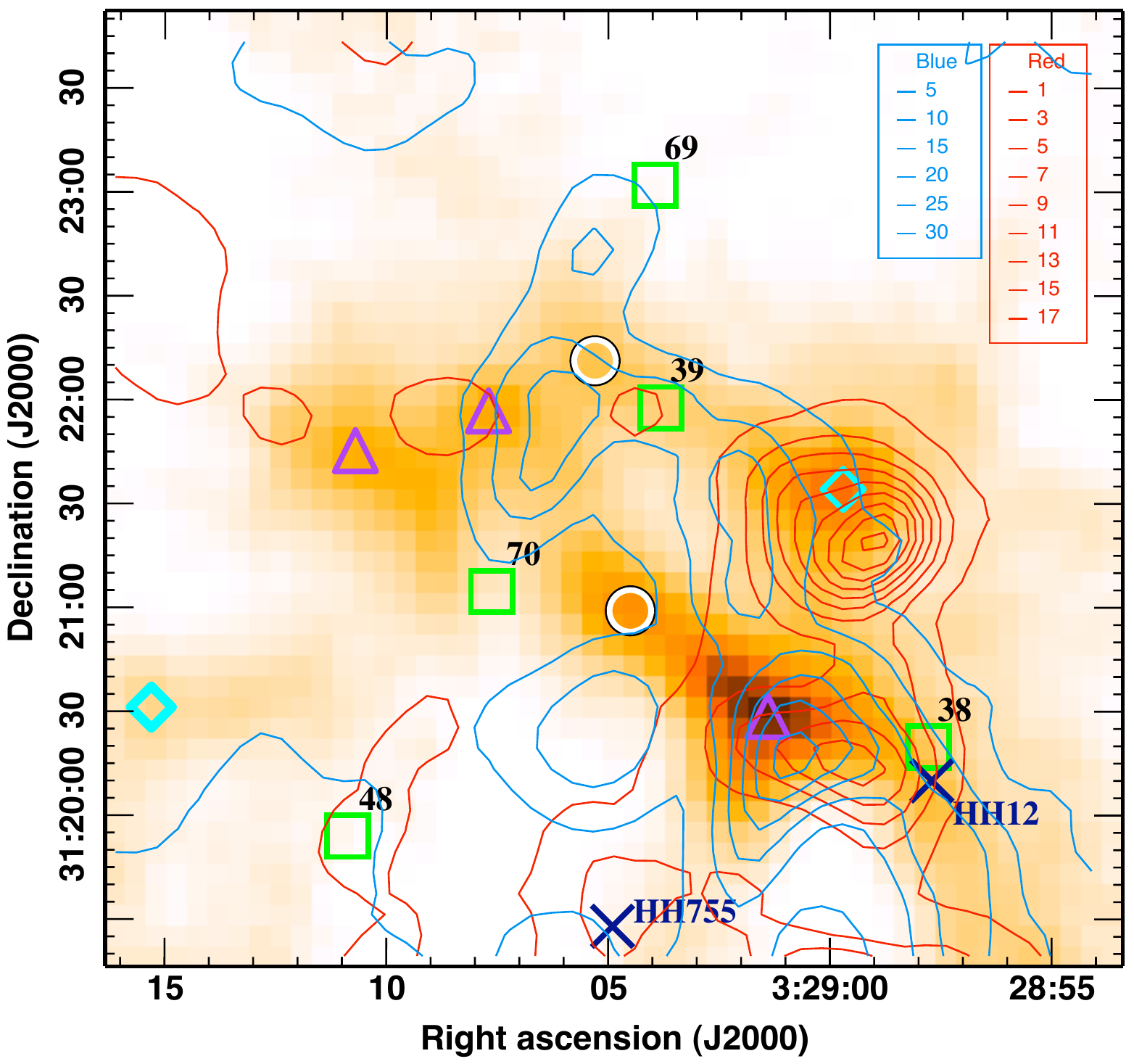}
\caption{HH12 as for Fig. \ref{fig:iras2_outflows}. Upper panel:
  High-velocity CO outflows, integrated from $-2$ to 3\,\kms\ (blue) and 15 to 20\,\kms\ (red). Lower panel: Low-velocity
  CO outflows, integrated from 3 to 5\,\kms\ (blue) and 12 to 14\,\kms\ (red).}
\label{fig:hh12_outflows} 
\end{center}
\end{figure}

\citetalias{knee00} propose that the horseshoe
of blue-shifted gas associated with HH12 (see Fig.
\ref{fig:hh12_outflows}) is the counterpart to the SVS13B outflow. Its
blue-shifted nature may result if the flow hits a density
gradient and is deflected. \citet{davis08} prefer an
explanation that infers an outflow driven by a source inside
HH12 (the most likely candidate being HRF45, a Class I protostar). Our data do not contribute much further, certainly there are a lot of
potential driving sources in the vicinity of HH12, some with
outflowing gas, although bipolar structures are not
obvious. In our analysis (Section \ref{sec:outflowevolution}), we attribute the horseshoe and
overlapping red-shifted emission to nearby HRF45.

\subsubsection{HH6}

\begin{figure} 
\begin{center}
\includegraphics[width=0.45\textwidth]{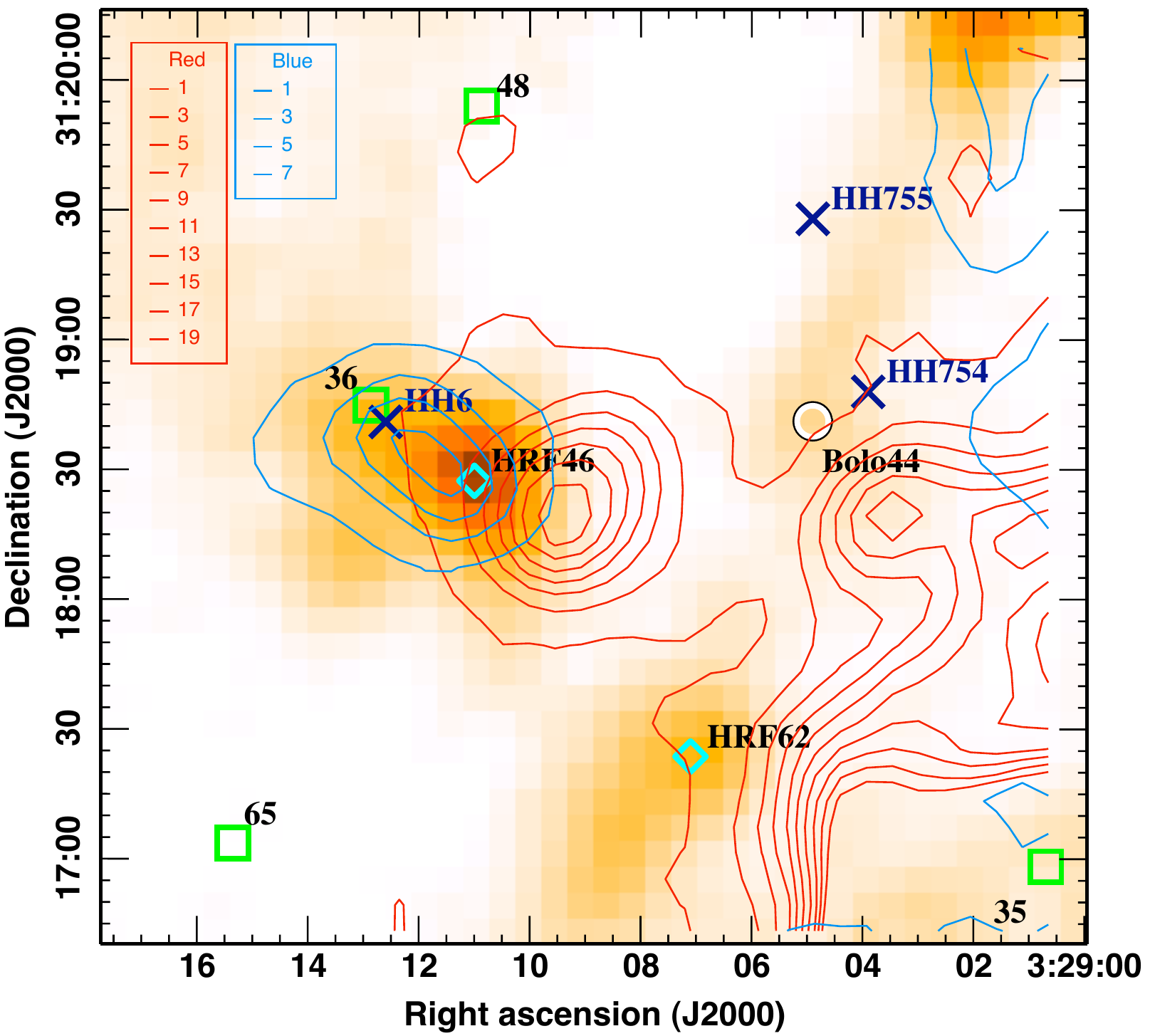}
\includegraphics[width=0.45\textwidth]{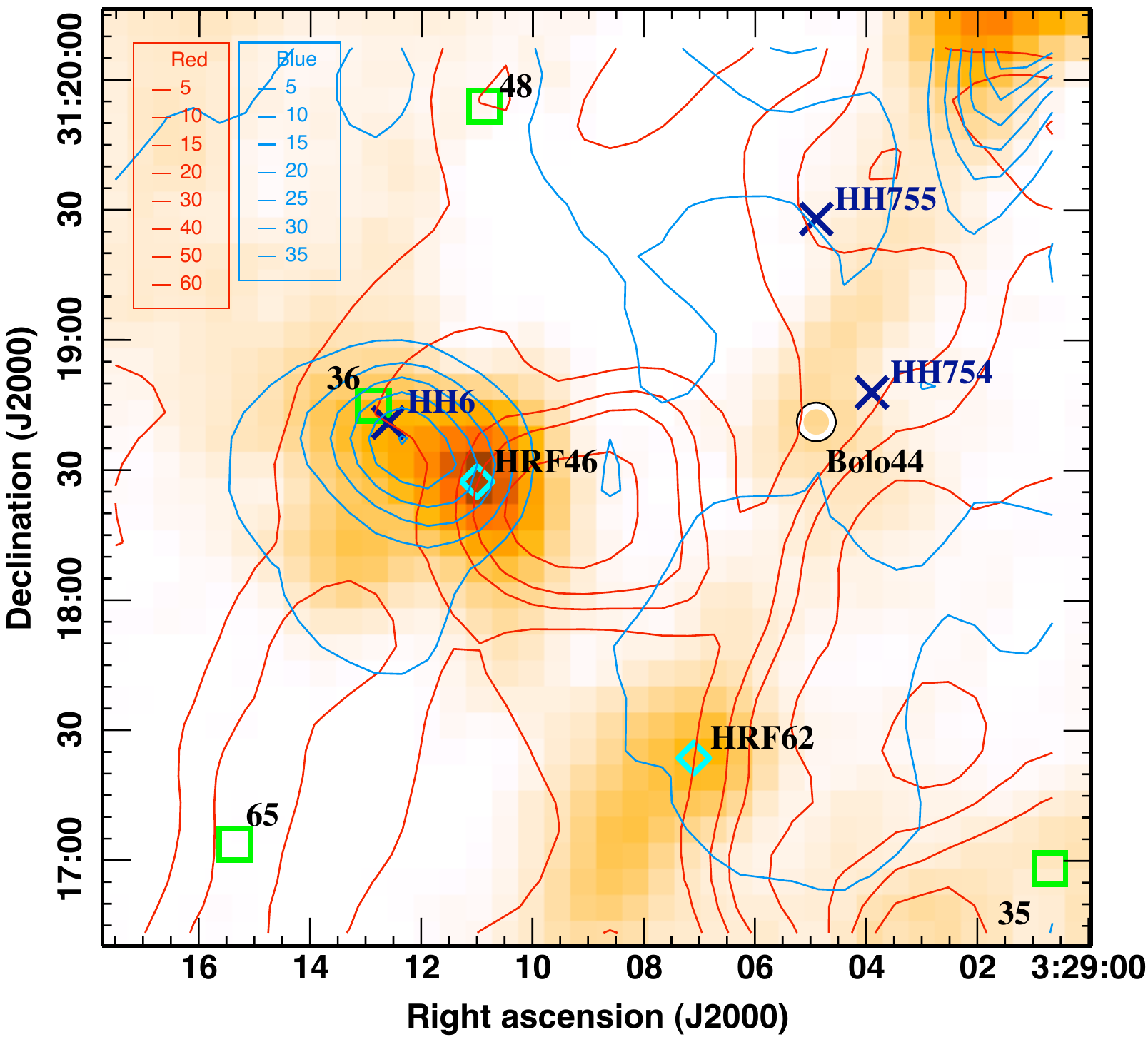}
\caption{HH6 as for Fig. \ref{fig:iras2_outflows}. Upper panel: High-velocity
  CO outflows, integrated from $-7$ to 0\,\kms\ (blue) and 15 to
  22\,\kms\ (red). Lower panel: Low-velocity CO outflows, integrated from 0 to 5\,\kms\ (blue) and 10 to 15\,\kms\ (red).}
\label{fig:hh6_outflows} 
\end{center}
\end{figure}

The HH6 jet (\citealp{lightfoot86}; \citealp*{cohen91}) and associated outflow
\citep{liseau88} to the north-east of SVS13 and
east of the central cavity are driven
by the Class 0 protostar NGC\,1333-IRAS7 (\citealp{jennings87};
HRF46). Again, it is not a single submillimetre core but two
\citep{sandell01}, both protostellar \citep{jorgensen07}. Either one
(both unresolved as HRF46 in \citealt{hatchell07a}) could drive the
main flow in Fig. \ref{fig:hh6_outflows}.
In addition, there is a long
orthogonal flow, detected in the near IR (\citealp{cohen91}; \citealp*{aspin94}), thought
to extend down the eastern side of the map from HH5 to the most
northerly \htwo\ feature in Fig. \ref{fig:ngc1333_outflows} \citep{davis08}. \citetalias{knee00}
note red-shifted CO from this flow in the south (as we do),
but no blue-shifted to the north where they suggest there is only an
infrared jet. We see the red-shifted CO gas continuing along the entire
jet extent suggested by \citet{davis08}. Indeed it is this trail that
\citet{difrancesco01} may confuse with a deflection north of the
IRAS4A outflow. 

\subsubsection{Elsewhere}

\begin{figure*} 
\begin{center}
\includegraphics[width=\textwidth]{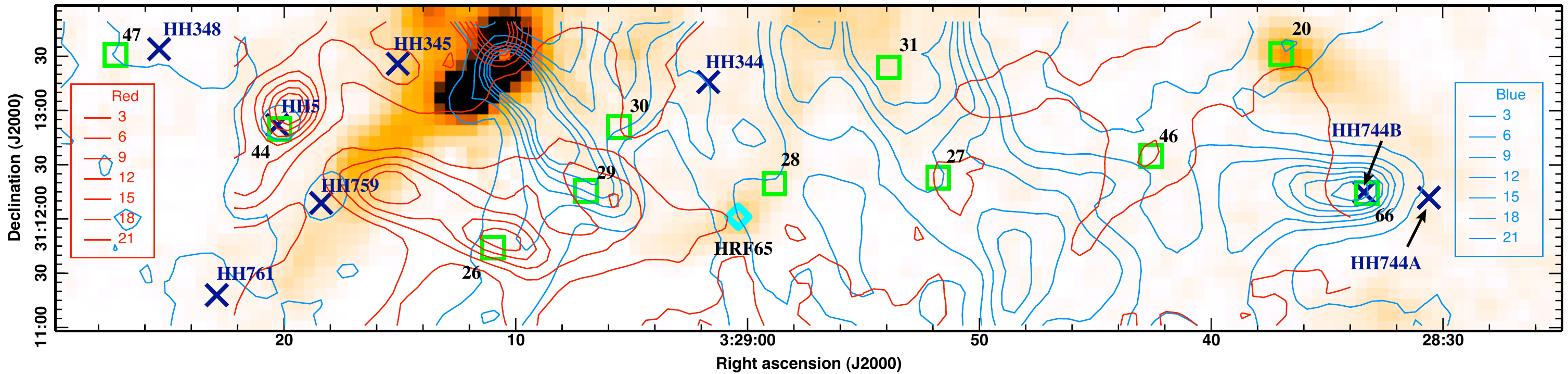}
\caption{HRF65 as for Fig. \ref{fig:iras2_outflows}. Integration limits: $-2$ to 5\,\kms\ (blue) and 9 to 15\,\kms\ (red). }
\label{fig:65_outflows} 
\end{center}
\end{figure*}

Very few of the remaining cores drive clear bipolar outflows, despite the vast majority having
high-velocity gas (see Section \ref{sec:outflow_detection_rates}). One
further distinct detection comes from HRF65 (see
Fig. \ref{fig:65_outflows}). It drives a long
collimated flow clear in Fig. \ref{fig:ngc1333_outflows}, just to the
south of \citetalias{knee00}'s field. Its blue lobe crosses the
southbound outflows from the SVS13 ridge and IRAS2 via numerous \htwo knots, before terminating
in the \htwo knot 66 \citep{davis08} and HH744A/B. The red lobe
follows a more curved path, possibly extending towards HH5, although
its structure in the \htwo\ images suggests it is a bow shock from the
north, possibly originating from HRF46 (see above). \citet{hatchell09}
also find a new outflow from HRF65, although their smaller map is not
able to follow this highly-collimated outflow along such a large extent.  

\subsection{IC348} \label{sec:outflows_ic348}

The centre of the IC348 cluster is devoid of star formation activity
with the youngest protostars concentrated in a ridge 
10\,arcmin south-west (e.g.\ \citealp{muench07}), around an IR object
(IC348-IR) first identified by \citet{strom76}. A number
of dense cores were found in \ammonia\ and CO by \citet*{bachiller87},
before \citet*{mccaughrean94} discovered HH211, a molecular
jet that has been the target of numerous interferometric studies
(e.g.\ \citealp{gueth99,chandler01}). The largest existing outflow
survey is by \citet*{tafalla06}, in CO \twotoone\ with the \iram\ 30-m telescope at 11\,arcsec
resolution. \citetalias{hatchell07b} find many more
distinct outflows although some overlap, causing confusion. 

\begin{figure} 
\begin{center}
\includegraphics[width=0.5\textwidth]{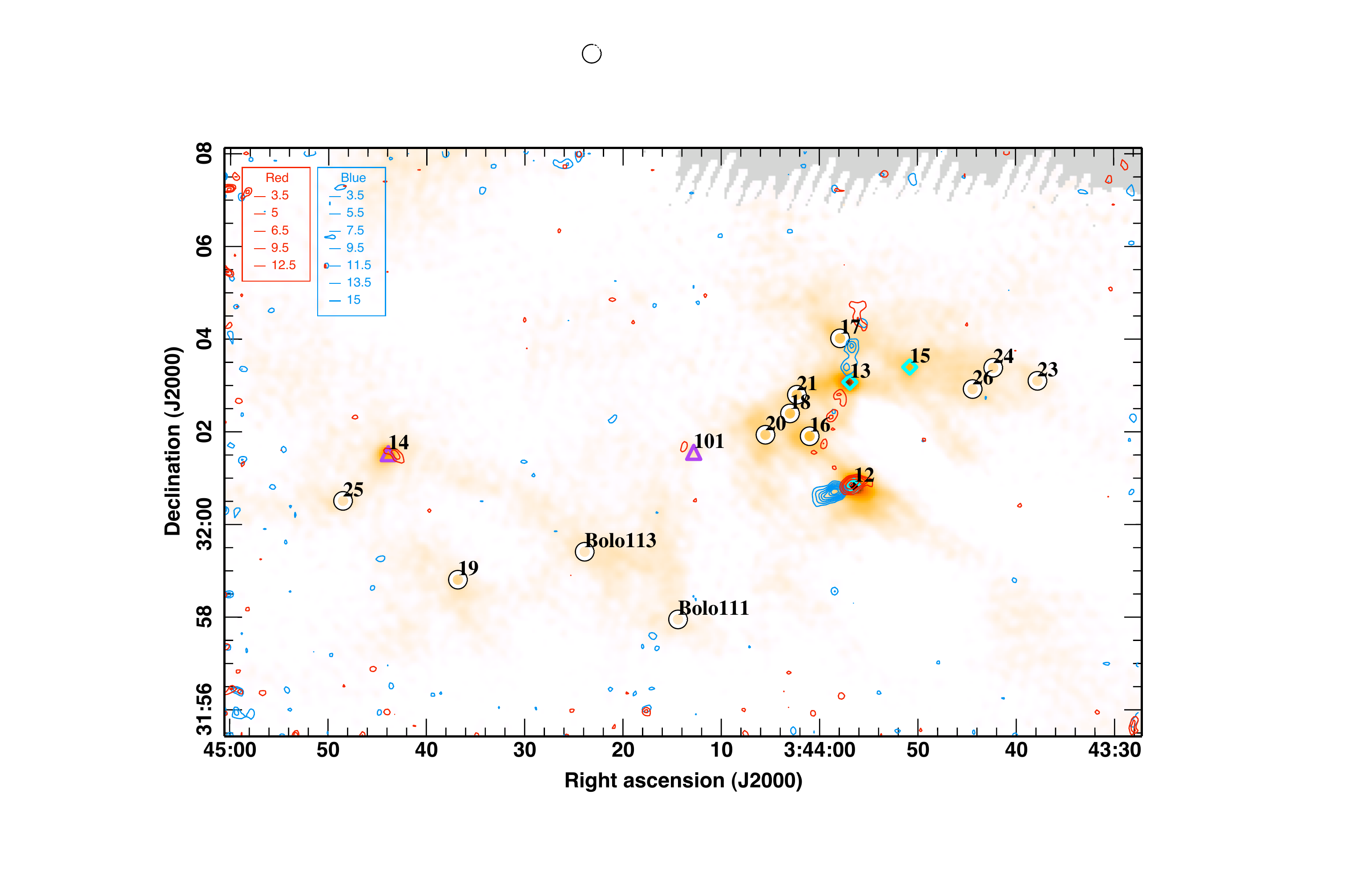}
\includegraphics[width=0.5\textwidth]{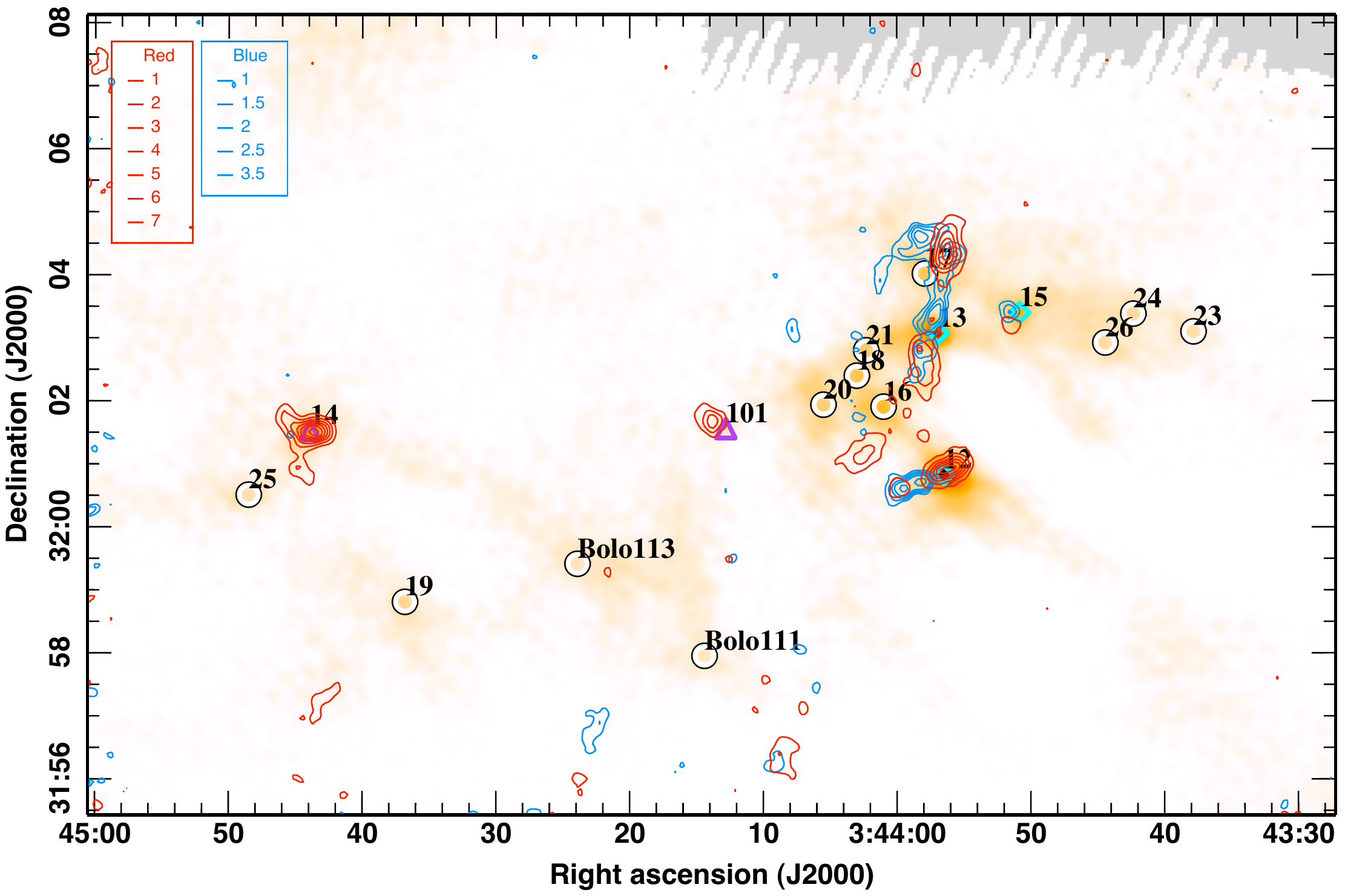}
\includegraphics[width=0.49\textwidth]{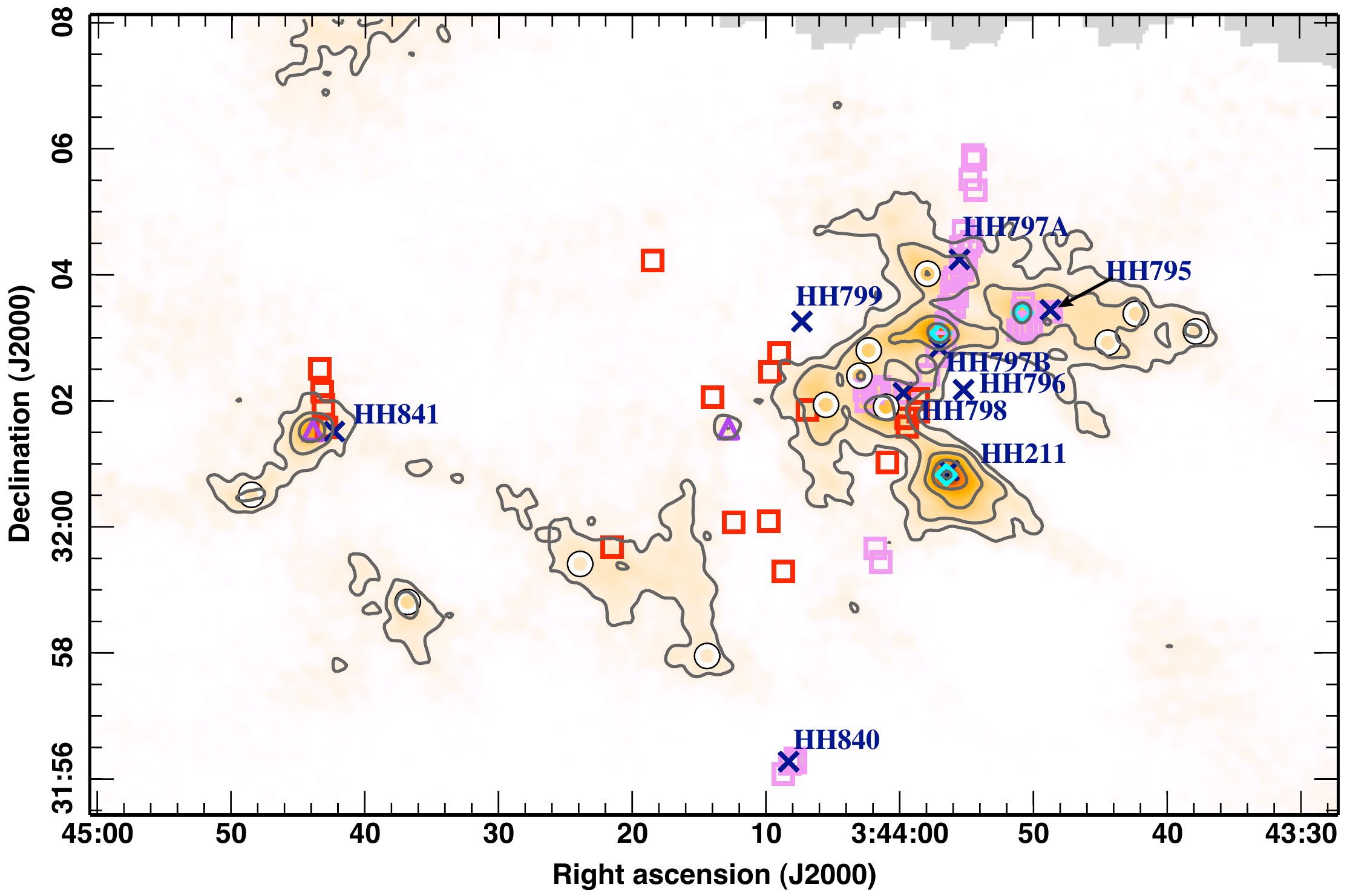}
\caption{CO \threetotwo\ outflows in IC348 as for
  Fig.\ \ref{fig:ngc1333_outflows}. Top: High-velocity
  CO outflows, integrated from $-7.4$ to
  $3.6$\,\kms\ (blue) and 13.6 to 23.6\,\kms\ (red). Middle: Low-velocity
  CO outflows, integrated from 3.6 to 5.6\,\kms\ (blue) and 11.6 to
13.6\,\kms\ (red). In the bottom panel squares denote H$_2$ knots identified by
\citet{eisloffel03} (pink) and \citet{walawender06} (red). }
\label{fig:ic348_outflows} 
\end{center}
\end{figure}

We present maps of the outflows in Fig.\ \ref{fig:ic348_outflows}. The two most
prominent outflows are the north-south flow driven
by HRF13, also known as IC348-MMS \citep{eisloffel03} or IC348-SMM2
\citep{walawender06,tafalla06} and HH211 driven by
HRF12. In the slow gas, the weak bipolar outflow discovered by
\citet{tafalla06} from HRF15 or IC348-SMM3 can be discerned with
further one-sided flows around HRF14 and HRF101, which are also noted
by \citet{hatchell09}.  

\citet{eisloffel03} suggest the IC348-MMS outflow extends some 10\,arcmin,
encompassing the H$_2$ knots in a line to its north along with the
group of knots south, next to HH840. IC348-MMS drives a complex bipolar
flow, in the north there is
red-shifted gas in the same place and beyond the blue-shifted
lobe. Furthermore, slow gas in the blue lobe appears to bend
around on itself. There are a number of different explanations for such; first the anomalous red emission might simply come
from another outflow. A likely driving source is HRF17, classed as
starless by \citet{hatchell07a}. \citetalias{hatchell07b} report a tentative outflow detection towards it,
although they note the potential confusion with the stronger
flow. However, the outflowing gas bends around HRF17 instead of
obviously emanating from the its position. \citet{tafalla06} suggest that this superposition of
red- and blue-shifted gas could result if the IC348-MMS outflow lies close to the plane of the sky
(e.g.\ \citealp*{cabrit88}) or if the outflow has broadened after
hitting a lower density medium. The arc of blue-shifted emission is
more perplexing, given that the outflow wind must continue in a
straight-line to encompass the northern H$_2$ jets. Again,
\citet{tafalla06} explain this simply as an unrelated gas, flowing into
a circular hole, evacuated by the cluster. In our data the blue arc is
part of a larger structure between 4 and 6\,\kms\ that extends to the
north-east, into a lower density area of the \scuba\ map. There is some
evidence of a collimated filament of molecular material, coincident with the southern H$_2$ knots attributed
to the same IC348-MMS flow by \citet{eisloffel03}, in blue-shifted velocity
channels between 4 and 6\,\kms\ and the red-shifted ones between 11 and 13\,\kms.

Elsewhere, there are a number of red-shifted flows with
no blue counterpart. HRF101 has a small red-shifted flow, that lies
in line with a number of H$_2$ knots \citep{walawender06}. The only
outflow activity in the east resides around HRF14,
whose red lobe encompasses HH841, an optical knot embedded in a reflection nebula
\citep{walawender05}. There is evidence of a weak blue lobe on
the opposite side of HRF14, prominent in the 4--5\,\kms\ channel. 

\subsection{L1448}

L1448 harbours a rich cluster of Class 0 protostars. \emph{IRAS} found three sources: L1448-IRS1 to IRS3. L1448-IRS1, to the west
and slightly north of the \scuba\ cores (at
03$^\mathrm{h}$25$^\mathrm{m}$09$^\mathrm{s}$, $+30^\circ$46$'$21$''$), has been identified as a Class I
protostar \citep{eisloffel00} but contains no compact 850\,\micron\ dust
emission and therefore was not examined by \citet{hatchell07a}. \citet{bally08} note it has a visual counterpart
\citep{cohen79} and optical reflection nebula association (RNO13) and
is thus likely to be a Class II source. Additionally, the c2d survey
do not detect it in \emph{Spitzer}'s {\sc IRAC}\footnote{Infrared
  Array Camera observing at 3.6, 4.5, 5.8 and 8\,\micron.}  bands
\citep{jorgensen07}, cataloging it as `star+dust'. However,
\citet{davis08} suggest this non-detection is due to saturation and
associate IRS1 with a number of \htwo\ knots and HH194. L1448-IRS2 is a
Class 0 protostar that \citet{wolf-chase00} suggested is actually
a binary (HRF30 and HRF31), driving two CO outflows. The third
source, L1448-IRS3, consists of three or more Class 0 protostars
(e.g.\ \citealp{olinger06,looney00}): L1448C (HRF29 also known as
L1448-mm) which drives the best studied outflow in the region,
L1448N:A,B (HRF28, a protobinary separated by $\sim 7$\,arcsec,
\citealp{curiel90}) and L1448NW (HRF27). 

\begin{figure} 
\begin{center}
\includegraphics[width=0.49\textwidth]{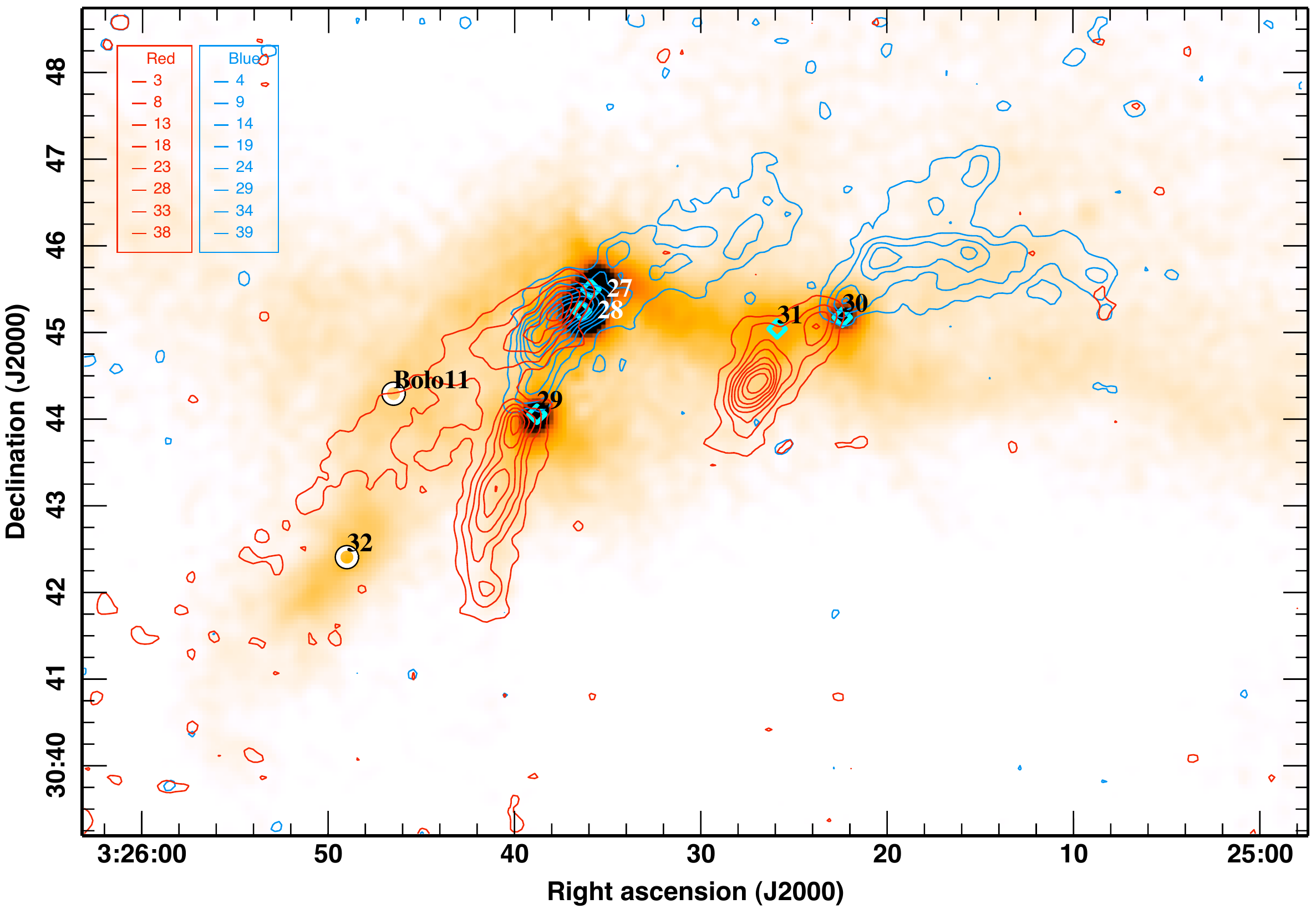}
\includegraphics[width=0.49\textwidth]{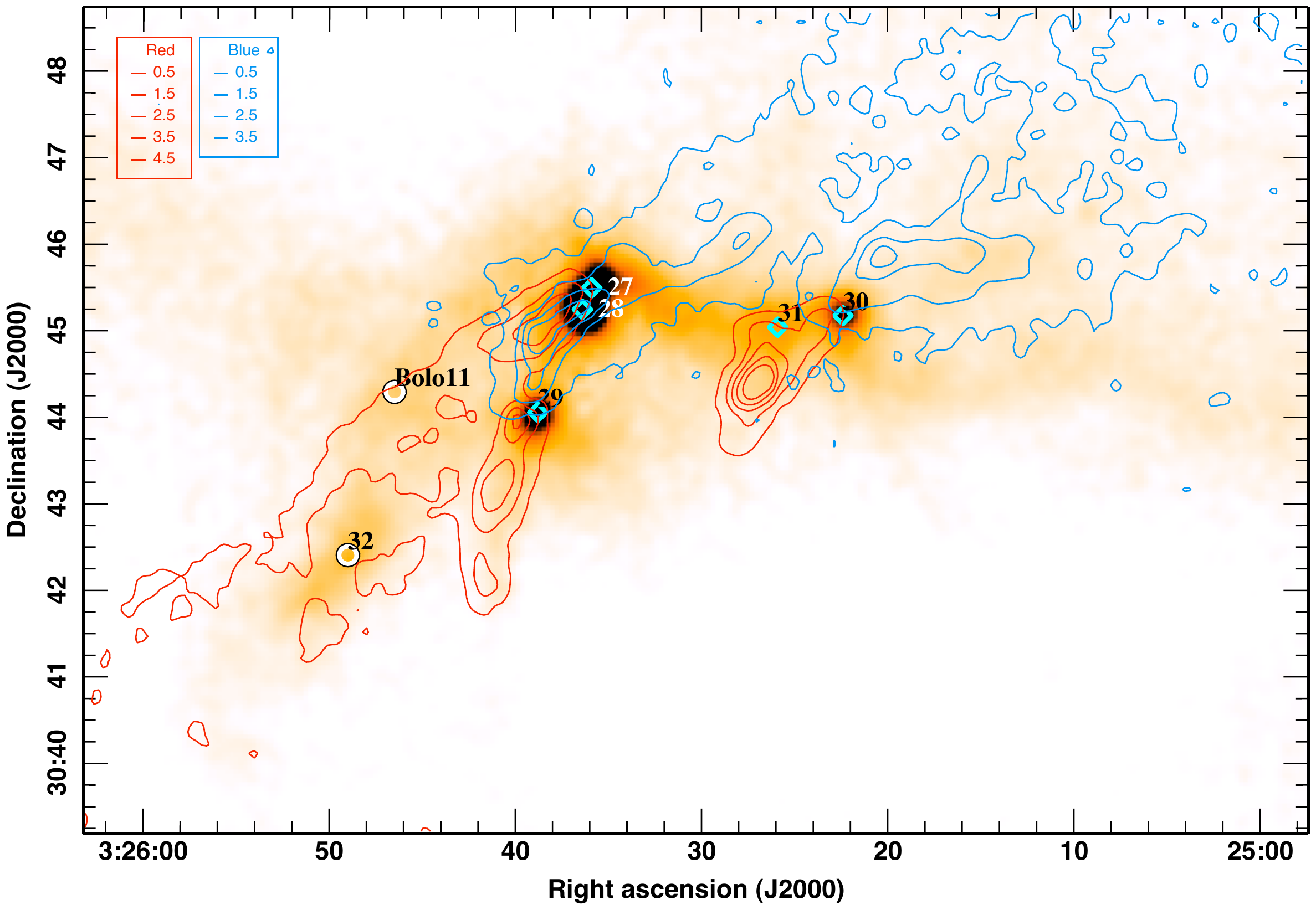}
\includegraphics[width=0.49\textwidth]{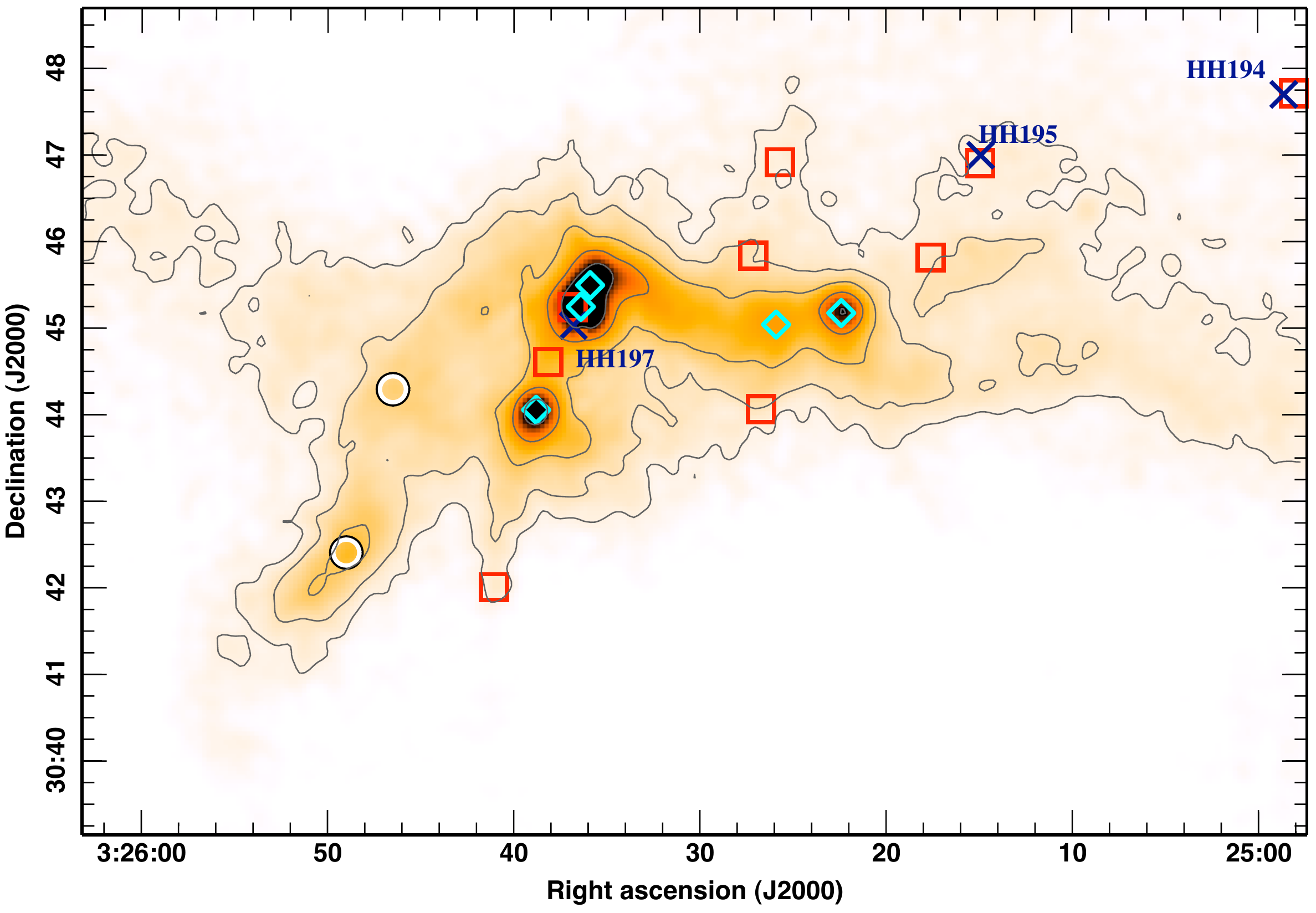}
\caption{CO \threetotwo\ outflows in L1448 as for
  Fig.\ \ref{fig:ngc1333_outflows}. Top: High-velocity outflows, integrated from $-25$ to
$0$\,\kms\ (blue) and 8 to 25\,\kms\ (red). Middle: Low-velocity outflows, integrated from  $-5$ to $2$\,\kms\ (blue) and 7 to 14\,\kms\ (red).}
\label{fig:l1448_outflows} 
\end{center}
\end{figure}

In Fig.\ \ref{fig:l1448_outflows}, we present our CO outflows. L1448C (HRF29)
drives the highly collimated flow first discovered by
\citet{bachiller90}: one of the youngest
and highest velocity flows known at the time (out to $\pm
70$\,\kms, \citealp{bachiller90,bachiller95}). A chain of \htwo\ knots extends
$\sim2$\,arcmin south of L1448C \citep{eisloffel00}. Towards the
north, the blue-shifted flow possibly continues beyond the other peaks, deflected by the ammonia core containing
the protobinary L1448N:A/B \citep{curiel99} terminating at HH267 (to
the north and west of our map, \citealp{wolf-chase00}). The second outflow
cuts across L1448C's and is associated with multiple Class 0
objects: L1448N:A/B (HRF28) and L1448NW (HRF27). Previously, one (from
L1448N:B \citealp{bachiller90,bachiller95}), two (from L1448N:A and B
\citealp{wolf-chase00}) or
three flows \citep{eisloffel00,davis95} have been suggested for this area, with sensitivity and
angular resolution a severe limitation. Our maps do not
add anything to this discussion;
one or more of the sources drives an outflow
with PA 140\,deg. Interferometric observations by
\citet{kwon06} find two flows: from L1448N:A at PA 155\,deg and from
L1448N:B at 110\,deg. In channel maps there is a hint of two
distinct flows, the red-shifted emission between 7 and 12\,\kms\ shows separate trails originating at
the L1448N core. 

The IRS2 core (HRF30 and HRF31) is thought to drive two outflows
\citep{wolf-chase00}. This is implied by the `V'-shaped blue lobe, although red counterparts are not
obvious. HRF31 seems to drive the red lobe oriented north-south
but it has no symmetric blue lobe. \citet{davis08} find
only one flow from this core, encompassing this red lobe out to HH195. They suggest the doubted protostar, IRS1, also
powers a roughly parallel outflow, reaching to
HH194. Our observations indicate that IRS1 is not a young protostar
and the outflowing gas nearby comes from flows powered by
two Class 0 objects in IRS2. A plausible morphology is 
HRF31 driving an outflow at a PA of 100\,deg and HRF30 one at 130\,deg. 

\subsection{L1455}

\begin{figure} 
\begin{center}
\includegraphics[width=0.49\textwidth]{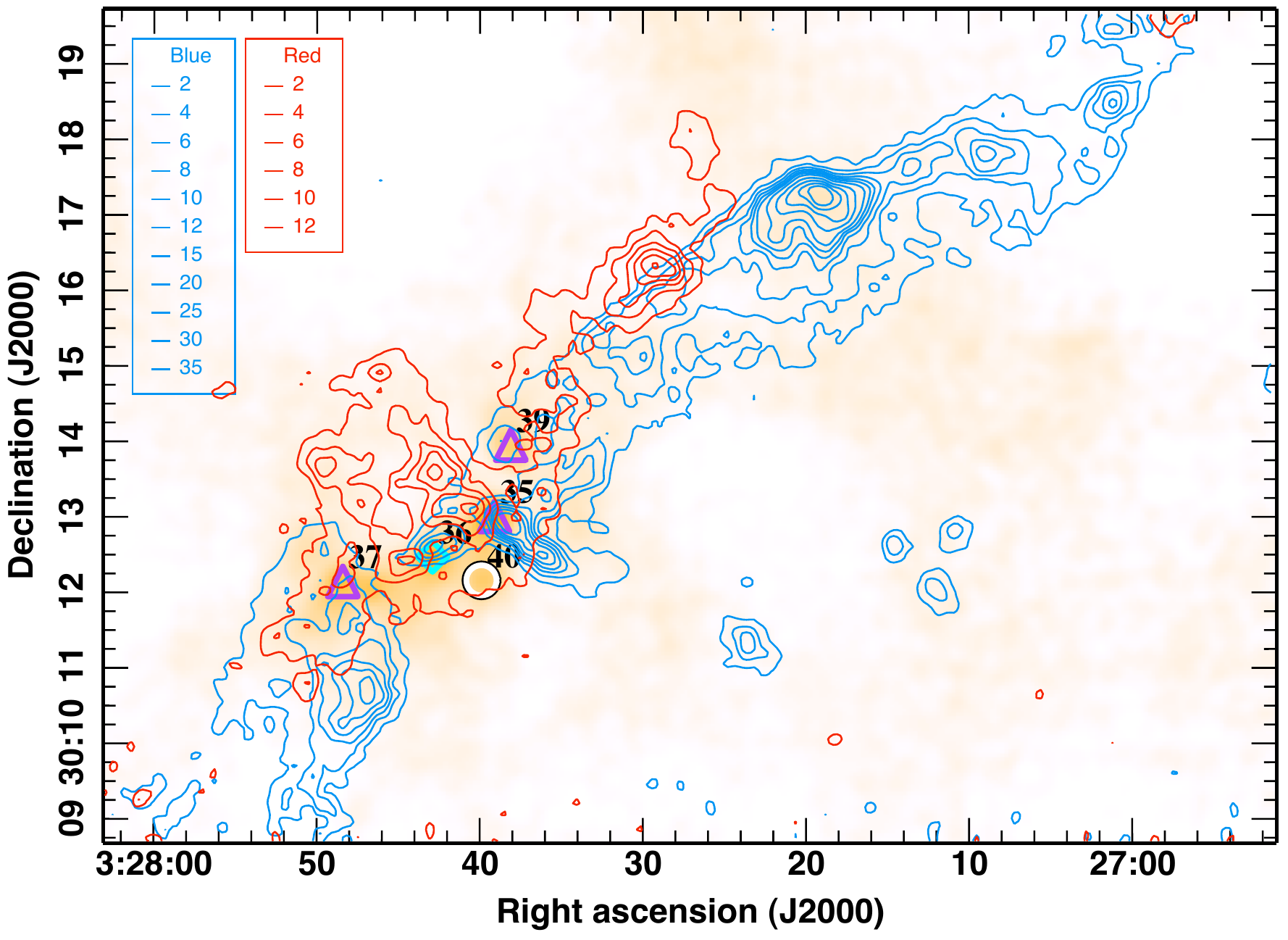}
\includegraphics[width=0.49\textwidth]{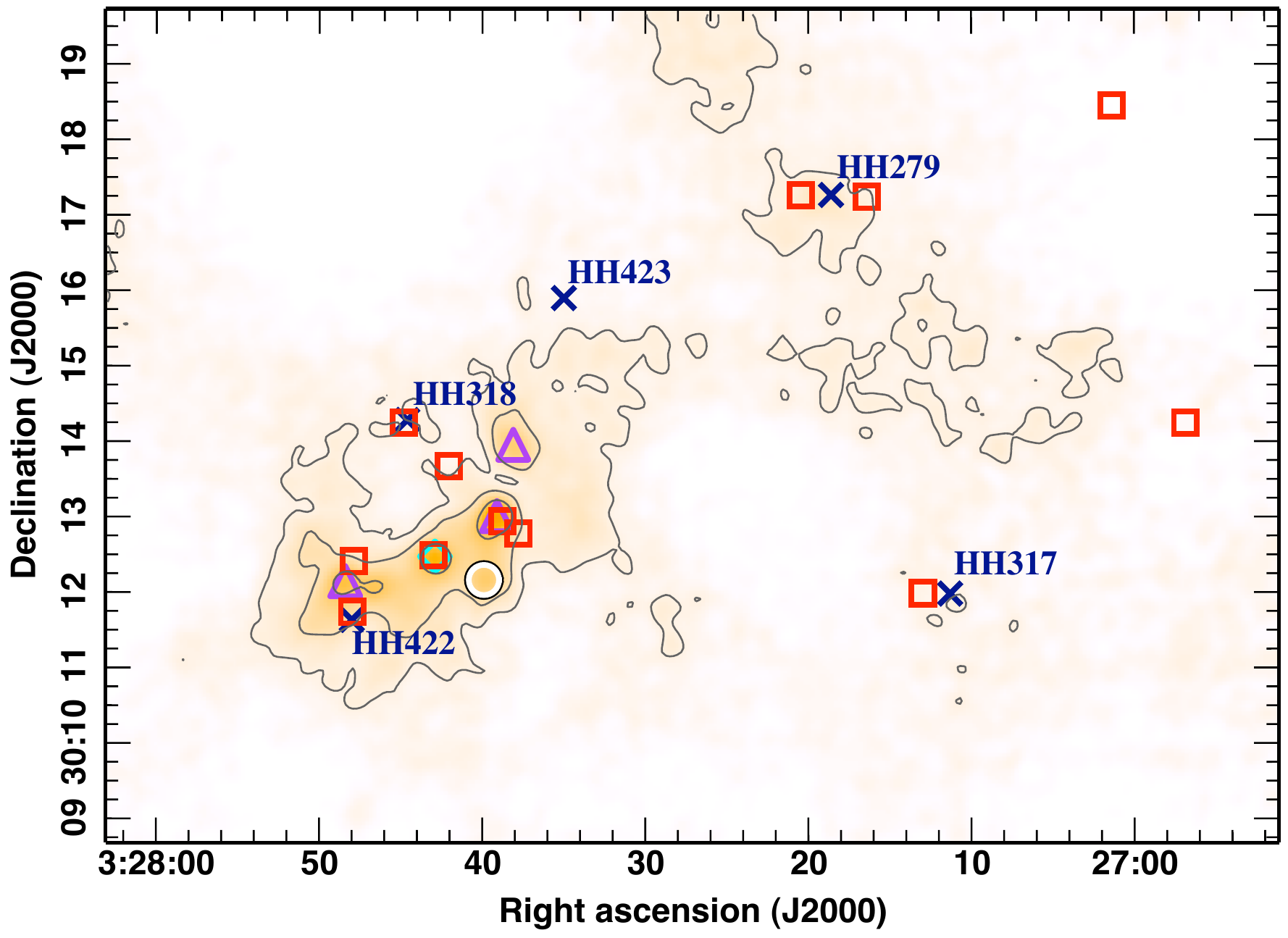}
\caption{CO \threetotwo\ outflows in L1455 as for
  Fig. \ref{fig:ngc1333_outflows}. Upper panel: integration from $-5$ to
$3$\,\kms\ (blue) and 9 to 16\,\kms\ (red). }
\label{fig:l1455_outflow} 
\end{center}
\end{figure}

Outflowing CO gas was first detected in L1455 by
\citet{frerking82b} with the \onetozero\ line later mapped by
\citet{goldsmith84} and \citet{levreault88}. Subsequently,
\citet{davis97} mapped a small area around one of the protostars,
RNO15-FIR in CO \threetotwo\ with the \jcmt, while
\citet{bally97} examined the whole region optically. Our outflows are depicted in Fig.\ \ref{fig:l1455_outflow}. All four protostars (HRF35-39) appear to be
driving outflowing gas of some description. Starless HRF40 has
red-shifted gas in its vicinity which does not have a distinct outflow
shape, so may just be confused with other flows. 

The brightest \scuba\ core is HRF35 (RNO15-FIR), a Class I protostar
identified in the \emph{IRAS} point-source catalogue (IRAS
02245+3002). It drives
the clearest bipolar outflow in the map (PA 42\,deg) which
\citet{davis97} found to have subtle variations in direction possibly due to the
presence of a binary source. HRF36 drives an orthogonal
outflow. However, the rest of the map is complicated: there is a wide,
nearly north-south outflow from HRF37, but
most perplexing is the outflow in the north-west which seems to require
a driving source far from the existing protostars. This
difficulty was noted by \citet{levreault88}, who found no
candidate and derived very low luminosity limits on such a source if
it exists. Subsequent submillimetre and IR surveys have not increased
the number of known protostars greatly; \citet{jorgensen07} and
\citet{hatchell07a} both found four protostars clustered around the three IR sources 
\citeauthor{levreault88} knew. The brightest \htwo\ emission and HH279 both tally with this flow and it would seem to be driven by
a low luminosity source or is a flow from
the known protostars to the south-east (probably from either HRF37 or
HRF39). This latter explanation is promoted
by \citet{davis08} who correlated the \wfcam\ \htwo\ images with the outflows in the CO \threetotwo\ survey of \citetalias{hatchell07b},
although their CO data do not extend so far north. 

\section{Spectra and parameters of individual outflows}
\label{appendix:outflowspectra}

The full version of this appendix is available as Supporting
Information to the online version of this article.

\begin{figure*}
\begin{center}
\begin{tabular}{lll}
\includegraphics[height=0.12\textheight]{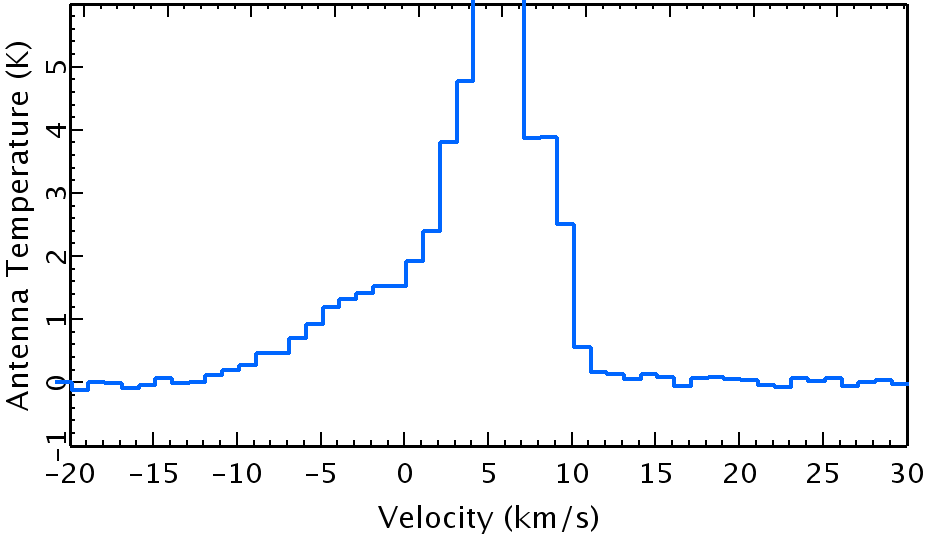} &
\includegraphics[height=0.12\textheight]{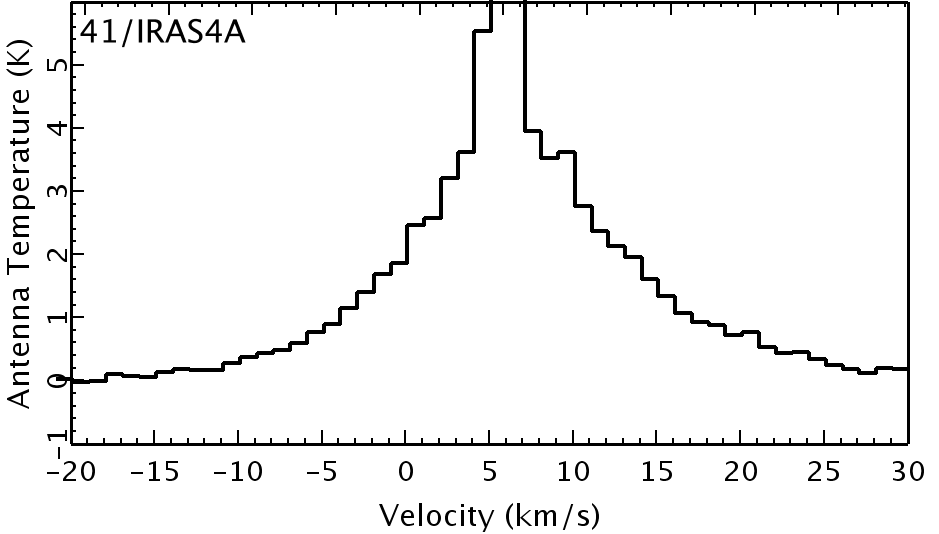} &
\includegraphics[height=0.12\textheight]{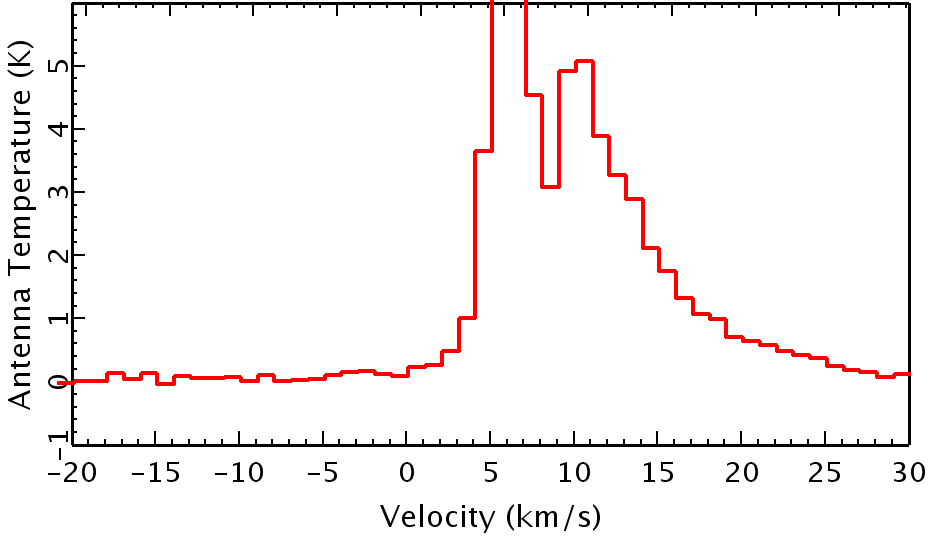} \\
\includegraphics[height=0.12\textheight]{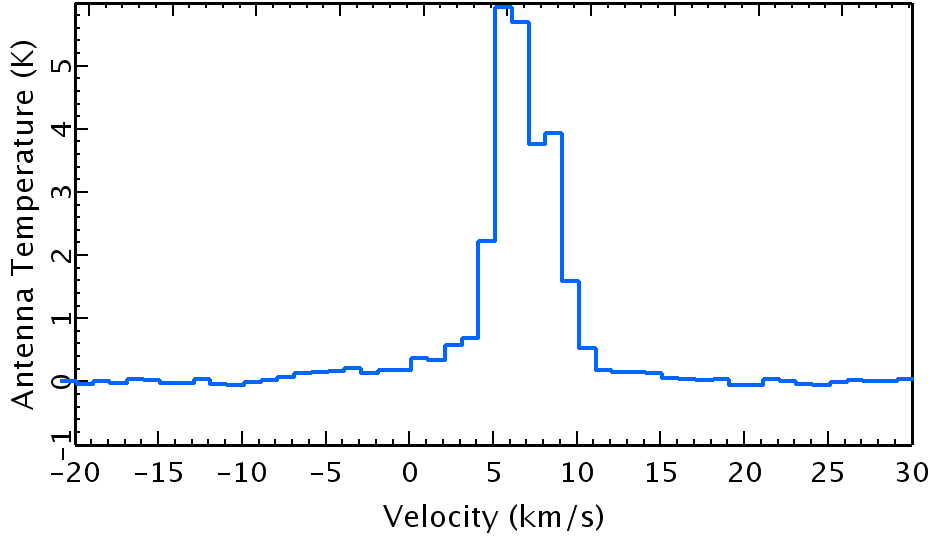} &
\includegraphics[height=0.12\textheight]{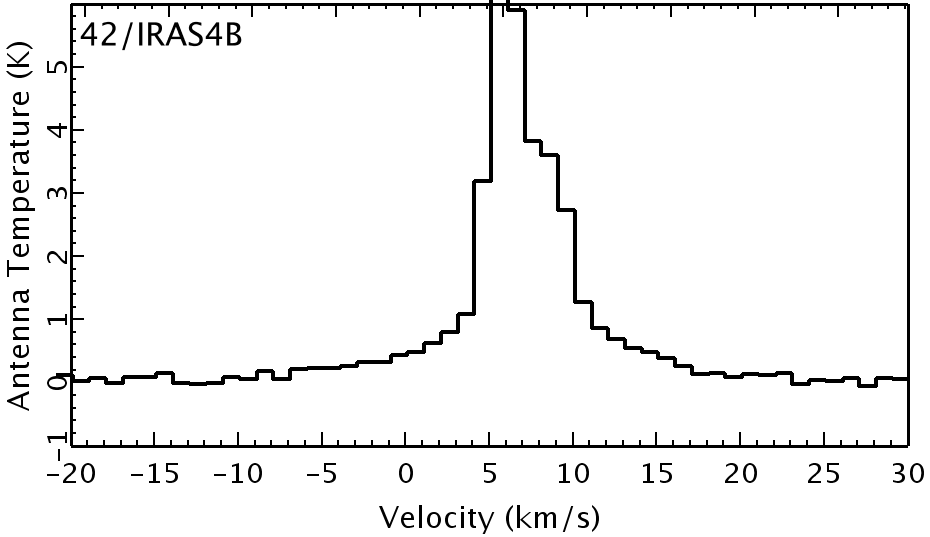} &
\includegraphics[height=0.12\textheight]{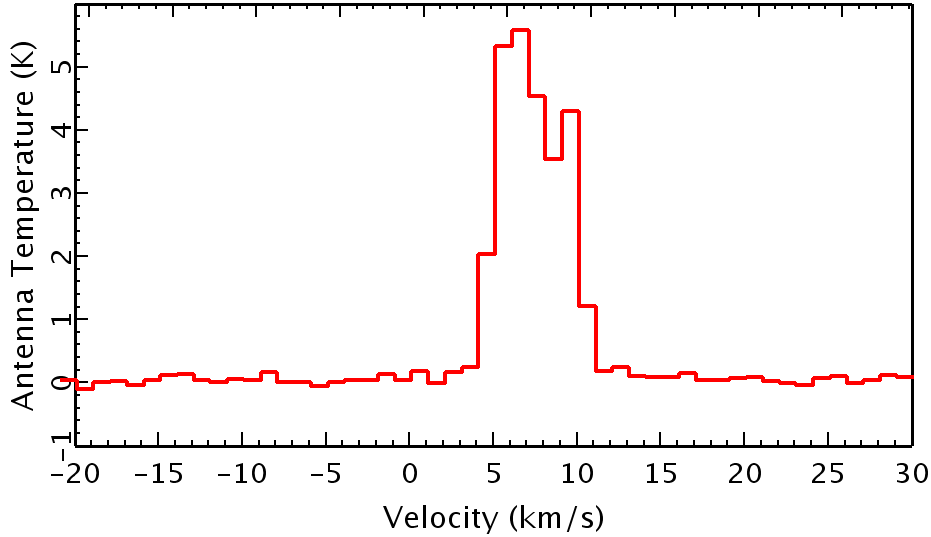} \\
... \\
\end{tabular}
\caption{Spectra of outflows in \ngc\ for sources in Hatchell et
  al.\ (2007)a. Spectra are towards the core peak (black) and the peak
  of the integrated outflow emission (from $v_0 + 2$ to $v_0 +
  v_\mathrm{max}$\,\kms; values of $v_0$ and $v_\mathrm{max}$ are
  given for each outflow in Table \ref{table:outflow_params1}) towards the red-shifted and
  blue-shifted outflow lobes (red and blue spectra respectively), unless distinct
  lobes are not apparent. Similar spectra for all the SCUBA cores
  examined are available as Supporting Information to the
online version of this article.}
\label{fig:outflows_ngc1333}
\end{center}
\end{figure*}

\begin{table*}
\caption{Observational parameters of the identified outflows. The full version of this table is available as Supporting Information to the
online version of this article.}
\label{table:outflow_params1}
%\begin{small}
\begin{tabular}{lccccccccc}
\hline
No. & Driving~$^\mathrm{a}$  & Class~$^\mathrm{b}$ &  $v_0$~$^\mathrm{c}$  & Lobes?~$^\mathrm{d}$ & Axis PA & \multicolumn{2}{c}{Lobe Length~$^\mathrm{e}$} & \multicolumn{2}{c}{$v_\mathrm{max}$ (\kms)~$^\mathrm{f}$} \\

 & Source & & (\kms) & & E of N & R & B & R & B \\
\hline
\multicolumn{10}{c}{\textbf{--- NGC\,1333 ---}}\\
1 & HRF41 & 0 & 7.11 & 2 & 37 & 260 & 145 & 27 & 26 \\      
2 & HRF42 & 0 & 7.52 & 2 & 56 & 169 & 33 & 8 & 14 \\         
3 & HRF43 & I & 8.24 & 2 & 143 & 112 & 115 & 31 & 32 \\      
4A$^1$ & HRF44 & 0 & 7.62 & 2 & 111 & 103 & 131 & 22 & 28\\ 
4B$^1$ & HRF44 & 0 & 7.62 & 2 & 18 & 120 & 460 & 32 & 20 \\ 
... \\
\hline
\end{tabular}\\
\begin{flushleft}
$^\mathrm{a}$~Driving source from \citet{hatchell07a}.\\
$^\mathrm{b}$~Source classification \citep{hatchell07a}: 0=Class 0 and I=Class
I. \\
$^\mathrm{c}$~Source \ceighteeno\ \threetotwo\ velocity where
available or the map median where not. \\
$^\mathrm{d}$~Number of lobes: R for red- and B for blue-shifted
only. \\
$^\mathrm{e}$~Maximum CO lobe length. \\
$^\mathrm{f}$~Maximum CO velocity (relative to $v_0$). \\
$^1$~Perpendicular outflows from the IRAS2A protobinary: A approximately east-west and B north-south.  \\
\end{flushleft}
%\end{small}
\end{table*}

\begin{table*}
\caption{Derived outflow parameters, see text for details. The full version of this table is available as Supporting Information to the
online version of this article.}
\label{table:outflow_params2}
\begin{tabular}{lcccccc}
\hline
Driving & $M_\mathrm{out}$ & $p_\mathrm{out}$~$^\mathrm{a}$ & $E_\mathrm{out}$~$^\mathrm{b}$ & $\tau_\mathrm{d}$~$^\mathrm{c}$ & $F_\mathrm{CO}$~$^\mathrm{d}$ & $L_\mathrm{out}$~$^\mathrm{e}$\\
Source & ($10^{-2}$~M$_\odot$) & ($10^{-1}~$M$_\odot$~km~s$^{-1}$) &
(10$^{36}$~J) & (10$^{4}$~yr) &
(10$^{-5}$~M$_\odot$~km~s$^{-1}$~yr$^{-1}$) & (10$^{-2}$~L$_\odot$)\\
\hline
\multicolumn{7}{c}{\textbf{--- NGC1333 ---}}\\
HRF41 &     7.05 & 5.64 & 12.05 & 0.93 & 6.07 & 10.57\\
HRF42 &     1.13 & 0.58 & 0.73 & 1.11 & 0.53 & 0.53\\
HRF43 &     36.30 & 30.79 & 69.32 & 0.44 & 70.29 & 128.91\\
HRF44A & 9.37 & 5.65 & 9.85 & 0.57 & 9.93 & 14.11\\
HRF44B & 42.16 & 26.05 & 41.00 & 1.36 & 19.22 & 24.64\\
... \\
\hline
\end{tabular}\\
\begin{flushleft}
$^\mathrm{a}$~Momentum. \\
$^\mathrm{b}$~Kinetic energy. \\
$^\mathrm{c}$~Dynamical timescale. \\
$^\mathrm{d}$~Momentum flux. \\
$^\mathrm{e}$~ Mechanical luminosity, i.e.\ $E_\mathrm{out}/\tau_\mathrm{d}$.
\end{flushleft}
\end{table*}

\label{lastpage}

\end{document}